\documentclass[10pt]{article}

\def\hybrid{\topmargin -20pt    \oddsidemargin 0pt
        \headheight 0pt \headsep 0pt
        \textwidth 6.25in       
        \textheight 9.5in       
        \marginparwidth .875in
        \parskip 5pt plus 1pt   \jot = 1.5ex}

\hybrid
\usepackage{amsmath}
\usepackage{bbm}
\usepackage{rotate}
\usepackage{rotating}
\usepackage[hypertex]{hyperref}

\makeatletter \@addtoreset{equation}{section} \makeatother

\newcommand{\be}{\begin{equation}}
\newcommand{\ee}{\end{equation}}
\newcommand{\bea}{\begin{eqnarray}}
\newcommand{\eea}{\end{eqnarray}}
\newcommand{\rmE}  {{\rm{E}}}

\newcommand{\refeq}[1]{(\ref{#1})}

\newcommand{\rmT}  {{\rm{T}}}
\newcommand{\Xb} {{\Bar{X}}}
\newcommand{\Yb} {{\Bar{Y}}}
\newcommand{\Fb} {{\Bar{F}}}

\newcommand{\zb} {{\Bar{z}}}
\newcommand{\lambdaB} {{\Bar{\lambda}}}
\newcommand{\chiB} {{\Bar{\chi}}}
\newcommand{\etaB} {{\Bar{\eta}}}
\newcommand{\epsilonB} {{\Bar{\epsilon}}}
\newcommand{\es}{{\xi}}
\newcommand{\esB}{{\Bar{\xi}}}
\newcommand{\esv}{{\eta}}
\newcommand{\esvB}{{\Bar{\eta}}}

\renewcommand{\Vec}[2]{\begin{pmatrix} {#1}\\{#2} \end{pmatrix}}
\newcommand{\VecT}[2]{\begin{pmatrix} {#1} & {#2} \end{pmatrix}}
\newcommand{\Matrix}[4]{\begin{pmatrix} {#1} & {#2} \\ {#3} & {#4}
 \end{pmatrix}}
\newcommand{\Cases}[2]{\left\{\begin{matrix} {#1}\\[.1in]{#2}
 \end{matrix}\right.}

\renewcommand {\slash}[1]{{#1}\!\!\!/}
\newcommand   {\ds}      {\slash{\partial}}

\newcommand{\forth}{\frac{1}{4}}
\newcommand{\half} {\frac{1}{2}}
\newcommand{\Imag} {{\Hat{\mbox{\rm i}}}}
\newcommand{\id}   {{\mathbbm{1}}}
\newcommand{\Lag}{\mathcal{L}}
\newcommand{\lb}{\lambdaB}
\newcommand{\eb}{\Bar{\epsilon}}
\newcommand{\gmu}{\gamma_\mu}
\newcommand{\ComplexI}{i}
\newcommand{\cN}{\mathcal{N}}

\newcommand{\amet}{a_{IJ}(\sigma)}
\newcommand{\Com}{\mathbbm{C}}
\newcommand{\Hom}{\mathbbm{H}}
\newcommand{\Kom}{\mathbbm{K}}
\newcommand{\Zom}{\mathbbm{Z}}
\newcommand{\R}{\mathbbm{R}}
\newcommand{\Som}{\mathbbm{S}}
\newcommand{\umat}{\id}

\newcommand{\rE}{{\rm E}}
\newcommand{\ft}[2]{{\textstyle\frac{#1}{#2}}}
\def\der{\partial}
\newcommand{\mscr}[1]{\mbox{\scriptsize #1}}

\newcommand{\CC}{{\sf C}}

\let\Bbb\relax
\newfont{\Bb }{msbm10 scaled 1000}
\newfont{\Bbb}{msbm10 scaled 1200}
\newfam\eufam%
\font\euzw=eufm10 scaled 1200%
\font\euac=eufm9%
\textfont\eufam=\euzw\scriptfont\eufam=\euac%
\renewcommand{\gg}{\mathfrak{g}} 

\def\bt{\begin{thm}}
\def\et{\end{thm}}
\def\bp{\begin{prop}}
\def\ep{\end{prop}}
\def\bc{\begin{cor}}
\def\ec{\end{cor}}
\def\bl{\begin{lemma}}
\def\el{\end{lemma}}
\def\bd{\begin{dof}}
\def\ed{\end{dof}}

\def\square{\kern1pt\vbox
                 {\hrule height 0.6pt\hbox{\vrule width 0.6pt\hskip 3pt
      \vbox{\vskip 6pt}\hskip 3pt\vrule width 0.6pt}\hrule height 
0.6pt}\kern1pt}

\def\pf{\noindent{\it Proof:\ }}
\def\qed{\hfill\square}

\def\a{\alpha}

\def\g{\gamma}

\def\o{\omega}
\def\q{\theta}

\def\G{\Gamma}

\def\O{\Omega}

\newfont{\mcal}{eusm10 scaled \magstep1}
\def\ca{{\cal A}}

\newfont{\goth}{eufm10 scaled \magstep1}

\def\gg{\mbox{\goth g}}

\def\gu{\mbox{\goth u}}

\def\n{\nabla}
\def\Re{\mbox {\rm Re\,}}
\def\Im{\mbox {\rm Im\,}}
\newtheorem{thm}{Theorem}
\newtheorem{prop}{Proposition}
\newtheorem{cor}{Corollary}
\newtheorem{lemma}{Lemma}
\newtheorem{dof}{Definition}

\def\p{\partial}

\def\ol{\overline}
\def\ra{\rightarrow}

\def\be{\begin{equation}}
\def\ee{\end{equation}}

\def\re#1{(\ref{#1})}
\def\arr{\begin{array}{rlll}}
\def\ea{\end{array}}
\def\bea{\begin{eqnarray}}
\def\eea{\end{eqnarray}}

\begin{document}

\begin{titlepage}
\begin{center}

\hfill hep-th/0312001\\
\hfill FSU-TPI-12/03\\
\vskip 1cm {\large \bf  Special Geometry of Euclidean
Supersymmetry I: Vector Multiplets}\footnote{Work supported by the 
`Schwerpunktprogramm Stringtheorie' of the 
DFG.}

\vskip .5in

{\bf Vicente Cort\'es} \\

{\em Institut de Math\'ematiques \'Elie Cartan\\ 
Universit\'e Henri Poincar\'e - Nancy I, 
B.P.\ 239, F-54506 Vand{\oe}uvre-l\`es-Nancy, France} \\

\vskip 0.5cm

{\bf Christoph Mayer, Thomas Mohaupt and Frank Saueressig}  \\

{\em Theoretisch-Physikalisches Institut,
Friedrich-Schiller-Universit\"{a}t
Jena, Max-Wien-Platz 1, \\ D-07743 Jena, Germany}\\
\vskip 0.5cm

\end{center}
\vskip 1.5cm

\begin{center} {\bf ABSTRACT } \end{center}

\noindent 
We construct the general action for Abelian vector multiplets
in rigid 4-dimensional Euclidean (instead of Minkowskian) 
${\cal N}=2$ supersymmetry, {\it i.e.},\ 
over space-times with a positive definite instead of a Lorentzian metric. 
The target manifolds for the scalar fields turn out to be 
para-complex manifolds endowed with a particular kind of special geometry, 
which we call affine special para-K\"ahler geometry. 
We give a precise definition and develop the mathematical theory
of such manifolds. 
The relation to the affine special 
K\"ahler manifolds appearing in Minkowskian ${\cal N}=2$ 
supersymmetry is discussed. Starting from the general 
5-dimensional vector multiplet action we consider 
dimensional reduction over time and space in parallel,
providing a dictionary between the resulting Euclidean 
and Minkowskian  theories.
Then we reanalyze supersymmetry in four dimensions and find
that any (para-)holomorphic prepotential defines a supersymmetric
Lagrangian, provided that we add a specific four-fermion term,
which cannot
be obtained by dimensional reduction.
We show that the Euclidean action and supersymmetry
transformations, when written in terms of para-holomorphic coordinates,
take exactly the same form as their Minkowskian counterparts.
The appearance of a para-complex and complex structure in the 
Euclidean and Minkowskian theory, respectively, is traced back
to properties of the underlying R-symmetry groups.
Finally, we indicate how our work will be extended to other types of multiplets
and to supergravity in the future and 
explain the relevance of this project for the study of instantons,
solitons and cosmological solutions in supergravity and M-theory.
\end{titlepage}

\tableofcontents

\begin{section}{Introduction, Summary, Conclusions and Outlook}

\subsection{Introduction}

Most of the knowledge about the non-perturbative properties
of string theory and M-theory relies on dualities, which re-interpret
the strong coupling behaviour of particular limits of M-theory
in terms of a dual, weakly coupled 
description \cite{Pol}. 
Although string dualities have passed various highly non-trivial
tests, they still have the status of conjectures. Moreover, they 
address non-perturbative physics only indirectly. 
Therefore the further development of non-perturbative methods
in string and M-theory is desirable. In particular one would like
to have the analogue of the instanton calculus used in gauge theories.
The first important step in this direction was the discovery of
the D-instanton \cite{D-Inst} in IIB string theory. It was then
realized that instanton effects in M-theory compactifications  
correspond to 
Euclidean wrappings of p-branes on (p+1)-cycles of the 
internal manifold \cite{BBS} (see  also \cite{Kir} for a review and 
more  references). 
Similar to string and M-theory solitons, these instantons 
can be described in terms of the low energy effective supergravity 
action. In this formulation they are instanton solutions, 
{\it i.e.}, solutions of the Euclidean theory which have a finite action. 
The IIB supergravity solution corresponding to the D-instanton
was found in \cite{GibGrePer}. 
There are other solutions corresponding 
to  instantons in Calabi-Yau compactifications in type II
string theory \cite{MohEtAl,TheVand} and in 11-dimensional 
M-theory \cite{GutSpa}.

Let us outline the problems which we will address in this
paper. One particular question arising in the context 
of D-instantons and their generalizations is how to 
define the Euclidean supergravity 
action. In their supergravity description of the D-instanton
\cite{GibGrePer} G.W. Gibbons, M.B. Green and M.J. Perry
invoked  an elegant but somewhat mysterious rule, which requires to replace
factors of $i$ in the Lagrangian and in the supersymmetry rules
by a formal factor  $e$, satisfying
$\overline{e} = -e$ and $e^2 = 1$. (The  formal factor $e$ should be 
interpreted as an imaginary unit in the algebra of para-complex
numbers.) 
The same Euclidean Lagrangian was obtained in \cite{GreGut}
by first Hodge-dualizing the IIB-axion into
a tensor field, then performing a Wick rotation, and dualizing 
 the tensor field back to a scalar afterwards. Due to properties
of the Hodge star operator in Euclidean signature, the Euclidean
action with a tensor field is definite, {\it i.e.}, \ bounded from above (or 
from below, depending on the choice of overall sign), while the dual action 
with a scalar field is indefinite.\footnote{Notice that an indefinite
target metric for the scalar fields implies an indefinite action. The 
converse is not true. Equivalently, a definite action implies a
(positive or negative) definite target metric.}    
The D-instanton can be described in terms of both 
actions, but the use of the indefinite action makes it
particularly simple to find explicit solutions  
\cite{GibGrePer}.  
Of course, instanton corrections
should be computed using the  definite version of the action, 
as was recently emphasized by \cite{TheVand} in the context of
type II Calabi-Yau compactifications.

String and M-theory instantons and solitons are 
related to one another by dimensional
reduction and T-duality transformations. In fact,
instanton solutions can be used
to generate solitons by `dimensional oxidation' \cite{NullSigma}, see
\cite{Stelle} for a review.
In this approach
one compactifies all world-volume directions of the brane, including
time.
The dimensional reduction over time has the effect
that the metric of the scalar manifold becomes indefinite, see formulae
(\ref{FlatReal}), (\ref{DRtime}) below.
Instanton solutions are given by harmonic maps from the transverse space 
of the brane
into  totally geodesic and totally isotropic 
submanifolds (such as null geodesics) of the scalar manifold. They can be 
re-expressed in terms of the quantities of the original higher-dimensional 
theory to obtain the corresponding soliton. This technique goes
by the name of dimensional oxidation. Obviously one needs to know
explicitly how the fields of the lower-dimensional theory are
related to those of the higher dimensional one. This is one
of the reasons for studying Euclidean theories using 
dimensional reduction over time.

Finally, scalar manifolds of the same type as in Euclidean theories
also occur in non-standard
Minkowski signature supergravity and string theories, 
which are obtained by T-duality transformations over time.
Particular examples are the type II${}^*$ string 
theories introduced in \cite{HullCosmo}. These theories have interesting 
cosmological solutions, which are supersymmetric and asymptotically 
de Sitter.\footnote{However, as discussed in \cite{HullCosmo},
unitarity and stability of these theories 
are unclear.}
In \cite{BehCve} it was shown that asymptotic de Sitter solutions 
of non-standard gauged supergravities in 4 and 5 dimensions
can be obtained by massive time-like T-duality transformations
from instanton solutions of ungauged supergravity. These solutions
are believed to be related to type II${}^*$ supergravity theory by 
dimensional reduction.

The geometrical structures of the scalar manifolds appearing in 
the above examples deserve a closer analysis. For example,
in \cite{GibGrePer} 
the $i \rightarrow e$ substitution rule
gives the desired result, a description of the
D-instanton in terms of supergravity, but it leads to the immediate
question: {\em what
is the geometrical meaning of the $i \rightarrow e$ substitution
rule?} More generally, we can ask what characterizes 
the geometries corresponding
to supersymmetric theories obtained by time-like dimensional reduction 
and time-like T-duality. In theories with 32 or 16 supercharges
the scalar manifolds are symmetric spaces, which are fixed 
uniquely by the matter content. These manifolds can be found
case by case \cite{CreEtAl,HulJul}. The scalar geometry becomes
richer when reducing the number of supersymmetries.
In this paper we will consider theories which have
${\cal N}=2$ supersymmetry, {\it i.e.},  eight real supercharges.\footnote{
We are counting in units of 4-dimensional Minkowskian
supersymmetry. Note that some of the theories we refer to
as ${\cal N}=2$ for terminological convenience are actually 
the minimal supersymmetric theories in their dimension and
signature.} 
The scalar manifolds of ${\cal N}=2$ and ${\cal N}=1$ 
theories are not fixed by the matter content. But whereas
the scalar manifolds of ${\cal N}=1$ theories are 
(for Minkowskian space-time) generic K\"ahler manifolds,
those of ${\cal N}=2$ are subject to more specific
conditions.
The resulting geometries are therefore
called {\em special geometries}. The precise type of geometry
depends on (i) whether supersymmetry is a rigid or
local symmetry, (ii) the type of supermultiplet and 
(iii) the number of space-time dimensions. We refer
to \cite{VanPro} 
for an overview of special geometries and their mutual relations. 
In this context we can now
rephrase our above question as follows:
{\em what are the special geometries of 
${\cal N}=2$ theories with Euclidean signature?} 
The present paper is the first in a series
which will answer this question. 
For technical simplicity we will start with rigid supersymmetry
and develop the geometry of Euclidean ${\cal N}=2$ vector
multiplets in 4 dimensions. In subsequent work we will extend this
to other multiplets and dimensions as well as to supergravity. 
The results will be used to study
instantons and solitons in string and M-theory compactifications.

There are two methods which can be applied to find a Euclidean
action. The first method is the dimensional reduction of a 
higher-dimensional Minkowskian action over time. This has 
been used in \cite{blau,BelVanVan} for 4-dimensional
${\cal N}=2$ Yang-Mills theory and in \cite{CreEtAl} for
Euclidean supergravity theories. The second approach is 
to continue the 4-dimensional Minkowskian theory analytically
to imaginary time. Here there are
two different versions. The first version was used in 
\cite{Schwinger,zumino}
to construct 4-dimensional Euclidean gauge theories. More 
recently, this construction has been
generalized to a continuous Wick rotation by \cite{Mehta,vN,vN2}.
As in the treatment of IIB supergravity
in \cite{GibGrePer}, one obtains 
an indefinite supersymmetric Euclidean action.
Moreover, the action obtained by 
the continuous Wick rotation agrees with the one obtained by
dimensional reduction over time \cite{blau,BelVanVan}. 
The second method using analytic continuation is based on the 
Osterwalder-Schrader formulation of Euclidean field theories,
and has been studied for 4-dimensional 
supersymmetric theories in \cite{Nicolai}. 
In this approach one uses reality constraints
which differ from those in the continuous Wick rotation,
as for fermions Hermitian 
conjugation is combined with Euclidean time reflection. Moreover,
the Euclidean action is definite. The Euclidean
actions obtained in the Osterwalder-Schrader approach are certainly
the correct actions to be used in the path integral quantization
and in lattice studies
of supersymmetric field theories. However, they are not suitable
for the applications we are interested in. To find  
generalizations of D-instantons we need to study the Euclidean
versions of multiplets which fundamentally are vector-tensor 
multiplets rather than vector multiplets, in analogy to the 
situation in IIB string theory discussed above.
In order to construct solitons by dimensional oxidation, we need 
Euclidean theories which are obtainable from higher-dimensional
theories by dimensional reduction over time. 
The relation between the two types of Euclidean continuations
of supersymmetric actions has been discussed in \cite{vN,Waldron},
and we plan to further analyze it in a future publication.

In this paper we construct the action of 4-dimensional
Euclidean vector multiplets by dimensional 
reduction.\footnote{A  generalization of the continuous Wick 
rotation of \cite{vN}
will be discussed in a separate paper.}
We consider the reduction of the general
5-dimensional action for Abelian vector multiplets 
over a time-like and a space-like dimension in parallel.
This way we obtain a dictionary between the 4-dimensional
theories in both signatures, and we can compare with 
the results \cite{DLP,DDKV} for vector multiplets in Minkowski
signature.
The action obtained by dimensional reduction is not the most general one.
In 5 dimensions the presence of a Chern-Simons term forces
the prepotential, which encodes the whole Lagrangian,
to be a cubic polynomial. This constraint is absent in 
4 dimensions, where the only condition is that the
prepotential  is a (para-)holomorphic function. The 
general Lagrangian is obtained by reanalyzing the supersymmetry
transformations for a general (para-)holomorphic prepotential,
with the result that a particular four-fermion term has to
be added. In the Minkowskian case the resulting Lagrangian 
agrees with the general vector multiplet Lagrangian
of \cite{DLP,DDKV}.\footnote{We only consider Lagrangians 
which contain at most second derivatives of the fields and
no terms of order higher than four in the fermions.}

Our action for 4-dimensional
Euclidean vector multiplets is real but 
indefinite. Similar to the case of type IIB supergravity discussed 
above \cite{GibGrePer,GreGut}, a dual action, which is 
both real and definite can be obtained by Hodge-dualizing the ${\cal N}=2$
vector multiplets into ${\cal N}=2$ 
vector-tensor multiplets \cite{Soh}. As is known from 
\cite{Sie}, this dualization 
is only possible for a restricted class of vector multiplet couplings.
However, for the applications we have in mind
it is guaranteed that the dualization is possible, because the vector 
multiplets relevant for string instantons are those which have been
obtained by dualizing vector-tensor multiplets.
The most important case is 
heterotic ${\cal N}=2$ compactifications, where the dilaton
sits in a vector-tensor multiplet, which is then 
dualized into a vector multiplet \cite{DKLL}. The dilaton controls
the quantum corrections to the vector multiplet part of the effective 
action. So far only perturbative corrections have been computed
directly, but the instanton corrections 
are predicted by the duality 
to type II compactifications on Calabi-Yau threefolds \cite{Het-II}.
There are also heterotic compactifications with more than one vector-tensor 
multiplet, namely
toroidal compactifications of six-dimensional string 
vacua with tensor multiplets \cite{SeiWit,LSTY}.

In the next subsection we will
discuss the effects of dimensional reduction over time
in a simple, but instructive example.

\subsection{Summary and Conclusions \label{EucAct}}

One way to obtain a $d$-dimensional Euclidean action is by dimensional 
reduction of a Minkowskian theory in dimension $(1,d)$ over time,
$(1,d) \rightarrow (0,d)$.\footnote{We say that
a space-time has dimension $(t,s)$, if it has $t$ time-like
and $s$ space-like dimensions. We will consider the case $d=4$, but the
remarks in this paragraph apply to general $d$.} In general, 
dimensional reduction modifies the geometry of the
scalar manifold ${\cal M}$, if (i) components
of tensor fields or gauge fields become scalars or if (ii) one obtains
a field strength of rank $d-1$. In the latter case one can
Hodge-dualize the field strength into a field strength
of rank 1, or, in other words, the derivative of a scalar field. 
Both phenomena are well known for dimensional reduction
over space, and dimensional reduction over time adds an interesting
twist.

Let us give a simple  example for case (i), which is the
mechanism relevant for the main part of this paper.
For definiteness, we start with one real scalar field $\sigma$ and one Abelian
gauge field $A_\mu$ in dimension $(1,4)$. Then the dimensional reduction
of the Lagrangian\footnote{We use the `mostly plus'
convention where the metric is negative definite in the 
time-like directions. With this convention standard kinetic terms
for tensor gauge fields always have a minus sign in front.}
\be
{\cal L}^{(1,4)} = - \frac{1}{2} \der_{\mu} \sigma \der^{\mu} \sigma
- \frac{1}{4} F_{\mu \nu} F^{\mu \nu} 
\label{FlatReal}
\ee
over space gives
\be
{\cal L}^{(1,3)} = - \frac{1}{2} \der_{\mu} \sigma \der^{\mu} \sigma
- \frac{1}{2} \der_{\mu} b \der^{\mu} b - \frac{1}{4} F_{mn} F^{mn} \;.
\ee
Here $F_{mn}$ is the field strength
of the $(1,3)$-dimensional Abelian gauge field $A_m$ obtained by decomposing 
$A_\mu = (A_m,A_5)$. The component $A_5$ gives rise
to the additional scalar field $b$.    Defining 
$z = \sigma + i b$, this can be rewritten as
\be
{\cal L}  = -  \frac{1}{2} \der_{\mu} z \der^{\mu} \overline{z}
- \frac{1}{4} F_{mn} F^{mn} \;.
\label{FlatCompl}
\ee
This shows that the complex coordinate $z$ defines 
a {\em complex structure}
on the scalar manifold,  with respect to which 
the target metric defined by the scalar couplings is {\em K\"ahlerian}.   

The same mechanism carries over to interacting theories, in 
particular to those involving non-linear sigma models,
where the scalar fields can be interpreted as maps from
space-time into a Riemannian manifold ${\cal M}$.
If the theory is supersymmetric, the geometry of ${\cal M}$
is subject to restrictions, the details of which depend on
the underlying supermultiplet. The minimal supersymmetric
extension of \refeq{FlatReal} is a theory 
of Abelian vector multiplets in dimension $(1,4)$, which 
reduces to an ${\cal N}=2$ supersymmetric theory of
vector multiplets in dimension $(1,3)$. The allowed target manifolds 
${\cal M}$ for the latter theories   
are the {\em affine special 
K\"ahler manifolds}  
\cite{Gates,ST,it,CRTV,Freed,ACD}.\footnote{These manifolds are also called
rigid special K\"ahler manifolds.}

Dimensional reduction of \refeq{FlatReal} over time gives instead
\be
{\cal L}^{(0,4)} = - \frac{1}{2} \der_{\mu} \sigma \der^{\mu} \sigma
+ \frac{1}{2} \der_{\mu} b \der^{\mu} b - \frac{1}{4} F_{mn} F^{mn} \;.
\label{DRtime}
\ee
Introducing an object $e$ with the properties 
$e^2 = 1$ and $\overline{e} = - e$ \cite{GibGrePer},
and defining 
$z = \sigma + e b$, we see that
\refeq{DRtime}
takes the same form as \refeq{FlatCompl}, but
with a different interpretation of the `complex' scalar field $z$.
As we will explain in detail in the main part of the paper, 
the scalar target space of the Euclidean 
$(0,4)$-dimensional  
theory carries a {\em para-complex structure}, for which $z$ 
is a {\em para-holomorphic coordinate}. 

The main objective of this paper is to generalize the above
simple example to supersymmetric models with a 
curved scalar manifold. In order to characterize the allowed
target manifolds, we introduce in section 2 
the notion of a  (rigid or affine) 
{\em special para-K\"ahler manifold} and investigate systematically the
corresponding geometry. Our fundamental result about such manifolds 
is that any simply connected affine 
special para-K\"ahler manifold $M$ admits a
para-K\"ahlerian Lagrangian immersion into a symplectic para-complex
vector space $V$ endowed with a compatible para-complex conjugation. 
The immersion induces the special geometric structures on $M$ and is unique 
up to an affine transformation of $V$ which preserves the 
symplectic structure and the para-complex 
conjugation. As a corollary, we obtain that any affine 
special para-K\"ahler manifold $M$ of para-complex dimension $n$  
is locally described by a para-holomorphic {\it prepotential} $F$ of 
$n$ para-complex variables, the Hessian of which has to satisfy a 
certain non-degeneracy condition. In contrast with the  
para-K\"ahlerian Lagrangian immersion, which is unique up to an affine
transformation, the prepotential depends nonlinearly on the choice of a 
compatible Lagrangian splitting of $V$ (choice of coordinates and 
momenta in the symplectic para-complex vector space $V$). 

In section 3 we discuss various properties of fermions and of supersymmetry
algebras in dimensions $(1,4), (1,3)$ and $(0,4)$, which we need for the 
dimensional reduction. We consider this both from a physicist's and  
from a mathematician's perspective by combining \cite{con1} with
\cite{AC,ACDV03}.
Particular attention is payed to
symplectic Majorana spinors and to the R-symmetry groups of
the relevant supersymmetry algebras. These play a crucial 
role, since symplectic Majorana spinors allow us to write
the fermionic terms in dimensions $(1,4), (1,3)$ and
$(0,4)$ in a uniform way, while 
R-symmetry is related to the existence of a (para-)complex
structure on the scalar manifold.

In section 4 we construct 
the general Lagrangian of rigid vector
multiplets in dimension (1,4) 
by adapting the results obtained in  \cite{con2} for 
superconformal vector multiplets. The couplings 
are encoded in a real (not necessarily homogeneous) cubic prepotential, and 
the metric of the scalar manifold is the Hessian of this 
function, as was also found in \cite{5dgauge}. 
This provides the definition of an 
{\em affine very special real manifold}, which is the affine version 
of the very special real manifolds considered in 
\cite{GST,deWvP,ACDV}.   

In section 5 we perform the 
dimensional reduction of the 
$(1,4)$-dimensional Lagrangian over space and time in parallel.
While the reduction
over space gives an affine special K\"ahler manifold, 
the reduction over time results in an affine special para-K\"ahler manifold.
The mapping of scalar manifolds induced by the dimensional
reduction $(1,4) \rightarrow (1,3)$ is known as the r-map  
\cite{deWvP}. Here we encounter a new mapping induced by dimensional reduction
over time $(1,4) \rightarrow (0,4)$, which we call {\em the temporal 
r-map}. It associates an affine special para-K\"ahler manifold to any
affine very special real manifold. 
To get the general 4-dimensional Lagrangian we proceed in two 
steps. First we observe that although the explicit Lagrangian 
and supersymmetry rules obtained
by dimensional reduction depend on a cubic prepotential,
all these formulae have a well defined geometrical 
meaning for general (para-)holomorphic prepotentials.
In particular, the prepotential
is the generating function of a (para-)holomorphic Lagrangian immersion 
which induces the special geometric structures on the scalar manifold. 
For Euclidean space-time signature 
this follows from the results of section 2. 
Therefore we can consider 4-dimensional Lagrangians defined by a
a general (para-)holomorphic prepotential.
However, these 
Lagrangians are not supersymmetric anymore, because supersymmetry
variations now generate terms which contain the fourth derivative
of the prepotential. Therefore the second step consists in 
adding further terms to the Lagrangian,
which restore its invariance under supersymmetry.
It turns out that it is sufficient (in the off-shell formulation) to
add one particular four-fermion term, which contains
the fourth derivative of the prepotential. For the case of
Minkowski space-time signature the general ${\cal N}=2$ vector multiplet 
Lagrangian is of course well known \cite{DLP,DDKV,it}.
We check our result by comparing it to \cite{DLP,DDKV,bc} and find
complete agreement. The advantage of our formalism is that
the Minkowskian and Euclidean Lagrangian and supersymmetry 
transformations take the same form,
when written in holomorphic and para-holomorphic coordinates,
respectively. Therefore it follows
immediately that every para-holomorphic prepotential 
defines a supersymmetric Euclidean Lagrangian, irrespective
of whether this theory can be obtained by dimensional reduction or not. 
Moreover, we see 
that we have found the general Euclidean vector multiplet Lagrangian.

Given the mathematical results of section \ref{SectPCGeom}, we 
could have approached this result more directly, without
invoking dimensional reduction. All what is needed in addition to
section \ref{SectPCGeom} is the geometric interpretation of 
the fermions and gauge fields as bundle-valued sections of the
para-complexified tangent bundle of the target manifold. We preferred
to proceed through dimensional reduction for two reasons,
which were already mentioned above: first, in various
applications we are interested in knowing explicitly how
theories are related by dimensional reduction or oxidation.
Second, dimensional reduction generates
a dictionary between the Minkowskian and Euclidean theory.
The geometrical structures can be read off easily and it 
is obvious how to bring the Lagrangians of both space-time
signatures to the same form.

The result that every para-holomorphic prepotential defines a
Euclidean supersymmetric Lagrangian does not imply that every
such theory can be obtained from a Minkowskian theory by
analytic continuation. The reason is that not every
para-holomorphic function can be obtained by analytic continuation
of a holomorphic function. This is not even possible for para-holomorphic 
functions which are real-valued on real points. In fact, a holomorphic
function is automatically  real-analytic, while a para-holomorphic
function need not be real-analytic.\footnote{We assume throughout
that the prepotential is $C^{\infty}$.} 
The Euclidean theories which can be obtained by analytic
continuation of Minkowskian theories therefore form a subset. 
They are defined by a para-holomorphic 
prepotential which is real-valued and real-analytic on real points.  
There are two obvious ways to find explicit examples of such
pairs of theories. The first approach is
dimensional reduction. We have already seen that theories with
a cubic prepotential come in pairs. This procedure can
be generalized by considering the full 5-dimensional theory in 
the background $M_4 \times S^1$, where $S^1$ is a space-like or
time-like circle of finite radius. In this case the massive
Kaluza-Klein modes do not decouple completely and induce effective
interactions in the action of the massless modes. 
Gauge theories and string theories on $M_4 \times S^1$, with space-like $S^1$, 
have been studied in the literature, see for example 
\cite{4d5drigid}, \cite{AFT}, \cite{Eguchi}, \cite{LMZ}.
The second approach is more general and intrinsically 4-dimensional. 
One can adapt and generalize the continuous Wick 
rotation of \cite{vN,vN2} 
to ${\cal N}=2$ theories with suitable non-trivial scalar manifolds.  
We will discuss this approach, and its relation to dimensional
reduction, in a future publication.

Finally, in section 5.4,  we show that the occurrence of a para-complex 
structure
in Euclidean space-time signature is a consequence of the 
non-compactness of the Abelian factor of the Euclidean R-symmetry
group. We also see that the general vector multiplet
Lagrangian in either signature is only invariant under a 
discrete $\Zom_2$-subgroup of this factor. This is as expected
from the fact that the continuous Abelian factor of the
R-symmetry group is anomalous, and therefore
generically broken by quantum effects.

We have organized the material such that 
the focus is on the mathematical aspects in sections 2 and 3,
and on the physical aspects in  
sections 4 and 5. The sections are as 
self-contained as possible.  Readers who want to approach the physical 
aspects directly can continue reading at section 4 and then go back to sections
2 and 3 for the underlying mathematics.

\subsection{Outlook}

As we have seen above,  
the geometrical structure underlying the $i \rightarrow e$ substitution
rule is the {\em replacement of a complex by a para-complex structure.}
In this paper 
the case of rigid vector multiplets is treated in detail, and 
it is already clear how this result is to be extended in various directions.
In the following we will indicate the most immediate 
extensions in the context of theories with ${\cal N}=2$ 
supersymmetry (8 supercharges). But before doing so, let us note
that para-complex structures also appear in other situations,
in particular in the first example for the $i \rightarrow e$
substitution rule, 10-dimensional IIB supergravity \cite{GibGrePer}. 
Here the
complex manifold ${\rm SL}(2,\R)/{\rm SO}(2)$ is replaced by the 
para-complex manifold ${\rm SL}(2,\R)/{\rm SO}(1,1)$.\footnote{In fact, these
symmetric manifolds are projective special K\"ahler and projective special
para-K\"ahler, respectively.}

The next step in our program will be the further reduction
to 3 dimensions. For dimensional reduction over space,
$(1,3) \rightarrow (1,2)$, it is well known that the resulting
Lagrangian can be expressed in terms of hypermultiplets
\cite{deWvP,CecFerGir}. This uses that one can dualize
a 3-dimensional gauge field into a scalar, which is a
particular case of the mechanism (ii) mentioned above. 
Supersymmetry requires that the scalar geometry of
hypermultiplets in rigid supersymmetry must be
hyper-K\"ahler \cite{AlvGauFre}. Dimensional reduction induces
a map which assigns to each affine special K\"ahler manifold 
a hyper-K\"ahler manifold. This is known as the c-map
\cite{deWvP,CecFerGir}. Dimensional reduction over time
yields a version of it, which we call the temporal c-map.
As we will explain in detail in a future publication,
dimensional reduction over time has the effect that one
obtains one complex and two para-complex structures, 
leading to {\em para-hyper-K\"ahler manifolds.}

The obvious next step is then  to consider vector and hypermultiplets coupled
to supergravity. The geometry of locally supersymmetric
vector multiplets in dimension (1,3) is known as projective 
special K\"ahler geometry 
\cite{deWvP2,Strominger,it2,it,CRTV,Freed,ACD}.\footnote{This is also called
local special K\"ahler geometry in the physics literature.} As indicated by 
the name,
such manifolds can be obtained from affine special K\"ahler manifolds
(with suitable homogeneity properties) by projectivization. This
can also be understood from the physical point of view in terms of the
conformal calculus. Here one first constructs a superconformally 
invariant theory and then eliminates the so-called conformal 
compensators by imposing gauge conditions. This gauge fixing
amounts to the projectivization of the scalar manifold underlying
the superconformal theory. Based on the results obtained in this paper,
one can adapt this construction to the case of Euclidean
signature and construct {\em projective special para-K\"ahler manifolds}.
Similar remarks apply to hypermultiplets coupled to supergravity.
In Minkowski signature the coupling to supergravity implies that
the scalar geometry is quaternionic-K\"ahler instead of hyper-K\"ahler
\cite{BagWit}. The relation between these two kinds of geometries
can again be understood as projectivization, 
because every quaternionic-K\"ahler
manifold can be obtained as the quotient of a hyper-K\"ahler cone \cite{Swan}. 
This relation is natural from the viewpoint of conformal
calculus \cite{deW:HKC}, and it is possible to adapt the construction
to theories with Euclidean signature in order to construct
{\em para-quaternionic-K\"ahler manifolds}. In fact, such manifolds have
already occurred in the context of instantons in IIB string theory
compactified on a Calabi-Yau threefold. 
Here it was found that in Euclidean signature 
the Hodge-dualization of the 
universal double-tensor multiplet gives a 
hypermultiplet with scalar manifold
is ${\rm SL}(3,\R)/({\rm SL}(2,\R) \otimes {\rm SO}(1,1))$ \cite{TheVand}. 
This is a
symmetric para-quaternionic-K\"ahler manifold.
\end{section}

\section{Para-complex geometry \label{SectPCGeom}}
\subsection{Para-complex manifolds}
\bd Let $V$ be a finite dimensional (real) vector space. A 
{\sf para-complex structure} 
on $V$ is a nontrivial involution 
$I \in {\rm End}\, V$, i.e.,\ 
$I^2 = {\rm Id}$ and $I \neq {\rm Id}$, such that the two eigenspaces 
$V^{\pm} := {\rm ker} ({\rm Id} \mp I)$ of $I$ are of the same dimension. 
A {\sf para-complex vector space} is a vector space  endowed 
with a para-complex structure. A {\sf homomorphism} from a para-complex 
vector space $(V,I)$ into a  para-complex 
vector space $(V',I')$ is a linear map $\phi : (V,I) \ra (V',I')$ 
satisfying  $\phi I = I' \phi$. 
\ed 
    
\noindent 
Let $I$ be a para-complex structure on a vector space $V$. Then there exists
a basis $(e_1^+, \ldots ,e_n^+, e_1^-, \ldots ,e_n^-)$ of $V$, $\dim V = 2n$,  
such that $Ie_i^\pm = \pm e_i^\pm$ 
and we can identify $I$ with the diagonal matrix 
\[ I = {\rm diag} (1, \ldots , 1, -1, \ldots , -1) = \left( \begin{array}{cc}
\mbox{\bf 1}_n& 0\\
0& -\mbox{\bf 1}_n
\ea \right) \, .\] 
It follows that the automorphism group ${\rm Aut}(V,I) := 
\{ L \in {\rm GL}(V)| \; L IL^{-1} = I\}$ of $(V,I)$, which 
coincides with the centralizer $Z_{{\rm GL}(V)}(I)$ of $I$ in 
${\rm GL}(V)$, is given by
\[ {\rm Aut}(V,I) = Z_{{\rm GL}(V)}(I) := 
\{ L \in {\rm GL}(V)| \; [L, I] = 0\} =  
\left\{ \left. \left( 
\begin{array}{cc}
A& 0\\
0& D
\ea 
\right) \right| A,D \in {\rm GL}_n(\R )\right\}
\, . \]

\bd An {\sf almost para-complex structure} on a smooth manifold $M$
is an endomorphism field $I \in \G ({\rm End}\,  TM)$, $p\mapsto I_p$,
such that $I_p$ is a para-complex structure on $T_pM$ for all
$p\in M$. In other words: 
\begin{enumerate} 
\item[(i)] $I \neq {\rm Id}_{TM}$ is an involution, {\it i.e.} 
$I^2 = {\rm Id}_{TM}$ and 
\item[(ii)] the two eigendistributions  
$T^{\pm}M := {\rm ker} ({\rm Id} \mp I)$ 
of $I$ have the same
rank.  
\end{enumerate} 
An almost para-complex structure $I$ is called {\sf integrable} if the
distributions $T^{\pm}M$ are both integrable. An integrable 
almost para-complex structure is called a {\sf para-complex structure}. 
A manifold $M$ endowed with a para-complex structure is called a 
{\sf para-complex manifold}. The {\sf Nijenhuis tensor} $N_I$ 
of an almost para-complex structure $I$ is the tensor defined by the
equation:
\[ N_I(X,Y) := [X,Y] + [IX,IY] -I[X,IY] -I[IX,Y] \, ,\]
for all vector fields $X$ and $Y$ on $M$.   
\ed

\bp An almost para-complex structure $I$ is integrable if and only if
$N_I = 0$. 
\ep 

\pf Consider the two projections $\pi_{\pm}: TM \ra T^{\pm}M$, $\pi_{\pm} 
:= \frac{1}{2}({\rm Id} \pm I)$. Then, by the Frobenius theorem, 
the integrability of $T^+M$ and $T^-M$ is equivalent to, respectively,  
\[ \pi_-[\pi_+ X, \pi_+ Y ] = 0 \quad \mbox{and} \quad 
\pi_+[\pi_- X, \pi_- Y ] 
= 0 \,,\]  
for all vector fields $X$ and $Y$. The sum and the difference of these  
expressions are proportional to $N_I(X,Y)$ and $IN_I(X,Y)$ respectively. 
\qed
  
\noindent 
{\bf Example 1:} Any para-complex vector space $(V,I)$ can be considered as 
a para-complex manifold, with constant para-complex structure. 

\noindent 
{\bf Example 2:} The Cartesian product $M\times N$ of two  para-complex 
manifolds $(M,I_M)$ and $(N,I_N)$ is a para-complex manifold with the
para-complex structure $I_{M\times N} := I_M \oplus I_N$. Here we have
used the identification $T(M\times N) = TM \oplus TN$.

\noindent 
{\bf Example 3:} Let $M = M_+ \times M_-$ be the Cartesian product
of two smooth manifolds $M_+$ and $M_-$ of the same dimension. 
We can identify $T_{(p_+,p_-)}M = T_{p_+}M_+ \oplus T_{p_-}M_-$ and
define a para-complex structure $I$ on $M$ by $I|T_{p_\pm}M_\pm :=
\pm {\rm Id}$. The next result shows that any para-complex manifold is 
locally of this form.

\bp \label{paracxprop} 
Let $(M,I)$ be a para-complex manifold of dimension $2n$. 
Then for every point in $M$ there is an open neighborhood $U$ and a 
diffeomorphism $\phi : U \cong M_+ \times M_-$ onto the product of two
manifolds $M_\pm \cong \R^n$ such that $\phi$ maps the leaves of the 
foliation $T^+M$ to the submanifolds $M_+ \times \{ p_-\}$, $p_-\in M_-$,   
and the leaves of $T^-M$ to the submanifolds $\{ p_+\}\times M_-$, 
$p_+\in M_+$. 
\ep

\pf By the Frobenius theorem applied to the distribution 
$T^-M$, there exists an open neighborhood $U$ of $p$
and functions $z^i_+$, $i= 1, \ldots , n$,  on $U$, which are 
constant on the leaves of $T^-M$ and such that the differentials 
$dz^i_+$ are (point-wise) linearly independent. Similarly, by restricting 
$U$, if necessary, we can find functions $z^i_-$, $i= 1, \ldots , n$,  
on $U$ constant on the leaves of $T^+M$ and such that the differentials 
$dz^i_-$ are linearly independent. {}From the transversality of the
two foliations $T^+M$ and $T^-M$ we conclude that 
$(z^1_+, \ldots, z^n_+, z^1_-,\ldots, z^n_-)$ is a system of local 
coordinates. Now we can define $\phi := (\phi_+ ,\phi_- ) : U \ra 
\R^n \times \R^n$ by $\phi_\pm = (z^1_\pm, \ldots, z^n_\pm)$. By 
restricting $U$, if necessary, we can assume that $\phi$ is a
diffeomorphism onto a product $M_+\times M_-$ of two open sets 
$M_\pm \cong \R^n$ in $\R^n$.  \qed 

Coordinates $(z^1_+, \ldots, z^n_+, z^1_-,\ldots, z^n_-)$ as in the
proof of Proposition \ref{paracxprop} will be called {\it adapted} to
the para-complex structure.  We put 
\[ x^i := \ft{z^i_+ +z^i_-}{2}\quad \mbox{and} \quad 
y^i := \ft{z^i_+ - z^i_-}{2}\, .\] 
We wish to think of $x^i$ and $y^i$ as `real' and `imaginary' parts of
a system of local `para-holomorphic' coordinates $z^i$, $i=1,\ldots ,n$. 

In order to make this analogy precise, we introduce the algebra $C$ 
of {\it para-complex numbers}. It is the (real) algebra generated by
$1$ and the symbol $e$ subject to the relation $e^2 = 1$. If we think of 
$e$ as a unit vector in a one-dimensional vector space with a 
negative definite scalar product, then $C$ is just the corresponding
Clifford algebra $C = C\! \ell_{0,1} \cong \R \oplus \R$, the same as 
$C\! \ell_{1,0} \cong \Com$ gives the field of complex numbers.  
The map $\ol{\cdot} : C \ra C$, 
$x+ey \mapsto x-ey$, where $x,y \in \R$, is called the 
{\it para-complex conjugation}. It is a $C$-antilinear involution: 
$\ol{ez} = -e\ol{z}$. The real numbers $x = \Re z := (z + \ol{z})/2$ and 
$y = \Im z := e(z - \ol{z})/2$ are called {\it real} and 
{\it imaginary} parts of 
$z = x+ ey$,  respectively. Note that $z\ol{z} = x^2 -y^2$. Therefore
$C$ is sometimes called the algebra of {\it hyperbolic} complex numbers.
Note also that the group $C^* \subset C$ of invertible elements has
4 connected components. In fact, any $z\in C^*$ can be written
as $z = \pm r \exp (et)$ or as  $z = \pm e r \exp (et)$, where $r>0$
and $t\in \R$. This is the analogue of the polar decomposition 
$z = r\exp (it)$ for $z\in \Com^* \cong \R^{>0} \times S^1$. The
circle $S^1 = \{ z = x + iy\in \Com| x^2 + y^2 = 1\}$ is replaced by 
the four hyperbolas $\{ z = x + ey\in C| x^2 - y^2 = \pm 1\}$.        
 
A {\it C-valued smooth function} on a smooth manifold is a smooth map 
$f: M \ra C \cong \R^2$. For such functions we can define the 
$C$-valued smooth function $\ol{f}$ and the real-valued functions
$\Re f$ and $\Im f$ on $M$. Obviously, $C$ is a para-complex vector space.  
Its para-complex structure is simply multiplication by $e$.  
Moreover, the eigenspaces $C^\pm = \R (1\pm e) \cong \R$ of $e$  
are precisely the two simple ideals of the algebra $C = C^+ \times C^-$.
More generally, the free $C$-module $C^n$ is a para-complex vector space.
Its para-complex structure is the scalar multiplication by $e\in C$.

\bd A smooth map $\phi : (M,I_M) \ra (N,I_N)$ between para-complex manifolds
is called {\sf para-holomorphic} (or just holomorphic, when no confusion is
possible), if  $d\phi I_M = I_N d\phi$. A para-holomorphic map
$f : (M,I) \ra C$ is called a {\sf para-holomorphic function}. 
The {\sf para-complex dimension} of a para-complex manifold
$M$ is the integer $\dim_C M := \dim M/2$.  
A system of para-holomorphic functions $z^i$, $i = 1, \ldots , n$, 
defined on some open subset
$U \subset M$ of a para-complex manifold is called
a {\sf system of local para-holomorphic coordinates} (or just system of 
local holomorphic coordinates) if 
$(x^1 = \Re z^1,\ldots , x^n = \Re z^n , y^1 = \Im z^1 ,,\ldots , y^n 
= \Im z^n)$ is a system of (real) local coordinates.   
\ed 
 
\noindent 
Notice that a collection of para-holomorphic functions 
$z^1,\ldots ,z^n$, defined in a neighborhood of a point $p\in M$, forms a 
system of local coordinates in some neighborhood of $p$ if and only  the  
differentials $dz^i_p, d\ol{z}^i_p$, $i=1,\ldots ,n$, 
form a basis of the free $C$-module
$T_p^*M \otimes C$. This is an easy consequence of the fact that
$dz^iI = edz^i$, {\it i.e.} 
\[ dx^iI = dy^i\quad  \mbox{and} \quad dy^iI = dx^i\, .\]  
   
A system of local para-holomorphic coordinates defined over $U \subset M$
defines an isomorphism from the para-complex manifold $U$ (with the 
para-complex structure induced by the inclusion) onto some open subset of 
the para-complex vector space $C^n$. 

\bp  Let $M$ be a smooth manifold endowed with an atlas
$\ca = \{ \phi :U \ra C^n = \R^{2n}\}$ such that the coordinate changes
are para-holomorphic. Then there exists a unique para-complex
structure $I_{\cal A}$ on $M$ such that all 
$\phi \in \ca$ are para-holomorphic. 
Conversely, any para-complex manifold admits such an atlas.   
\ep 

\pf Given a smooth manifold with an atlas as above, we can pull back 
the para-complex structure from $C^n$ to $M$. This gives a  well defined
para-complex structure on $M$,  
since the coordinate changes are para-holomorphic. 
Conversely, let $(M,I)$ be a para-complex manifold. Any adapted system
of coordinates $(z^1_+, \ldots, z^n_+, z^1_-,\ldots, z^n_-): U \ra \R^{2n}$
defines a system of para-holomorphic coordinates 
$(z^1, \ldots, z^n)$  by $\Re z^i := x^i = (z^i_+ +z^i_-)/2$ and 
$\Im z^i := y^i = (z^i_+ -z^i_-)/2$. In fact, from the definition of 
$z^i_\pm$ it follows easily that  $dz_\pm^i \circ I = \pm dz_\pm^i$. 
Using the above equations, this implies that  $dx^i\circ I = dy^i$,  
and hence $dz^i \circ I = (dx^i + edy^i) 
\circ I = dy^i + e dx^i = edz^i$.  This shows that the functions 
$z^i$ are indeed para-holomorphic. It is clear that their real
and imaginary parts form a (real) coordinate system, since the 
$z_\pm^i$
do. Now it suffices to observe that we can cover $M$ by coordinate 
domains $U$ as above and that the coordinate changes are para-holomorphic.  
\qed  

Notice that a $C$-valued function $f : M\ra C$ on  a para-complex
manifold $(M,I)$ is para-holomorphic if and only if it satisfies
the partial differential equations 
\[ \frac{\partial f}{\partial \ol{z}^i} := 
\frac{1}{2}(\frac{\partial f}{\partial x^i} - 
e\frac{\partial f}{\partial y^i}) 
= 0\, ,\] 
where the $(z^i)$ are local para-holomorphic coordinates. 
The general solution of this system is of the form
\[ {\rm Re}\, f = f_+ + f_- \, , \quad {\rm Im}\, f = f_+ -f_-\, ,
\quad \frac{\partial f_\pm}{\partial z_\mp^i} = 0 \, .\]
In other words $f_+$ is a function which, in adapted coordinates 
$(z_\pm^i)$, depends only on the $z_+^i = x^i + y^i$ and $f_-$ depends only
the $z_-^i = x^i -y^i$.

\subsubsection*{Para-holomorphic vector bundles}
\bd A {\sf para-holomorphic vector bundle} of rank $r$ 
is a (smooth) real
vector bundle $\pi : W\ra M$ of rank $2r$ whose total space $W$ and base 
$M$ are para-complex manifolds and whose projection $\pi$ is 
a para-holomorphic 
map. 
\ed 

\noindent
{}From the fact that the projection $\pi$ is para-holomorphic it
follows that the para-complex structure on $W$ induces on each 
fibre $W_p = \pi^{-1}(p)$, $p\in M$, the structure of a para-complex 
vector space of para-complex dimension $r$. In particular, we have a
canonical splitting $W = W^+\oplus W^-$, where $W_p^\pm$ are the
two eigenspaces of the para-complex structure on $W_p^\pm$, for all 
$p\in M$.  

\noindent
{\bf Example 4:} The tangent bundle $TM \ra M$ over any para-complex manifold 
$M$ is a para-holomorphic vector bundle and we have $(TM)^\pm = T^{\pm}M$.

\noindent
{\bf Example 5:} Sums and duals  of para-holomorphic vector 
bundles are again para-holomorphic vector bundles. 
Also  tensor products, over the commutative
algebra $C$, of para-holomorphic vector bundles over the same base $M$
are again para-holomorphic vector bundles over $M$.

The decomposition $T^*M = (T^*M)^+\oplus (T^*M)^-$, defined for any 
almost para-complex manifold, induces a 
bigrading on exterior forms:
\[ \wedge^k T^*M = \oplus_{p+q = k} \wedge^{p+,q-} T^*M\, .\]  
In particular, $\wedge^{1+,0-}T^*M = (T^*M)^+$ and 
$\wedge^{0+,1-}T^*M = (T^*M)^-$. We will say that a k-form
$\omega$ is of {\it type} $(p+,q-)$ if $\omega \in \wedge^{p+,q-} T^*M$. 
The corresponding decomposition of differential forms on $M$
will be denoted by
\[ \O^k(M) = \oplus_{p+q = k} \O^{p+,q-}(M)\, .\]  
If the almost para-complex structure is integrable, then the 
de Rham differential $d : \O^k(M) \ra \O^{k+1}(M)$ splits as
$d = \partial_+ + \partial_-$, where 
\[ \partial_+ :\O^{p+,q-}(M)\ra \O^{(p+1)+,q-}(M)\quad \mbox{and} \quad
\partial_- :\O^{p+,q-}(M)\ra \O^{p+,(q+1)-}(M)\, .\] 
Moreover, we have $\partial_+^2 = \partial_-^2 = \partial_+\partial_- +  
\partial_-\partial_+ =0$. As a consequence, the 
differential operator $\partial_+$ (or $\partial_-$) can be used
to define a real version of Dolbeault cohomology on any para-complex
manifold.  

Consider now the {\it para-complexification} $TM^C := TM\otimes C$ 
of the tangent bundle $TM$ of an almost para-complex manifold $(M,I)$. 
We extend the almost para-complex structure 
$I : TM \ra TM$ to a $C$-linear endomorphism field
$I : TM^C \ra TM^C$. Then for all $p\in M$ the free $C$-module 
$T_pM^C$ has the following canonical decomposition into the direct sum of
two free $C$-modules  
\[ T_pM^C =  T_p^{1,0}M  \oplus T_p^{0,1}M\, ,\]
where 
\[ T_p^{1,0}M := \{ X + eIX| X\in T_pM\} \quad \mbox{and} \quad
T_p^{0,1}M := \{ X -eIX| X\in T_pM\}\, .\] 
The subbundles $T^{1,0}M$ and  $T^{0,1}M$ are characterized as 
the $\pm e$-eigenbundles of $I : TM^C \ra TM^C$, 
in the sense that $I = e$ on  $T^{1,0}M$ 
and $I = -e$ on  $T^{0,1}M$. 
The canonical isomorphism $TM\ra T^{1,0}M$, $X\ra X^{1,0} := 
\frac{1}{2}(X +eIX)$,  of real 
vector bundles is compatible with the para-complex structures on the 
fibres; $(IX)^{1,0} = eX^{1,0}$. In particular, it is an
isomorphism of para-holomorphic vector bundles if the
almost para-complex structure $I$ happens to be integrable. 
Note that the isomorphism $TM\ra T^{0,1}M$, $X\ra X^{0,1} := 
\frac{1}{2}(X -eIX)$,  of real 
vector bundles maps $I$ to $-e$; $(IX)^{0,1} = -eX^{0,1}$. 
Consider now the $\pm e$-eigenbundles of $I^* : T^*M^C \ra T^*M^C$:
\[ \wedge^{1,0}T^*M := \{ \alpha + eI^*\alpha | \alpha \in T^*M\}
\quad \mbox{and} \quad  \wedge^{0,1}T^*M := \{ \alpha - 
eI^*\alpha | \alpha \in T^*M\}\, .\]
The decomposition $T^*M^C := \wedge^{1,0}T^*M \oplus  \wedge^{0,1}T^*M$ 
induces a bigrading on $C$-valued exterior forms: 
\[ \wedge^k T^*M^C = \oplus_{p+q = k} \wedge^{p,q} T^*M\, .\]
We will say that a $C$-valued $k$-form
$\omega$ is of {\it type} $(p,q)$ if $\omega \in \wedge^{p,q} T^*M$. 
The corresponding decomposition of $C$-valued differential forms on $M$
will be denoted by
\[ \O_C^k(M) = \oplus_{p+q = k} \O^{p,q}(M)\, .\] 
Notice that the real vector bundles $\wedge^{p,0}T^*M$ are para-holomorphic
with the para-complex structure on the fibres defined by $e$, if the
almost para-complex structure $I$ is integrable. 
Suppose now that $I$ is integrable. 
Then the  ($C$-linearly extended) de Rham differential $d : \O_C^k(M) \ra 
\O_C^{k+1}(M)$ splits as
$d = \partial + \overline{\partial}$, where 
\[ \partial :\O^{p,q}(M)\ra \O^{p+1,q}(M)\quad \mbox{and} \quad
\ol{\partial} :\O^{p,q}(M)\ra \O^{p,q+1}(M)\, .\]  
Moreover,  
$\partial^2 = \ol{\partial}^2 = \partial\ol{\partial} +  
\ol{\partial}\partial = 0$. As a consequence, the 
differential operator $\ol{\partial}$ can be used
to define a para-complex version of Dolbeault cohomology.

\subsection{Para-K\"ahler manifolds}
\bd Let $(V,I)$ be a para-complex vector space. A pseudo-Euclidean
scalar product $g$ on $V$ is called {\sf para-Hermitian} 
if $I$ is an anti-isometry for $g$, i.e.,
\[ I^*g = g(I \cdot , I\cdot ) = -g\, .\]  
A {\sf para-Hermitian vector space} is a para-complex vector space 
endowed with a para-Hermitian scalar product. The pair $(I,g)$  is called
a {\sf para-Hermitian structure} on the vector space $V$. 
\ed 

\noindent 
{\bf Example 6:} Consider the vector space $\R^{2n} = \R^n \oplus \R^n$ with 
its standard basis $(e_i^\pm)$, where $e_i^+ := e_i \oplus 0$ and 
$e_i^- = 0 \oplus  e_i$, and its standard para-complex structure,  
given by $Ie_i^\pm = \pm e_i^\pm$.  We can define a para-Hermitian 
scalar product $g$ by $g(e_i^\pm,e_j^\pm) = 0$ and $g(e_i^\pm,e_j^\mp) = 
\delta_{ij}$. We will call $(I,g)$ the {\it standard para-Hermitian 
structure} of $\R^{2n}$. 

\noindent
{\bf Example 7:} Consider the vector space $C^n = \R^n \oplus e\R^n$ 
with its standard basis\linebreak 
$(e_1, \ldots, e_n, f_1, \ldots, f_n)$, 
where $f_i = ee_i$, and its standard para-complex structure,  
given by $Ie_i = f_i$ and $If_i = e_i$. We can define a 
para-Hermitian scalar product $g$ by $g(e_i,e_j) = -g(f_i,f_j) = \delta_{ij}$
and $g(e_i,f_j) = 0$. We will call $(I,g)$ the {\it standard para-Hermitian 
structure} of $C^n$. 

Any two para-Hermitian vector spaces $(V,I,g)$ and $(V',I',g')$ 
of the same dimension are isomorphic, {\it i.e.}, there exists a  
linear para-holomorphic isometry $\phi : V \ra V'$. In particular,
$\R^{2n} = \R^n \oplus \R^n$ and $C^n$ are isomorphic as para-Hermitian 
vector spaces. An isomorphism $\phi : R^{2n} \ra C^n$ is given by 
\[ \phi e_i^\pm = \frac{1}{\sqrt{2}}(e_i\pm f_i)\, .\] 

The model $\R^{2n}$ has the advantage that $I$ is diagonal in the standard
basis, whereas $g$ is diagonal in the standard real basis of $C^n$. 

\bd 
Let $V$ be  a  para-Hermitian vector space  with
para-Hermitian structure $(I,g)$. The {\it para-unitary group} of $V$
is the automorphism group 
\[ {\rm U}^\pi(V) := {\rm Aut}(V,I,g) = \{ L\in {\rm GL}(V)|\; 
[L,I] = 0 \quad \mbox{and} \quad L^*g = g\}\, .\] 
\ed

\bp \begin{enumerate} 
\item[(i)] The para-unitary group of the para-Hermitian vector space 
$\R^{2n} = \R^n \oplus \R^n$ is given by 
\[ {\rm U}^\pi(\R^{2n}) = \left\{ \left. \left( 
\begin{array}{cc}
A& 0\\
0& A^{-t}
\ea 
\right) \right| A\in {\rm GL}_n(\R )\right\}
\cong {\rm GL}_n(\R )\, , \]
where $A^{-t} := (A^{-1})^t = (A^t)^{-1}$ is the contragredient matrix.  
\item[(ii)] 
The para-unitary group ${\rm U}^\pi(C^n) \cong {\rm U}^\pi(\R^{2n}) 
\cong {\rm GL}_n(\R )$ of the para-Hermitian vector space 
$C^n = \R^n \oplus e\R^n$ is given by 
\[ {\rm U}^\pi(C^n) = \left\{ \left. \left( 
\begin{array}{cc}
A& B\\
B& A
\ea 
\right) \right| A, B \in {\rm End}(\R^n)\, ,\; 
A^tA -B^tB = {\bf 1}_n\, , \; A^tB -B^tA = 0 \right\}\, . \]
\end{enumerate} 
\ep 

\pf (i) The centralizer of $I$ consists of block diagonal matrices 
$L = {\rm diag}(A,D)$, $A, D \in {\rm GL}_n(\R )$. It is sufficient to
check that $L^*g= g$ is equivalent to $A^tD = {\rm 1}_n$.\\ 
(ii) The centralizer of $I$ consists of matrices of the form
\[ L = \left( 
\begin{array}{cc}
A& B\\
B& A
\ea 
\right)\, .\] 
The equation $L^*g = g$ is now equivalent to $A^tA -B^tB = {\bf 1}_n$ and
$A^tB -B^tA = 0$. Since the para-Hermitian vector spaces $\R^{2n}$ and $C^n$ 
are isomorphic, the corresponding para-unitary groups are isomorphic as well.
An explicit isomorphism ${\rm U}^\pi(\R^{2n}) \stackrel{\sim}{\ra} 
{\rm U}^\pi(C^n)$ is 
\[ {\rm diag}(A,D) = \left( 
\begin{array}{cc} 
A& 0\\
0&D  \ea 
\right)
\ra \left( 
\begin{array}{cc}
\frac{A+D}{2}& \frac{A-D}{2}\\
\frac{A-D}{2}& \frac{A+D}{2}
\ea 
\right)\, ,\]
where $D=A^{-t}$. 
\qed   
  
\noindent
Notice that the para-complex structure $I$ does not belong to
the para-unitary group, simply because $I^*g = -g \neq g$. However, 
it belongs to the {\it para-unitary Lie algebra} $\gu^\pi(V) = 
Lie\, {\rm U}^\pi(V)$ and generates a closed non-compact 
central subgroup 
$\{ \exp tI| t\in \R\} \cong \R^{>0}$ of the 
para-unitary group. If $V$ is a para-complex vector space of 
para-complex dimension 1, {\it i.e.}, $V \cong C \cong \R^2$, then
${\rm U}^\pi(V) = \{ \pm \exp tI| t\in \R\} \cong {\rm GL}_1(\R) = \R^*$.

\bd An {\sf almost para-Hermitian manifold} $(M,I,g)$ is an almost  
para-complex manifold $(M,I)$ endowed with a pseudo-Riemannian
metric $g$ such that $I^*g = -g$. If $I$ is integrable, we
say that $(M,I,g)$ is a {\sf para-Hermitian manifold}. 
The two-form $\o := g (I\cdot ,\cdot ) = -g (\cdot , I\cdot )$
is called the {\sf fundamental two-form} of the almost para-Hermitian 
manifold $(M,I,g)$. 
\ed 
 
Notice that the fundamental two-form satisfies $I^*\o = -\o$,  
and is hence of type $(1+,1-)$ or, equivalently, of type $(1,1)$
(when considered as $C$-valued two-form).

\bd A {\sf para-K\"ahler manifold} $(M,I,g)$ is an almost
para-Hermitian manifold $(M,I,g)$  such that $I$ is parallel with 
respect to the Levi-Civita connection $D$ of $g$, {\it i.e.}, $DI = 0$.  
\ed

\noindent 
The condition $DI = 0$ easily implies $N_I =0$ and $d\o =0$. Therefore any
para-K\"ahler manifold $(M,I,g)$ is a 
para-Hermitian manifold with closed fundamental form $\o$. The 
symplectic form $\o$ will be called the 
{\it para-K\"ahler form}. 

The next theorem characterizes para-K\"ahler manifolds as para-Hermitian
manifolds with closed fundamental form. 
\bt \label{dwNKaehlerThm}  Let $(M,I,g)$ be a para-Hermitian
manifold with closed fundamental form $\o$. Then $(M,I,g)$ is a 
para-K\"ahler manifold. Conversely, any
para-K\"ahler manifold is a para-Hermitian manifold with closed 
fundamental form. 
\et 

\pf  Let $(M,I,g)$ be a para-Hermitian
manifold with closed fundamental form $\o$ and $D$ the
Levi-Civita connection of $g$. It is determined by the 
Koszul formula 
\[ 2g(D_XY,Z) = Xg(Y,Z) + Yg(X,Z) - Zg(X,Y) + g([X,Y],Z) - g([X,Z],Y)  
- g([Y,Z],X)\, ,\] 
which holds for all vector fields  $X, Y$ and $Z$ on $M$. We have to show that
$DI = 0$. Therefore we compute  
\begin{eqnarray*} 2g((D_XI)Y,Z) &=&  
2g(D_X(IY),Z) - 2g(ID_XY,Z) = 2g(D_X(IY),Z) + 2g(D_XY,IZ)\\
&=& Xg(IY,Z) + IYg(X,Z) - Zg(X,IY)\\
&& + g([X,IY],Z) - g([X,Z],IY)  
- g([IY,Z],X)\\
&& + Xg(Y,IZ) + Yg(X,IZ) - IZg(X,Y)\\
&&+ g([X,Y],IZ) - g([X,IZ],Y) - 
g([Y,IZ],X)\\ 
&=& (X\o( Y,Z) + Y\o (Z,X) + Z\o (X,Y)\\
&& - \o ([X,Y],Z) - 
\o ([Y,Z],X) - \o ([Z,X],Y)) + \o ([Y,Z],X)\\  
&& + IYg(X,Z) + Xg(Y,IZ) - IZg(X,Y)\\ 
&&+ g([X,IY],Z)
 - g([IY,Z],X) - g([X,IZ],Y) - g([Y,IZ],X)\\ 
&=& d\o (X,Y,Z) + \o ([Y,Z],X) + (X\o (IY,IZ) + IY\o (IZ,X)\\&& +IZ\o (X,IY)
- \o ([X,IY], IZ) - \o ([IY,IZ],X) - \o ([IZ,X],IY))\\
&& 
+\o ([IY,IZ],X) - g([IY,Z],X) - g([Y,IZ],X)\\
&=& d\o (X,Y,Z) + \o ([Y,Z],X) + d\o(X,IY,IZ)\\
&& 
+\o ([IY,IZ],X) - g([IY,Z],X) - g([Y,IZ],X)\\
&=&  d\o (X,Y,Z) + d\o(X,IY,IZ)\\
&& + g(-[Y,Z]-[IY,IZ] + I[IY,Z] + I[Y,IZ],IX)\\
&=&  d\o (X,Y,Z) + d\o(X,IY,IZ) -g(N_I(Y,Z),IX) = 0  \, .
\end{eqnarray*} 
\qed 

\bp Let $(M,I,g)$ be a para-K\"ahler manifold. Then the eigendistributions 
$T^\pm M$ of $I$ are\linebreak[4] 
$g$-iso\-tro\-pic. 
Their leaves are Lagrangian submanifolds
with respect to the para-K\"ahler form $\o$. In particular, $g$ has split 
signature $(n,n)$, where $n = \dim_C M$.      
\ep 

\pf Let $X\in T^\pm M$. Then, since I is an anti-isometry, we have  
\[ g(X,X) = - g(IX,IX) = -g(\pm X,\pm X) = -g(X,X) \, .\]
This shows that the $I$-invariant foliations $T^\pm M$ are 
$g$-isotropic and hence $\o$-Lagrangian. 
\qed

K\"ahler manifolds can be characterized as Riemannian manifolds with
holonomy group in the unitary group. The corresponding characterization
of para-K\"ahler manifolds is the following. 

\bp Let $(M,g)$ be a (connected) pseudo-Riemannian manifold of dimension $2n$. 
Then there exists a para-complex structure $I$ on $M$ such that
$(M,I,g)$ is a para-K\"ahler manifold if and only if the 
holonomy group of $(M,g)$ is a subgroup of the para-unitary group, 
{\it i.e.},  
if there exists a point $p\in M$ and a linear isometry 
$T_pM \cong \R^{2n}$ which identifies the holonomy group ${\rm Hol}_p(M,g)
\subset {\rm O}(T_pM,g_p)$ with a 
subgroup of the para-unitary group ${\rm U}^\pi(\R^{2n})$. 
\ep  

\pf If $(M,I,g)$ is a para-K\"ahler manifold, $\dim_C M = n$, 
then 
\[ {\rm Hol}_p(M,g)
\subset {\rm Aut}(T_pM,I_p,g_p) \cong {\rm U}^\pi(\R^{2n})\, .\] 
Conversely, let $(M,g)$ be a pseudo-Riemannian manifold of dimension $2n$
such that $\phi {\rm Hol}_p(M,g)\phi^{-1} \subset {\rm U}^\pi(\R^{2n})$,
where $\phi : T_pM \ra \R^{2n}$ is a linear isometry. Then we can define
a para-complex structure $I_p$ on the vector space $T_pM$ by 
\[ I_p :=  \phi^{-1}I_0\phi \, ,\]
where $(I_0,g_0)$ denotes the standard para-Hermitian structure
on $\R^{2n}$. Since $\phi$ is an isometry and $I_0$ is an anti-isometry 
with respect to $g_0$, $I_p$ is an anti-isometry 
with respect to $g_p$. As $I_p$ is  invariant under 
the holonomy group  ${\rm Hol}_p(M,g)$, it extends (by parallel
transport) to a parallel anti-isometric para-complex structure $I$ on $(M,g)$.
Thus $(M,I,g)$ is a para-K\"ahler manifold. 
\qed

\subsubsection*{The para-K\"ahler potential}
\bt Let $(M,I,g)$ be a para-K\"ahler manifold with para-K\"ahler
form $\o$. Then for any point $p\in M$ there exists a real-valued function 
$K$ defined on some open neighborhood $U \subset M$ of $p$ such that
$\o = \partial_- \partial_+ K$ on $U$. The function $K$ is unique 
up to addition of a real-valued function $f$ satisfying the equation
$\partial_- \partial_+ f = 0$. Any such function is of the form
$f = f_+ + f_-$, where $f_\pm: U \ra \R$ satisfies $\p_\mp f_\pm =0$. 
Conversely, let $(M,I)$ be a para-complex manifold and $K$  
a real-valued function defined on some open subset $U \subset M$  such that
the two-form $\partial_- \partial_+ K$ is non-degenerate. Then 
$g := \o (I\cdot ,\cdot )$ is a pseudo-Riemannian metric
such that $(U,I,g)$ is a para-K\"ahler manifold. 
\label{theorem2}
\et 

\pf Let $(z^i_\pm)$ be adapted coordinates defined on some open neighborhood 
$U$ of $p\in M$, which map $U$ onto the product of two simply connected
open sets $U^\pm \subset \R^n$, $n = \dim_C M$. We will also assume that
$z^i_\pm(p) = 0$.  In particular, the  
first cohomology of $U$ vanishes; $H^1(U,\R) = 0$. Therefore, since $\o$ is 
closed, there exists 
a one-form $\theta$ on $U$ such that $\o = d \theta$. We decompose
$\theta$ into its homogeneous components: $\theta = 
\theta^+ + \theta^-$, $\theta^+ \in \O^{1+,0-}(U)$, $\theta^-\in 
\O^{0+,1-}(U)$.  
This shows that
\[ d \theta = \partial_+\theta^+ + (\partial_-\theta^+ +
\partial_+\theta^-) + \partial_-\theta^- \, .\] 
{}From the fact that $\o$ is of type $(1,1)$ we obtain
the equations:
\[ \partial_\pm\theta^\pm = 0\quad \mbox{and} \quad  \partial_-\theta^+ +
\partial_+\theta^- = \o\, .\] 
\bl (Para-Dolbeault Lemma) Under the above topological assumptions on 
$U\subset M$, the equation $\partial_\pm\theta^\pm = 0$ implies the
existence of a real-valued function $K^\pm$ such that 
\[ \q^\pm = \p_\pm K^\pm \, .\]   
The function $K^\pm$ is unique up to addition of a real-valued function 
$f_\mp$ which satisfies $\p_\pm f_\mp =0$. 
\el 

\pf The uniqueness statement is obvious. To prove the existence, 
suppose e.g.\ that $\partial_+ \theta^+ = 0$ on $U\cong U^+\times U^-$. 
Then we can define a function $K^+$ on $U^+\times U^- \cong U$ by 
\[ K^+(z_+,z_-) := \int_{(0,z_-)}^{(z_+,z_-)}\theta^+\, ,\] 
where $z_\pm := (z_\pm^1, \ldots , z_\pm^n)$ and the integration is
over any path from $(0,z_-)$ to $(z_+,z_-)$ contained in 
$U^+\times \{ z_-\}$. The condition  $\partial_+\theta^+ = 0$
ensures that the integral is path independent. In fact, it implies that  
the one-form ${\theta^+|}_{U^+\times \{ z_-\}}$ is closed
and hence exact, since $U^+$ is simply connected. \qed

In virtue of the lemma we can choose two real-valued functions $K^\pm$
such that $\p_\pm K^\pm = \q^\pm$. Putting $K := K^+ - K^-$, we 
obtain
\[ \partial_- \partial_+ K = \partial_- \partial_+ K^+ + 
\partial_+ \partial_- K^- = \partial_- \q^+ + \partial_+\q^- = \o\, .\] 
It is clear that the function $K$ is unique up to adding a solution of 
$\partial_- \partial_+ f = 0$. We have to show that any solution is of the 
form $f = f_+ + f_-$, 
where $\p_\mp f_\pm = 0$. Expanding 
\[ \partial_+ f = \sum f_i^+dz_+^i\, , \quad \mbox{with} \quad f_i^+ := 
\frac{\p f}{\p z_+^i}\, ,\] 
we get 
\[ 0 = \partial_-\partial_+ f = \sum \frac{\p f_i^+}{\p z_-^j}dz_-^j\wedge
dz_+^i\, .\] 
Thus $\frac{\p f_i^+}{\p z_-^j} = 0$ and the functions $f_i^+$ 
depend only on the `positive' coordinates $z_+ = (z_+^1, \ldots , 
z_+^n)$; $f_i^+ = f_i^+(z_+)$. This implies that
\[ f(z_+,z_-) = \sum \int_0^{z_+} f_i^+(\zeta )d\zeta^i + f_-(z_-)\, ,\]
where $\zeta = (\zeta^1,\ldots , \zeta^n)$ and $f_-(z_-) = f(0,z_-)$  
(the `constant of integration') is a
function of the `negative' coordinates $z_- = (z_-^1, \ldots , 
z_-^n)$ alone. 
The path integral is well defined since $U^+$ is simply connected. 
Setting $f_+(z_+) :=\sum \int_0^{z_+} f_i^+(\zeta )d\zeta^i$ we 
obtain the desired decomposition $f(z_+,z_-) = f_+(z_+) + f_-(z_-)$. 

Now we prove the converse. Let $K$ be a real-valued function 
on some open subset $U \subset M$ such that $\o := \partial_- \partial_+ K$
is a non-degenerate two-form. Since $\o$ is automatically closed it is
a symplectic structure. Moreover, it is of type $(1,1)$, which is equivalent 
to $I^*\o = -\o$. This implies that $g := \o (I\cdot , \cdot )$ is symmetric, 
\[ g(X,Y) = \o (IX,Y) = (I^*\o)(X,IY) = - \o (X,IY) = \o (IY,X) = g(Y,X)\, ,\]
and hence a pseudo-Riemannian metric. Moreover, $I^*g = -g$, {\it i.e.},   
$g$ is para-Hermitian. Now the theorem follows from Theorem 
\ref{dwNKaehlerThm}. 
\qed

\subsection{Affine special para-K\"ahler manifolds \label{subsec2.3}}
\bd An {\sf (affine) special para-K\"ahler manifold} $(M,I,g,\n )$ is a 
para-K\"ahler manifold $(M,I,g)$ endowed with a flat torsion-free
connection $\n$ such that
\begin{enumerate}
\item[(i)] $\n$ is symplectic, i.e.,\ $\n \o = 0$ and 
\item[(ii)] $\n I$ is a symmetric (1,2)-tensor field, i.e.,\ 
$(\n_XI)Y = (\n_YI)X$ for all $X,Y$.   
\end{enumerate} 
\label{Def9}
\ed 

We will not discuss {\it projective} special para-K\"ahler manifolds
in this paper. Therefore, in the following, we will do without the
adjective `affine' and speak simply 
of  {\it special para-K\"ahler manifolds}. 
Now we give an extrinsic construction of special para-K\"ahler manifolds
from certain para-holomorphic immersions into the para-complex 
vector space $V = T^*C^n \cong C^{2n}$. 

The cotangent bundle $N = T^*M$
of any para-complex manifold $M$ carries a canonical non-degenerate exact  
$C$-valued two-form $\O$ of type $(2,0)$ which is para-holomorphic, 
in the sense
that it defines a para-holomorphic section of the para-holomorphic
vector bundle $\wedge^{2,0}T^*N$. We will now describe $\O$ explicitly. 
If $(z^1,\ldots ,z^n)$ are local 
para-holomorphic
coordinates on $U \subset M$, then any point of 
$T_p^*M  = {\rm Hom}_{\R} (T_pM,\R ) \cong {\rm Hom}_C(T_pM,C)$,
$p\in M$, is of   
the form $\sum w_idz^i|_p$. Here ${\rm Hom}_C(T_pM,C)$ denotes the 
vector space of homomorphisms from the para-complex vector space $(T_pM,I_p)$ 
into the para-complex vector space $C$. As usual, the $z^i$ and $w_j$ 
can be considered as locally defined (para-)holomorphic functions
on $T^*M$. They form a system of para-holomorphic coordinates on the bundle 
$T^*M|_U$. The functions $w_j$ induce a system of linear para-holomorphic 
coordinates on each fibre $T_p^*M$ for all $p\in U$. In the 
para-holomorphic coordinates $(z^i,w_j)$ the two-form $\O$ is given by
\[ \O = \sum dz^i\wedge dw_i = - d(\sum w_idz^i)\, .\]
One can check that the one-form $\sum w_idz^i$ does not depend on the 
choice of coordinates. Therefore, $\O$ is also coordinate independent. 
We will call it {\it the symplectic form} of $T^*M$.  

In the following, $V$ will always denote the para-holomorphic 
vector space $V = T^*C^n \cong C^{2n}$ endowed with its standard
para-complex structure $I_V$, its symplectic 
form $\O$ and with the para-complex conjugation $\tau : V \ra V$,
$v \mapsto \tau (v) := \ol{v}$, whose fixed point set is 
$V^\tau := T^*\R^n \cong 
\R^{2n}$. We can choose a system of  linear para-holomorphic 
coordinates $(z^i)$ 
on $C^n$ which are real-valued on the subspace
$\R^n \subset C^n = \R^n \oplus e\R^n$.  The corresponding 
para-holomorphic  coordinates  $(z^i,w_j)$ on $V$ are linear 
and real-valued on the subspace $V^\tau$. The algebraic data
$(\O,\tau )$ on $V$ induce a para-Hermitian scalar product
$g_V$ on $V$ by 
\[ g_V(v,w) := {\rm Re} \big( e\O (v,\tau (w))\big) 
\quad \forall v, w\in V\, .\]
$(V,I,g_V)$ is a flat para-K\"ahler manifold, whose para-K\"ahler form 
$\o_V$ is given by
\[ \o_V(v,w) := g_V(I_Vv,w) = {\rm Im}\big( e\O (v,\tau (w))\big)  
\quad \forall v, w\in V\, . \]

\noindent
{\bf Remark 1:} The combination $\gamma_V := g_V + e\o_V = e\O 
(\cdot , \tau \cdot )$ is a {\it para-Hermitian form}, in the sense that 
it  is $C$-sequilinear and 
$\gamma (w,v) =  \overline{\gamma (v,w)}$ for all $v,w\in V$. It is obviously 
non-degenerate.

Let $(M,I)$ 
be a (connected) para-complex manifold of para-complex dimension $n$. 

\bd A para-holomorphic immersion $\phi : M\ra V$ is called 
{\sf para-K\"ahlerian} if $g := \phi^*g_V$ is non-degenerate 
and {\sf Lagrangian} if $\phi^*\O = 0$.   
\ed    

The following proposition is an easy consequence of Theorem 
\ref{dwNKaehlerThm}, since the pull-back of a closed form 
is closed. 
\bp
Any para-K\"ahlerian immersion $\phi : M\ra V$ induces on $M$ 
the structure of a para-K\"ahler manifold $(M,I,g)$ with
para-K\"ahler form $\o = g (I\cdot , \cdot ) = \phi^*\o_V$. 
\ep

\noindent
{\bf Remark 2:}  For any para-K\"ahler manifold $(M,I,g)$, with
para-K\"ahler form $\o$, the combination $\gamma = g +e\o$ is a field of 
non-degenerate para-Hermitian forms. For a 
para-holomorphic immersion $\phi : M\ra V$ the following conditions are
equivalent:\\
(i) $\phi$ is para-K\"ahlerian, {\it i.e.}, $g = \phi^*g_V$ is non-degenerate\\
(ii) $\o = \phi^*\o_V$ is non-degenerate and\\
(iii) $\phi^*\gamma_V$ is non-degenerate.\\   
Under these assumptions, $\gamma = g +e\o = \phi^*\gamma_V$.

\bl Let $\phi : M\ra V$ be a para-K\"ahlerian Lagrangian immersion. Then 
the para-K\"ahler form $\o = g(I\cdot , \cdot ) = - g(\cdot ,I\cdot)$  of $M$ 
is given by 
\[ \o = 2 \sum d\tilde{x}^i\wedge d\tilde{y}_i \, ,\]
where $\tilde{x}^i := {\rm Re}\, \phi^*z^i$ and 
$\tilde{y}_i := {\rm Re}\, \phi^*w_i$. 
\el

\pf 
Since 
$\O = \sum dz^i\wedge dw_i =  \sum (dz^i\otimes dw_i - dw_i\otimes dz^i)$, the
para-Hermitian form $\g_V = e\O (\cdot , \tau \cdot )$ is given by 
\be \g_V =   e \sum (dz^i\otimes d\ol{w}_i - dw_i\otimes  
d\ol{z}^i)\, . \label{gVEq}
\ee 
Decomposing the functions $z^i$ and $w_i$ into their real
and imaginary parts,   
\[ z^i = x^i +eu^i\, ,\quad w_i = y_i+ev_i\, ,\]
we obtain 
\begin{eqnarray*} \o_V &=& {\rm Im}\, \g_V 
= \frac{1}{2} (\sum (dz^i\otimes d\ol{w}_i - dw_i\otimes 
d\ol{z}^i) + \sum (d\ol{z}^i\otimes dw_i - d\ol{w}_i\otimes 
dz^i))\\ 
&=& \sum ( dx^i\wedge dy_i - du^i\wedge dv_i)\, .
\end{eqnarray*}  
On the other hand, 
\[ {\rm Re}\, \O = \sum ( dx^i\wedge dy_i + du^i\wedge dv_i)\, .\]
{}From the assumption $\phi^*\O = 0$ we obtain 
\[ 0 = \phi^*({\rm Re}\, \O) = \phi^*(
\sum ( dx^i\wedge dy_i + du^i\wedge dv_i))\]
and hence
\[ \o = \phi^*\o_V =  \phi^*(
\sum ( dx^i\wedge dy_i - du^i\wedge dv_i)) = 2\phi^* (\sum  dx^i\wedge dy_i)
= 2\sum d\tilde{x}^i\wedge d\tilde{y}_i\, .\] 
\qed  
 
\bc Let $\phi : M\ra V$ be a para-K\"ahlerian Lagrangian immersion. Then 
there exists a canonical flat torsion-free connection $\n$ on $M$. 
It is characterized by the condition that $\n ({\Re}\, \phi^*df) =0$ for
all complex affine functions $f$ on $V$. 
\ec 

\pf The lemma implies that any point $p\in M$ has an open neighborhood
$U$ on which the functions $(\tilde{x}^1, \ldots ,\tilde{x}^n,
\tilde{y}_1, \ldots ,\tilde{y}_n)$ form a system of local coordinates
and, hence, define a flat torsion-free connection $\n^U$ on $U$. 
Obviously, $\n^{U_1} = \n^{U_2}$ on the overlap\linebreak[3] $U_1\cap U_2$
of two such coordinate domains. Therefore, there exists a unique 
flat torsion-free connection $\n$ on $M$ which is the common extension of
the locally defined connections $\n^U$. It is characterized by
the system of equations $\n ({\Re}\, \phi^*dz^i) = 
\n ({\Re}\, \phi^*dw_i) = 0$, $i = 1,2, \ldots, n$. \qed   

\bt \label{mainThm} 
Let $\phi : M\ra V$ be a para-K\"ahlerian Lagrangian immersion with
induced geometric data $(I, g, \n )$. Then  $(M, I, g, \n )$ is a
special para-K\"ahler manifold. Conversely, any  simply connected 
special para-K\"ahler manifold $(M, I, g, \n )$
admits a para-K\"ahlerian Lagrangian immersion inducing the 
special geometric data $(I, g, \n )$ on $M$. The para-K\"ahlerian Lagrangian 
immersion $\phi$ is unique up to an affine transformation of $V$
whose linear part belongs to the group ${\rm Aut}_C(V, \O , \tau ) = 
{\rm Aut}_{\R} (V, I_V, \O , \tau ) = 
{\rm Sp}(\R^{2n})$.  
\et 

\pf Let $\phi : M\ra V$ be a para-K\"ahlerian Lagrangian immersion with
induced geometric data $(I, g, \n )$. We have to check that
i) $\n \o = 0$ and that ii) $\n I$ is symmetric. The first condition
is satisfied since $\o$ has constant coefficients with respect to 
the\linebreak[3] 
$\n$-affine local coordinates $(\tilde{x}^1, \ldots ,\tilde{x}^n,
\tilde{y}_1, \ldots ,\tilde{y}_n)$. In order to 
verify the second condition, we evaluate   
a $\n$-parallel\linebreak[3] $1$-form $\xi$ on the alternation of $\n I$: 
\[ \xi ((\n_XI)Y- (\n_YI)X) = (\n_X\xi\circ I)Y- (\n_Y\xi \circ I)X 
= d(\xi \circ I)(X,Y)\]
for all vector fields $X$ and $Y$ on $M$.  Here we have used that
$\n$ is torsion-free. Since $\n$ is flat it is sufficient to check that
the one-form $\xi \circ I$ is closed for all $\n$-parallel $1$-forms $\xi$.
Any $\n$-parallel $1$-form is a linear combination with constant coefficients 
of the differentials $d\tilde{x}^i$ and $d\tilde{y}_i$. So we have 
to show that the one-forms $d\tilde{x}^i\circ I$ and $d\tilde{y}_i\circ I$
are closed. 

The functions $z^i$ and $w_i$ on $V$ are para-holomorphic. Since the map 
$\phi : M \ra V$ is para-holomorphic, this implies that the one-forms
$\phi^*dz^i$ and $\phi^*dw_i$ on $M$ are of type $(1,0)$. Thus
\begin{eqnarray*} \phi^*dz^i  &=& d\tilde{x}^i +ed\tilde{x}^i\circ I\\
\phi^*dw_i &=& d\tilde{y}_i + e d\tilde{y}_i\circ I\, .
\end{eqnarray*}
These equations show that the one-forms $d\tilde{x}^i\circ I$ 
and $d\tilde{y}_i\circ I$ are closed, which proves ii). 

Now we prove the converse statement. Let $(M, I, g, \n )$ be a 
simply connected special para-K\"ahler manifold. We have to construct
a para-K\"ahlerian Lagrangian immersion $\phi : M \ra V$, which induces
the special geometric structures on $M$. Since the symplectic
structure $\o$ is $\n$-parallel, for any point $p\in M$ there exists a 
system of local
$\n$-affine coordinates  $(\tilde{x}^1, \ldots ,\tilde{x}^n,
\tilde{y}_1, \ldots ,\tilde{y}_n)$ defined on some simply connected open 
neighborhood $U\subset M$ of $p$ such that $\o = 
2 \sum d\tilde{x}^i\wedge d\tilde{y}_i$.  The coordinate system 
near $p$ is unique up to an affine symplectic transformation. 
We define $2n$ one-forms $\o^i$ and $\eta_i$ of type $(1,0)$ by
\begin{eqnarray*} \o^i  &=& d\tilde{x}^i +ed\tilde{x}^i\circ I\\
\eta_i &=& d\tilde{y}_i + e d\tilde{y}_i\circ I\, .
\end{eqnarray*}
The symmetry of 
$\n I$ implies that these forms are closed, by the same calculation 
as above. It follows that $\o^i$ and $\eta_i$ are closed 
para-holomorphic sections of $\wedge^{1,0}T^*M$. As $U$ is simply connected,
there exists para-holomorphic functions $\tilde{z}^i$ and $\tilde{w}_i$ 
defined at $q\in U$ by 
\begin{eqnarray*} \tilde{z}^i(q)  &:=& \tilde{x}^i(p) + \int_p^q\o^i \,,\\
\tilde{w}_i(q) &:=&  \tilde{y}_i(p) + \int_p^q\eta_i \, ,
\end{eqnarray*} 
where the integral is over any path $c \subset U$ from $p$ to $q$. 
Thus $\o^i = d \tilde{z}^i$, $\eta_i = d\tilde{w}_i$ and 
$\tilde{x}^i = {\rm Re}\, \tilde{z}^i$, $\tilde{y}_i = {\rm Re}\,  
\tilde{w}_i$. 

Now we can define a para-holomorphic map $\phi_U : U \ra V \cong C^{2n}$ by
the equations
\[ \phi_U^*z^i := \tilde{z}^i\quad \mbox{and} \quad \phi_U^* w_i :=
\tilde{w}_i\, .\]
We claim that $\phi_U : (U, I, g, \n ) \ra V$ is a para-K\"ahlerian Lagrangian 
immersion with induced geometric data $(g, \n )$. To see that
$\phi_U$ is an immersion (and even an embedding), it suffices to remark that 
$\psi_U := {\rm Re}\, \phi_U : U \ra V^{\tau} \cong \R^{2n}$ is given by
\[ \psi_U^*x^i = \tilde{x}^i\quad \mbox{and} \quad \psi_U^*y_i = \tilde{y}_i
\, ,\]
which is obviously a diffeomorphism of $U$ onto its image. 
Using the formulas $\O = \sum dz^i\wedge dw_i$, 
$\phi_U^*z^i = d\tilde{x}^i +ed\tilde{x}^i\circ I$, 
$\phi^*dw_i = d\tilde{y}_i + e d\tilde{y}_i\circ I$ and 
$\o = 2 \sum d\tilde{x}^i\wedge d\tilde{y}_i$ we calculate
\[ \phi_U^*\O = \frac{1}{2}(\o + \o (I \cdot , I \cdot )) 
+ \frac{1}{2}(\o (I\cdot , \cdot ) + \o (\cdot, I \cdot )) = 0 \, ,\] 
which  shows that $\phi_U$ is Lagrangian. 
Similarly, using the formulas $\o_V = (dx^i\wedge dy_i - du^i\wedge dv_i)$, 
$\phi_U^*dx^i = d\tilde{x}^i$, $\phi_U^*du^i = d\tilde{x}^i\circ I$,
$\phi_U^*dy_i = d\tilde{y}_i$ and $\phi_U^*dv_i= d\tilde{y}_i\circ I$,  
we calculate 
\[ \phi_U^*\o_V = \sum (d\tilde{x}^i\wedge d\tilde{y}_i -  
(d\tilde{x}^i\circ I)\wedge (d\tilde{y}\circ I)) = 
\frac{1}{2} (\o - I^*\o ) = \o \, . \]
This shows that the immersion $\phi_U$ is para-K\"ahlerian with 
para-K\"ahler form $\phi_U^*\o_V = \o$ and para-K\"ahler metric 
$\phi_U^*g_V = g$. The 
flat torsion-free connection induced by the para-K\"ahlerian
Lagrangian immersion $\phi_U$ coincides with the special connection $\n$,
since the functions ${\rm Re}\, \phi_U^*z^i = \tilde{x}^i$ and
${\rm Re}\, \phi_U^*w_i = \tilde{y}_i$ form a system of $\n$-affine
coordinates on $U$. It is clear that the germ $\phi_p$ of the para-K\"ahlerian
Lagrangian immersion $\phi_U$ at $p$ is unique up to an affine transformation
of $V$ with linear part in ${\rm Sp}(\R^{2n})$ and that this germ
has a unique continuation $\phi_{c(t)}$ along any path 
$c : [0,1] \ra M$ from $p$ to $q$. 
Since $M$ is simply connected, the value $\phi_q := \phi_{c(1)}$  
of the above continuation at the endpoint is independent of $c$.  
Therefore, the para-K\"ahlerian Lagrangian immersion 
$\phi_U :U\ra V$ extends uniquely to a para-K\"ahlerian
Lagrangian immersion $\phi : M \ra V$ with the claimed properties. 
\qed

The theorem implies that any special para-K\"ahler manifold
can be locally constructed from a para-holomorphic prepotential, as
we shall explain now. 
For any para-holomorphic function $F : U\ra C$ on an open subset 
$U \subset C^n$
we can consider the para-holomorphic Hessian 
\[ \partial^2 F := (\ft{\partial^2F}{\partial z^i \partial z^j}) \, .\] 
\bc Let $F : U\ra C$ be a para-holomorphic function
defined on an open subset $U \subset C^n$ which satisfies  
the non-degeneracy condition $\det (\Im \partial^2F) \neq 0$. 
Then $\phi_F := dF : U \ra T^*C^n = V$ is a para-K\"ahlerian 
Lagrangian immersion
and hence defines a special para-K\"ahler manifold $M_F = (U,I,g,\n)$. 
Conversely, any special para-K\"ahler manifold is locally isomorphic
to a manifold of the form $M_F$. 
\label{Cor2}
\ec  

\pf It is clear that $\phi_F : U \ra V$ is a para-holomorphic 
Lagrangian immersion. In fact, its components 
\[ \phi_F^*z^i = z^i\quad \mbox{and} \quad \phi_F^*w_i = 
\frac{\partial F}{\partial z^i}\] 
are para-holomorphic functions and 
\[ \phi_F^*\O = \phi_F^*(-d\sum w_idz^i) = -d\phi_F^*( \sum w_idz^i) =
-ddF = 0
\, .\]
Let us check, with the aid of equation \re{gVEq},  
that the non-degeneracy condition implies that 
$\phi_F$ is para-K\"ahlerian:   
\begin{eqnarray}
\nonumber
 \phi_F^*\gamma_V &=& e \sum (dz^i\otimes d
(\frac{\partial \ol{F}}{\partial \ol{z}^i}) - 
d(\frac{\partial F}{\partial z^i})\otimes  
d\ol{z}^i) = 
e \sum (\ft{\partial^2\ol{F}}{\partial \ol{z}^i \partial \ol{z}^j}
- \ft{\partial^2F}{\partial z^i \partial z^j})dz^i\otimes 
d\ol{z}^j\\ {}  &=& - 2 \sum ({\rm Im}\, \ft{\partial^2F}{\partial z^i \partial z^j})
dz^i\otimes d\ol{z}^j\, .\label{Prep2Metric} \end{eqnarray}
This shows that $\phi_F$ is a para-K\"ahlerian Lagrangian immersion. 

Conversely, by Theorem \ref{mainThm}, we know that any  
special para-K\"ahler manifold is locally defined by a 
para-K\"ahlerian Lagrangian immersion $\phi_U: U\ra V$, 
where $U$ is (biholomorphic to) a simply connected 
open subset of $C^n$. Restricting $U$ and composing $\phi_U$ with 
an affine transformation of $V$ whose linear part is 
in ${\rm Sp}(\R^{2n})$, if necessary, we can assume
that the functions $z^i$ on $V$ induces a global system
of para-holomorphic coordinates $(\phi^*_Uz^i)$ on $U$. 
(Equivalently, we can choose the canonical coordinates $(z^i,w_j)$ 
of $V$ adapted to the immersion.) Then we can write the 
Lagrangian submanifold $\phi_U(U) \subset V$ as a graph: 
$w_i = F_i(z)$, $i = 1, 2, \ldots ,n$. The Lagrangian condition
$\phi^*_U\O = \sum dz^i \wedge dF_i = 0$ is equivalent to the para-holomorphic 
one-form $\sum F_idz^i$ on $U$ being closed and hence exact, since $U$ is 
simply connected. This proves the existence of a para-holomorphic 
function $F : U\ra C$ such that $\phi_U = \phi_F$. Finally, by the 
previous calculation, the K\"ahlerian condition on the holomorphic immersion 
$\phi_U = \phi_F$ is equivalent to $\det (\Im \partial^2F) \neq 0$. 
\qed   

\noindent 
{\bf Remark 3:} The above construction can be easily 
reformulated in purely real
terms without using para-complex numbers. In fact, an immersion 
$\phi : M\ra V$ of a real differentiable manifold of dimension 
$2n = \frac{1}{2}\dim_{\R} V$ satisfies $\phi^*\O = 0$ if and only if it is 
Lagrangian with respect to the two real symplectic structures 
\[ \O_\pm  = \frac{1}{2}\sum 
(dz^i_+\wedge dw_i^+ \pm dz^i_-\wedge dw_i^-)\, ,\]
where $z_\pm^i = x^i\pm u^i$ and $w_i^\pm = y_i\pm v_i$. The condition 
$\phi^*\O = 0$  implies immediately that $\phi$ induces a para-complex 
structure on  $M$ such that $\phi : M\ra V$ is para-holomorphic. However, 
it seems more natural to us to describe para-complex manifolds using
para-complex numbers.


\begin{section}{Fermions and Supersymmetry in 5 and 4 Dimensions
\label{Section3}}

In the following section we review the relevant properties of
Clifford algebras, fermions and supersymmetry algebras. 
We start in section 3.1 with those properties which can be discussed 
in arbitrary dimension and signature. In section 3.2 we
specialize to 5 dimensions. We derive realizations of the 
minimal supersymmetry algebra in terms of Dirac and of
symplectic Majorana spinors and show that the R-symmetry 
group is ${\rm SU}(2)$. In section 3.3 we obtain supersymmetry
algebras in dimensions $(1,3)$ and $(0,4)$ by dimensional
reduction over space and time, respectively. The R-symmetry
group now contains an additional Abelian factor, which is
${\rm U}(1)$ for Minkowski, but ${\rm SO}(1,1)$ for Euclidean space-time
signature.

Our discussion of Clifford algebras and spinors combines 
\cite{AC,con1,ACDV03}, where more details can be found.
This section should be 
readable for physicists as well as for mathematicians. 
Note that in this section and in \cite{AC,ACDV03}
spinors are taken to 
be commuting (complex-valued) objects, whereas in 
sections 4 and 5, and in \cite{con1} anticommuting
(Grassmann-valued) spinor fields\footnote{See subsection 
\ref{ExplGrass} for a mathematical definition.} are used,
as usual in supersymmetric
field theories.

\subsection{Clifford algebras, spinors and supersymmetry in $(t,s)$ dimensions}
\label{Cliffordsubsection} 
Let us first consider space-times with $t$ time-like and $s$ space-like 
dimensions. Space-time indices will be denoted by  
$\mu,\nu, \ldots$  More precisely, we consider the vector space 
$V := \R^n$, $n = t+s$, endowed
with the pseudo-Euclidean scalar product $\eta$ given by 
$\eta_{\mu\nu} := \eta (e_\mu , e_\nu )$~$=$~$\mbox{diag}(-1, \cdots , -1, +1, 
\cdots , +1)$, with respect to the standard basis $(e_\mu )$ of $\R^n$. 
The negative (positive) values correspond to time-like 
(space-like) directions. We
will say that such a space-time is $(t,s)$--dimensional. 

\subsubsection*{The Clifford algebra}   
The Clifford algebra $Cl_{t,s}$ can be defined as the 
algebra generated by an orthonormal basis $(e^\mu )$ of the dual
vector space $V^*$  subject to
the defining relations\footnote{In the mathematical literature 
one often finds a minus sign on the
RHS of equation (\ref{CliffAlg1}). If one simultaneously
defines the space-time metric with a minus sign relative to our choice,
as for example in \cite{LawMic}, then 
$Cl_{t,s}$ denotes the same algebra as in the present paper. There are, 
however, also authors who denote our $Cl_{t,s}$ by $Cl_{s,t}$.}
\begin{equation} \label{CliffAlg1} 
e^\mu e^\nu + e^\nu e^\mu = 2\,\eta^{\mu\nu}1 \, ,
\end{equation} 
where $\eta^{\mu\nu}$ is the inverse of $\eta_{\mu\nu}$ and $1$ is the
unit in the algebra. 
The gamma-matrices 
$\gamma^{\mu} = (\gamma^{\mu\;\;\beta}_{\;\;\alpha})$ are 
the matrices  which represent the generators $e^\mu$ of $Cl_{t,s}$ 
in an irreducible representation of the complex Clifford algebra
$\Com l_n = Cl_{t,s}\otimes \Com$. They are complex
square matrices of size $2^{[n/2]}$ satisfying the Clifford 
relations 
\begin{equation}
  \left\{\gamma^\mu, \gamma^\nu\right\} := \gamma^{\mu} \gamma^{\nu}
  + \gamma^{\nu} \gamma^{\mu} =
  2\,\eta^{\mu\nu}\id\;,
\label{CliffAlg}
\end{equation}
where $\id$ is the identity map $\Com^{2^{[n/2]}} \ra \Com^{2^{[n/2]}}$. 

\subsubsection*{The spin group and the spinor module} 
The group ${\rm Spin}(t,s) \subset Cl_{t,s}$ which consists of 
products of an even number of unit vectors  ({\it i.e.} $v\in V^*$ 
such that $\eta (v,v) = \pm 1$)  
is called the {\it spin group}. The maximal connected subgroup  
${\rm Spin}_0(t,s) \subset {\rm Spin}(t,s)$ is called the 
{\it connected spin group}. The ${\rm Spin}(t,s)$-representation 
which is obtained by restricting an irreducible $\Com l_n$-representation
to  ${\rm Spin}(t,s) \subset \Com l_n$, is called
{\it the complex spinor representation}. 
The corresponding representation space $\Som$ 
is called {\it the complex spinor module} and its elements 
$\lambda = (\lambda_{\alpha})$ are called {\it Dirac spinors}. 
The complex spinor module is an irreducible complex module if the 
space-time dimension 
$n=t+s$ is odd. It is the sum of two irreducible complex modules  
({\em Weyl spinors}) $\Som_+$ and $\Som_-$ if $n=t+s$ is even. 
 
The Lie algebra ${\rm spin}(t,s)
\subset Cl_{t,s}$ of the spin group is generated by
the commutators $[e^\mu , e^\nu ] = e^\mu e^\nu - e^\nu e^\mu$,
which are represented by the matrices 
\begin{equation}
  2\gamma^{\mu\nu} := 2\gamma^{[\mu} \gamma^{\nu]} := 
\gamma^\mu\gamma^\nu - \gamma^\nu\gamma^\mu = 
[\gamma^{\mu}, \gamma^{\nu}] 
\;.
\label{LorRep}
\end{equation}
This Lie algebra is canonically isomorphic to the Lorentz Lie algebra 
${\rm so}(t,s) = Lie\, {\rm SO}(t,s)$. In fact, there is a 
canonical double covering of Lie groups $R : {\rm Spin}(t,s) 
\ra {\rm SO}(t,s)$, $g \mapsto R(g)$, given by $R(g)v = gvg^{-1}\in 
\R^n$, for all $v\in \R^n$, where the product is multiplication 
in the Clifford algebra. 
We recall that,  
by definition, {\it the real spinor module} $S$  
of ${\rm Spin}(t,s)$ is the restriction 
of an irreducible  module of the real Clifford algebra $Cl_{t,s}$ 
to ${\rm Spin}(t,s) \subset Cl_{t,s}$. 
As a ${\rm Spin}(t,s)$-module it is either irreducible or a sum
of two irreducible {\it semi-spinor} modules $S_+$ and $S_-$ (which may
happen to be equivalent).    
 
\subsubsection*{The Dirac conjugate and the corresponding 
invariant sesquilinear form on 
$\Som$}  
Following \cite{con1} we  
take  the space-like and time-like gamma matrices to be 
Hermitian and anti-Hermitian, respectively.
One can then always find a matrix $A = (A^{\alpha \beta})$ 
with the property
\be\label{gammaAEqu} 
(\gamma^{\mu})^{\dagger} = (-1)^t A \gamma^{\mu} A^{-1} \;.
\ee 
An explicit realization of $A$ is
\be
A = \prod_{\tau} \gamma_{\tau} \;,
\ee
where $\gamma_{\mu} := \eta_{\mu \nu}
\gamma^{\nu}$. The product is over all time-like $\gamma$-matrices.
It is taken in lexicographic order.

The above equation implies that the matrix $A$ represents
an isomorphism of ${\rm spin}(t,s)$-modules 
from the complex spinor module (\ref{LorRep}) to the 
complex-conjugate of its dual module (in which the generators
$[e^\mu , e^\nu ]$ of the Lie algebra ${\rm spin}(t,s)$ act by
the matrices $-2(\gamma^{\mu\nu})^\dagger$).    
The matrix $A$ features in the definition of the {\it Dirac conjugate} 
spinor 
\begin{equation} \label{DiracconjEqu} 
\overline{\lambda}^{(D)} := \lambda^{\dagger} A\, .
\end{equation}  
The map  
\begin{equation}\label{sesquilinEqu}
\Som \times \Som \ni (\lambda , \chi ) \mapsto \overline{\lambda}^{(D)}\chi
= \lambda^{\dagger} A\chi  
=  \lambda^{*}_{\alpha} A^{\alpha \beta} \chi_{\beta}
\in \Com \end{equation} 
defines a non-degenerate ${\rm Spin}_0(t,s)$-invariant sesquilinear  
form $h_\Som$ on $\Som$, {\it i.e.}, it is antilinear 
in $\lambda$  and linear in $\chi$. Moreover,  
it is either Hermitian or skew-Hermitian depending on the value of $t$. 
This follows from the fact that 
\begin{equation} \label{AEqu} A^\dagger = (-1)^{\frac{t(t+1)}{2}}A\, .
\end{equation}

\noindent {\bf The Majorana conjugate and the corresponding invariant 
bilinear form $C$ on $\Som$}\\ 
Another operation considered in the physical literature 
is the {\it Majorana conjugation}   
\be
\lambda \mapsto \overline{\lambda} := \lambda^{T} \CC \;.
\label{MajConjEqu} 
\ee 
 The {\it charge conjugation matrix} $\CC=(\CC^{\alpha \beta})$ 
satisfies 
\begin{equation}\label{T1}
  \CC^\rmT = \sigma \,\CC\;,\quad
  \gamma^{\mu\,\rmT} = \tau \, \CC\gamma^\mu \CC^{-1}\;,
\end{equation}
where  $\sigma$ and $\tau$  
can take the values $\pm 1$, depending 
on the dimension and signature of space-time and on the choice of 
$\CC$, see \cite{AC,con1} for a list of the possible values  
$\sigma$ and $\tau$.\footnote{We use here the
notation of \cite{AC}, explained below, where $\sigma$ stands for symmetry and
$\tau$ for type, which correspond to $-\varepsilon$ and $-\eta$
in \cite{con1}.}   The Majorana conjugation   
relates the complex spinor representation (\ref{LorRep}) 
to its dual representation; it represents an  
isomorphism of ${\rm spin}(t,s)$-modules. It can be considered as a
non-degenerate complex bilinear form $C$ 
\begin{equation} \label{bilinEqu} 
\Som \times \Som \ni (\lambda , \chi ) \mapsto \overline{\lambda}\chi = 
\lambda^T\CC \chi = \lambda_{\alpha} \CC^{\alpha \beta} \chi_{\beta}
\in \Com \end{equation}  
on the complex spinor module. The Majorana conjugation is  
invariant under the complex spin group ${\rm Spin}(n,\Com ) \subset
\Com l_n$, $n = t+s$, which is the 
connected Lie group consisting of even products of unit vectors  
$v\in V^*\otimes \Com$, $\eta^\Com (v,v) = 1$, with respect to the 
complexified scalar product $\eta^\Com$. Note that ${\rm Spin}(n,\Com )
\supset {\rm Spin}_0(t,s)$. 
 
All bilinear forms on the 
(real or complex) spinor module invariant under the connected 
(real or complex) spinor group were described in \cite{AC}, where the
notion of an {\it admissible} bilinear form on the spinor module 
was introduced. For such a form $\beta$ two fundamental invariants, which take
the values $\pm 1$, are defined: 
the {\it symmetry} $\sigma (\beta )$ and  and the 
{\it type} $\tau (\beta )$. In the case of $\beta = C$ these invariants
correspond precisely 
to $\sigma$ and $\tau$ in (\ref{T1}). In general,
a bilinear form $\beta$ is symmetric if  $\sigma (\beta ) = +1$ and 
skew-symmetric if $\sigma (\beta ) = -1$. The operators $\gamma^\mu$ 
are symmetric with respect 
to $\beta$ if $\tau (\beta )= +1$ and skew-symmetric if $\tau (\beta ) = -1$. 
A bilinear form $\beta$ for which the invariants 
$\sigma (\beta )$ and $\tau (\beta )$
are defined is automatically ${\rm Spin}_0(t,s)$-invariant, as was shown in 
\cite{AC}. So admissibility implies ${\rm Spin}_0(t,s)$-invariance. 

A basis of the vector space of
${\rm Spin}_0(t,s)$-invariant bilinear forms on the spinor module 
which consists of non-degenerate admissible forms was constructed 
in \cite{AC} for all $(t,s)$. 
For even space-time dimension, the vector space of
${\rm Spin}_0(t,s)$-invariant (and hence ${\rm Spin}(n,\Com )$-invariant) 
complex bilinear forms on the 
complex spinor module is two-dimensional. This corresponds to the fact that 
there are two independent 
matrices $\CC$ which satisfy 
(\ref{T1}) with different values of $\sigma$ and $\tau$. An explicit
example can be found in the appendix.

\subsubsection*{Majorana and symplectic Majorana spinors} 

In many physical applications one uses Weyl or Majorana spinors
instead of Dirac spinors. Weyl spinors exist if the space-time
dimension is even, $n=t+s$, as already discussed above. They are 
obtained from Dirac spinors by imposing a chirality constraint,
and we will come back to them in section 3.3. Majorana
spinors exist if one can impose a reality constraint. 
If $t-s = 0, 1, 2, 6$ and $7\!\pmod  8$  
one can impose a reality constraint on a single Dirac spinor, 
leading to Majorana spinors and 
for $t-s = 0\!\pmod 8$  also to Majorana-Weyl spinors. In mathematical terms
this means that the complex spinor representation admits a 
${\rm Spin}(t,s)$-invariant real structure ({\it i.e.}, antilinear 
involution). 
In the remaining
cases, $t-s = 3, 4$ and $5\!\pmod 8$, the complex spinor representation
admits a ${\rm Spin}(t,s)$-invariant quaternionic structure 
({\it i.e.}, antilinear map which squares to $-\id$), leading to
symplectic Majorana spinors and for $t-s=4\!\pmod 8$ also to 
symplectic Majorana-Weyl
spinors. See \cite{ACDV03} for a proof of these facts, which are summarized
in Table 1 of that paper. As we will use symplectic Majorana spinors
to construct supersymmetry algebras and 
supersymmetric actions, we will give a detailed account in the following.

Reality constraints are formulated using 
$B := (\CC A^{-1})^{T}$. As a consequence of (\ref{gammaAEqu})  and
(\ref{T1}) this matrix satisfies
\be 
(\gamma^\mu)^* = 
\tau  (-1)^tB\gamma^\mu B^{-1}\, .
\label{gammacc}
\ee
One can also show that
\be
B^{\dagger} B = \id \;\;\;\mbox{and}\;\;\;
B^{*} B = \pm \id \;.
\ee
Depending on the sign in the last equation, one can define
Majorana or symplectic Majorana spinors.\footnote{Note that
the condition $t-s = 3,4,5\, \pmod 8$ is sufficient, but
not necessary for the existence of symplectic Majorana spinors.
In some cases, including $(t,s)=(1,3)$, both Majorana and
symplectic Majorana spinors exist. See the appendix for 
more details.}

If $B^* B = \id$, the map $\lambda \mapsto B^*\lambda^*$   
is a 
${\rm Spin}(t,s)$-invariant real structure  on the complex 
spinor module. Its real points satisfy the 
Majorana condition
\begin{equation} \lambda^{*} = B \lambda \;,
\label{MajCon}
\end{equation} 
which can also be expressed in the form 
$\overline{\lambda}^{(D)} = \overline{\lambda}$
where 
$\overline{\lambda}^{(D)}$
is the Dirac conjugate and $\overline{\lambda}$ is the
Majorana conjugate.

In the case $B^*B = -\id$, the map\footnote{The sign 
has been chosen in order to be consistent with the conventions of
\cite{con1}.} 
\be \label{jSEqu} \lambda \mapsto j_\Som\lambda :=- B^*\lambda^*
\ee 
is a ${\rm Spin}(t,s)$-invariant
quaternionic structure $j_\Som$ 
on the complex spinor module $\Som$. The invariance is a consequence of the
equation
\be \label{jgammaEqu} j_{\Som}(\gamma^\mu \lambda) = 
\tau  (-1)^t \gamma^\mu j_{\Som}(\lambda)\, ,
\ee 
which is equivalent to (\ref{gammacc}).
Notice that the quaternionic structure $j_\Som$ is invariant
under the group ${\rm Pin}(t,s)\subset Cl_{t,s}$ generated by unit vectors
$v\in V^*$, $\eta (v,v) = \pm 1$, if
$\tau  (-1)^t = +1$. In fact,  $j_\Som$ commutes with the  
operators $\gamma^\mu$ if 
$\tau  (-1)^t = +1$ and anti-commutes with the $\gamma^\mu$ if 
$\tau (-1)^t = -1$, which is the case e.g.\
for $(t,s) = (1,4)$.

If $B^{*} B = - \id$, one can 
impose a so-called symplectic
Majorana constraint on a pair of Dirac spinors $\lambda^i$, $i=1,2$, 
which reads
\be
(\lambda^i)^* = - B\lambda^j \epsilon_{ji} \;.
\label{SymMajCon}
\ee
Here, $\epsilon_{ij}$ is an antisymmetric two-by-two matrix, which we
take to be
\be\label{EpsMat}
(\epsilon_{ij}) =
 \left( \begin{array}{cc} 0 & 1 \\ -1 & 0 \end{array} \right)  \;.
\ee
The indices $i,j, \ldots=1,2$ are raised and lowered according 
to the so-called NW-SE convention:
$\lambda_i := \lambda^j\epsilon_{ji}$ and 
$\lambda^i = \epsilon^{ij} \lambda_j$, where $\epsilon^{ij} =
\epsilon_{ij}$ and therefore 
$\epsilon^{ik} \epsilon_{kj} = - \delta^i_j$.

In this formalism
one uses
double spinors $\lambda = (\lambda^1, \lambda^2)^T$ which are
elements of the complex 
${\rm Spin}(t,s)$-module
\[ \Som \otimes \Com^2 = 
\Som \otimes_\Com \Com^2 \cong \Som \oplus \Som \;.\]  
The tensor product 
of the quaternionic structure $j_{\Som}$ of $\Som$ 
with the standard quaternionic structure $j_{\Com^2}$ of 
$\Com^2 = \Hom$,
\[
j_{\Com^2} ( z^1 \vec{e}_1 + z^2 \vec{e}_2 ) =
\overline{z}^1 \vec{e}_2 - \overline{z}^2 \vec{e}_1 \;,
\]
defines 
a ${\rm Spin}(t,s)$-invariant real structure 
\be 
\label{realstructureEqu} \rho = j_\Som \otimes j_{\Com^2}
\ee 
on $\Som \otimes \Com^2$.
The real points of $\Som \otimes \Com^2$ are exactly the symplectic 
Majorana spinors defined by (\ref{SymMajCon}):
\be
\label{SympMaj2Dirac}
\rho \left[  \left( \begin{array}{c} \lambda^1 \\ \lambda^2 \\
\end{array}\right) 
\right]= 
\left( \begin{array}{c}
- j_{\Som}(\lambda^2) \\  j_{\Som}(\lambda^1) \\
\end{array} \right) =
\left( \begin{array}{c}
- B^* (\lambda^i)^* \epsilon_{i1} \\
- B^* (\lambda^i)^* \epsilon_{i2} \\
\end{array}\right) \;.
\ee

Symplectic  Majorana spinors form a real ${\rm Spin}(t,s)$-submodule
$\Som_{SM} \subset \Som \otimes \Com^2$, isomorphic to the complex
spinor module $\Som$. In fact, $\rho(\lambda)=\lambda$ if and only
if $\lambda^2 =  j_{\Som} (\lambda^1)$. Thus
under the above isomorphism 
$\Som \otimes \Com^2 \cong \Som \oplus \Som$, $\Som_{SM}$ corresponds to 
the graph $\{ (\psi, j_{\Som}(\psi) | \psi \in \Som \}$
of the quaternionic structure $j_\Som$ on $\Som$. 

It has often advantages to rewrite Dirac spinors as symplectic
Majorana spinors, {\it i.e.},  as pairs of
Dirac spinors subject to the symplectic Majorana condition.
For example, as we will see below, the R-symmetries of 5-
and 4-dimensional supersymmetry algebras only act on the
internal space $\Com^2$.

\subsubsection*{The induced bilinear form on the space of double spinors and 
its restriction to $\Som_{SM}$} 
The tensor product 
\be b = C \otimes \varepsilon \label{bilinformEqu} 
\ee 
of the 
bilinear form $C$  on $\Som$ introduced in (\ref{bilinEqu}) and
the antisymmetric bilinear form $\varepsilon(z,w) = z^i w^j \epsilon_{ji}$
on $\Com^2$ defines a 
non-degenerate bilinear form on the space $\Som \otimes \Com^2$ of double 
spinors: 
\be
(\lambda , \chi) \mapsto \overline{\lambda} 
\chi := \overline{\lambda}^i\chi_i = 
 \overline{\lambda}^i \chi^j \epsilon_{ji} = (\lambda^i)^T\CC  
\chi^j \epsilon_{ji} = \lambda_\alpha^i  
\chi_\beta^j \CC^{\alpha \beta}\epsilon_{ji}\;.  
\label{BilFor1}
\ee
Here $\lambda = (\lambda^i), \chi = (\chi^i) \in \Som \otimes \Com^2$.    
Without imposing the reality constraint (\ref{SymMajCon}), 
this bilinear form is 
obviously  invariant 
under the natural action of the group 
${\rm Spin}(n,\Com)\otimes {\rm Sp}(2,{\Com})$, thus
motivating the name symplectic Majorana spinor.  
It is symmetric if $\CC$ is skew-symmetric
and skew-symmetric if $\CC$ is symmetric. The group ${\rm Sp}(2,{\Com})$ is 
precisely the centralizer of ${\rm Spin}(n,\Com)$ in the group
of automorphisms of the complex bilinear form $b$. 
Either the restriction of $b$ or that of $ib$ 
to the subspace of symplectic Majorana spinors
$\Som_{SM} \subset \Som\otimes \Com^2$ is a real valued   
 bilinear form on 
$\Som_{SM}$. This follows from the fact that $\Som_{SM}$ is 
the set of fixed points of $\rho = j_\Som 
\otimes j_{\Com^2}$ and the following equations 
\begin{equation} \label{jCEqu} 
j_\Som^*C =  \pm \overline{C}\;, \;\;\; 
j_{\Com^2}^*\varepsilon = \overline{\varepsilon}\, , \;\;\;
\rho^*b = \pm \overline b \,,
\end{equation}  
where 
\[ (j_\Som^*C)(\lambda , \chi ) := 
C(j_\Som\lambda ,j_\Som\chi )\quad \mbox{and}\quad   
\overline{C}(\lambda , \chi ) := \overline{C(\lambda , \chi )}\, .\]
The first equation of (\ref{jCEqu}) is equivalent to 
\be B^\dagger C B^* = \pm C^*\, ,\label{matrixjCEqu}\ee 
which follows from (\ref{AEqu}) using the definition of $B$ and the property 
that $B^*B = -1$. The sign $\pm$ on the right hand side of 
equations (\ref{jCEqu}) and (\ref{matrixjCEqu}) 
equals $-\sigma (-1)^{t(t+1)/2}$, which is $-1$ for $(t,s) = (1,4)$,  
for example. Obviously, the restriction of $b$ to $\Som_{SM}$ is 
${\rm Spin}_0(t,s)\otimes {\rm SU}(2)$-invariant. Here ${\rm SU}(2)$ arises 
as the intersection ${\rm SU}(2) = {\rm Sp}(2,{\Com})\cap {\rm GL}(1,\Hom )$,
the subgroup of ${\rm Sp}(2,{\Com})$, which preserves the symplectic Majorana
constraint. The last equation of (\ref{jCEqu}) shows that the restriction 
of $b$ to symplectic Majorana spinors is either real or imaginary.

\subsubsection*{The induced sesquilinear form on the space of double spinors 
and its restriction to $\Som_{SM}$} 
Similarly, the tensor product 
\[ h_{\Som \otimes \Com^2} = h_{\Som} \otimes h_{\Com^2}\] 
of the 
${\rm Spin}_0(t,s)$-invariant sesquilinear form $h_{\Som}$ on $\Som$ 
defined in 
(\ref{sesquilinEqu}) and 
the standard sesquilinear form $h_{\Com^2}$ on $\Com^2$ 
represented by $(\delta_{ij})$ is a a non-degenerate sesquilinear 
form  on $\Som \otimes \Com^2$. It is given by:  
\be \label{sesquidoubleEqu} 
(\lambda , \chi )\mapsto 
\sum_{i=1}^2 (\lambda^i)^\dagger A \chi^i = 
\sum_{i,j=1}^2 (\lambda^i)^\dagger A \chi^j\delta_{ij}\;, 
\ee
where $\lambda = (\lambda^i)$ and $\chi = (\chi^i)$ are elements of 
$\Som \otimes \Com^2$. This form 
is Hermitian if $A$ is Hermitian and  
skew-Hermitian if $A$ is skew-Hermitian,  see (\ref{AEqu}). It is obviously   
invariant under the group ${\rm Spin}_0(t,s) \otimes {\rm U}(2)$.  
Moreover, depending on $(t,s)$, 
the sesquilinear form is real or purely imaginary on the 
subspace $\Som_{SM} \subset \Som \otimes \Com^2$ of symplectic 
Majorana spinors. This follows from
\be  j_\Som^*h_\Som =  \pm \overline{h}_\Som \quad  
\mbox{and} \quad j_{\Com^2}^* h_{\Com^2} = \overline{h}_{\Com^2}\, ,
\ee   
which is a consequence of (\ref{jCEqu}) 
(with the same sign $\pm$), since $h_\Som =
C(j_\Som \cdot ,\cdot )$ and 
$h_{\Com^2} = \varepsilon (j_{\Com^2}\cdot ,\cdot )$. This shows also that 
\be h_{\Som \otimes \Com^2} = b(\rho \cdot , \cdot )\, \quad
\mbox{and}\quad
\rho^* h_{\Som \otimes \Com^2} = \pm \overline{h}_{\Som \otimes \Com^2} 
 \ee 
(with the same sign $\pm$ as above). 
It follows that the restriction of $h_{\Som \otimes \Com^2}$ to 
symplectic Majorana spinors is a real (or purely imaginary) bilinear form 
invariant under the group  ${\rm Spin}_0(t,s) 
\otimes {\rm SU}(2)$, which is the
subgroup of ${\rm Spin}_0(t,s) \otimes {\rm U}(2)$, which preserves the
subspace of symplectic Majorana spinors.  
Note that ${\rm SU}(2) = {\rm U}(2) \cap {\rm GL}(1,\Hom ) =  
{\rm Sp}(2,{\Com})\cap {\rm GL}(1,\Hom )$  
is precisely the centralizer of 
${\rm Spin}_0(t,s)$ in the group of automorphisms of
that real bilinear form on $\Som_{SM}$. The centralizer of 
${\rm Spin}_0(t,s)$ in the larger group 
of automorphisms of the sesquilinear form on $\Som \otimes \Com^2$ 
is ${\rm U}(2) = {\rm U}(1)
\times {\rm SU}(2)$, the automorphism group of the standard Hermitian 
form $h_{\Com^2}$  on $\Com^2$.

Note that we have also 
\be
{\rm SU}(2) = {\rm Sp}(2,{\Com}) \cap {\rm U}(2) = {\rm USp}(2)\;.
\ee
Therefore the internal indices $i,j$ are often called
${\rm USp}(2)$ indices.

\subsubsection*{Admissible bilinear forms and super-Poincar\'e
algebras}

Above we discussed the notion of an admissible bilinear
form on the real spinor module $S$ of ${\rm Spin}(t,s)$.
We recall that these are bilinear forms $\beta$, which 
have a definite symmetry
$\sigma(\beta) = \pm 1$ and type $\tau(\beta) = \pm 1$, and that
such forms are automatically ${\rm Spin}_0(t,s)$ invariant.
The classification of all super-Poincar\'e algebras
for a given dimension $(t,s)$ is equivalent to finding all
admissible bilinear forms $\beta$, which satisfy the additional 
condition $\sigma(\beta) \tau(\beta) = +1$ \cite{AC}.
In fact, to classify super-Poincar\'e algebras,
one needs to classify the possible
Lie brackets $\Gamma: S\times S \rightarrow V$ which give
${\rm so}(V) + V + S$ the structure of a super-Poincar\'e 
algebra.\footnote{In other words one needs to
classify  the consistent anticommutation relations between
the supercharges.} 
Obviously $\Gamma$ has to be symmetric, because $S$ is in the odd part, 
and the Jacobi identity is equivalent to the ${\rm Spin}_0(t,s)$-equivariance
of $\Gamma$, since $V$ acts trivially on $S$.  
Any admissible bilinear form 
$\beta$ defines such a Lie bracket $\Gamma_{\beta}$ on
${\rm so}(V) + V + S$ by 
\be  \label{ACEqu} 
\langle e^\mu , \Gamma_\beta (\lambda , \chi )\rangle 
:= \beta (\gamma^\mu 
\lambda , \chi)\, .
\ee
Here $\lambda , \chi \in S$ and $\langle \cdot , \cdot \rangle$ is the 
natural pairing $V^* \times V \rightarrow \R$, $\langle e^\mu , e_\nu \rangle
= \delta^\mu_\nu$ for the dual basis $(e_\nu)$ of $V$. 
Since $\beta$ is admissible, the vector-valued bilinear form 
$\Gamma_{\beta}$ is ${\rm Spin}_0(t,s)$-equivariant. Moreover the
condition $\sigma(\beta) \tau(\beta) = +1$ implies that the
Lie bracket $S \times S \rightarrow V$ is symmetric, so that on
obtains a super Lie algebra. 
Conversely, it has been shown \cite{AC} that 
any super-Poincar\'e algebra with odd 
part $S$ is defined by a linear combination of such maps $\Gamma_\beta$.

\subsection{Minimal supersymmetry in 5 dimensions}

In this section we specialize to dimension (1,4)
and construct the minimal supersymmetry algebra in terms
of Dirac spinors and symplectic Majorana spinors. We also prove
that the R-symmetry group is ${\rm SU}(2)$.

\subsubsection*{The admissible bilinear forms on the spinor module of 
${\rm Spin}(1,4)$}  
So far our discussion applied to arbitrary $(t,s)$. We now
specialize to the case of 5-dimensional Minkowski space,
$(t,s)=(1,4)$. Since the total dimension $n = t+s = 5$ is odd, 
the complex vector space of ${\rm Spin}_0(1,4)$-invariant 
complex bilinear forms
on the complex spinor module $\Som$ is one-dimensional. This means that, 
up to scale, there is only one  non-zero invariant bilinear form $C$.  
For $n=5$ 
the symmetry and type of this form are given by $\sigma (C) = -1$ and
$\tau (C) = +1$  \cite{AC,con1}. Since  
$n = t+s = 5$ is odd and $t-s \equiv 5 \pmod 8$, the complex 
spinor module $\Som$ is an irreducible complex  ${\rm Spin}_0(1,4)$-module
of quaternionic type, see \cite{ACDV03} Table 1. In other words, there are
no Weyl spinors, and since $B^{*}B= - \id$ one cannot  
impose a Majorana condition. But it turns out to be convenient 
to rewrite Dirac spinors as symplectic Majorana spinors, {\it i.e.}, 
as pairs of Dirac spinors subject to the symplectic Majorana constraint.  

We remark that for $(t,s) = (1,4)$ the real spinor module $S$ 
is irreducible of quaternionic type and coincides with the complex 
spinor module $\Som$, see \cite{ACDV03} Table 1. The real vector space 
$(S^*\otimes S^*)^{{\rm Spin}_0(1,4)}$ of ${\rm Spin}_0(1,4)$-invariant 
bilinear forms on the real spinor module $S$ is four-dimensional
and admits a basis $(\beta_0 , \beta_1 ,\beta_2 ,\beta_3)$ 
consisting of non-degenerate admissible forms with the following 
invariants, see \cite{AC}:  
\begin{center}  
\begin{tabular} {|c   |c        |c      |c      |c | } \hline   
$i$ & $0$ & $1$ & $2$ & $3$\\ \hline
$(\sigma (\beta_i) , \tau (\beta_i))$ & $(+1,-1)$ & $(-1,-1)$ & $(-1,+1)$ &
$(-1,+1)$\\
\hline  
\end{tabular}  
\end{center}

This list shows that 
only one of the admissible forms 
$\beta_i$, namely $\beta_1$,  satisfies $\sigma (\beta) \tau (\beta ) = +1$ 
and gives hence rise to a super-Poincar\'e 
algebra, according to the algorithm of \cite{AC}. In other words, the vector
space $(\vee^2S^*\otimes V)^{{\rm Spin}_0(1,4)}$ of super-Poincar\'e 
algebra structures with odd part $S$ is one-dimensional 
and is generated by $\Gamma_{\beta_1}$:
\be (\vee^2S^*\otimes V)^{{\rm Spin}_0(1,4)} = \R \Gamma_{\beta_1}\, .\ee 
Here $\vee^2S^*$ denotes the
symmetric tensor square of the ${\rm Spin}(t,s)$-module $S^*$. 

It is easy to check that, up to scale, 
\be \beta_0 = {\rm Im}\, h_{\Som}\, ,\quad \beta_1 = {\rm Re}\, h_{\Som}\, , 
\quad \beta_2 = {\rm Im}\, C\, ,\quad  \beta_3 = {\rm Re}\, C\, , 
\ee 
where we are using the fact that $\Som = S$ for  $(t,s) = (1,4)$. 
The complex bilinear form $C$ and the sesquilinear form $h_\Som$ on $\Som$ 
were introduced in section \ref{Cliffordsubsection}. 

\subsubsection*{The admissible bilinear forms in terms of symplectic Majorana 
spinors} 
Next we describe the forms $\beta_i$ 
in terms of symplectic Majorana spinors 
$\Som_{SM} \subset \Som \otimes \Com^2$. 
We can consider the four real valued bilinear
forms ${\rm Re}\, b$, ${\rm Im}\, b$, ${\rm Re}\, h_{\Som \otimes \Com^2}$ and
${\rm Im}\, h_{\Som \otimes \Com^2}$ on the space of double spinors. The
complex bilinear form $b$ and the sesquilinear form $h_{\Som \otimes \Com^2}$ 
on $\Som \otimes \Com^2$ were introduced in section \ref{Cliffordsubsection}. 
For $(t,s) = (1,4)$ the bilinear form $b$ is symmetric and the 
sesquilinear form $h_{\Som \otimes \Com^2}$ is anti-Hermitian. This shows that
\be \sigma ({\rm Re}\, b) = +1\, , \quad \sigma ({\rm Im}\, b) = +1\, , 
\quad \sigma ({\rm Re}\, h_{\Som \otimes \Com^2}) = -1\, ,\quad 
\sigma ({\rm Im}\, h_{\Som \otimes \Com^2}) = +1\, .\ee
Moreover, the restriction $b_{SM}$ of $ib$ (or of  
$ih_{\Som \otimes \Com^2}$) to the
subspace $\Som_{SM} \subset \Som \otimes \Com^2$ is a non-degenerate 
real-valued bilinear form on $\Som_{SM}$. In particular, ${\rm Re}\, b = 
{\rm Re}\, h_{\Som \otimes \Com^2}$ vanish on $\Som_{SM}$. 
 Notice that $\sigma (b_{SM}) = +1$. Since, up to scale, 
there is only one ${\rm Spin}_0(1,4)$-invariant
real symmetric bilinear form on $S = \Som_{SM}$, we conclude  
that $b_{SM}$ is proportional to $\beta_0$. 
Notice that although 
$\Som_{SM} \subset \Som \otimes \Com^2 = \Som \oplus \Som$ is {\it not} a
Clifford submodule it carries nevertheless an intrinsic 
Clifford module structure due to the (real) 
isomorphism $\Som \cong \Som_{SM}$, 
\be \Som \ni \lambda \mapsto \lambda_{SM} := 
\lambda \oplus j_\Som {\lambda}\in 
\Som_{SM} \subset \Som \oplus \Som\, .
\ee
The intrinsic Clifford multiplication $\gamma_{SM}^\mu$ is given by 
\be \gamma_{SM}^\mu \lambda_{SM} = 
(\gamma^\mu \lambda )_{SM} = -i\gamma^\mu (i\lambda)_{SM} \neq \gamma^\mu 
 \lambda_{SM}\, ,\ee 
cf.\ (\ref{jgammaEqu}). 
To obtain the 
remaining admissible bilinear forms $\beta_1$, $\beta_2$ and $\beta_3$ 
it suffices to
consider the forms $b_{SM}(i_{SM}\cdot , \cdot)$, $b_{SM}(j_{SM}\cdot , 
\cdot)$ and $b_{SM}(k_{SM}\cdot , \cdot)$, respectively, 
where $i_{SM}$ is the intrinsic 
${\rm Pin}(1,4)$-invariant complex structure,  
$j_{SM}$ is the intrinsic ${\rm Spin}(1,4)$-invariant quaternionic 
structure  given by
\begin{eqnarray} i_{SM}\lambda_{SM} &=& (i_{\Som}\lambda)_{SM}\, ,\\
 j_{SM}\lambda_{SM} &=& (j_{\Som}\lambda)_{SM} \;,
\end{eqnarray}    
and $k_{SM} = i_{SM}j_{SM}$. 
 
\subsubsection*{The Clifford algebra $Cl_{1,4}$} 
It follows from the classification of Clifford algebras that
$Cl_{1,4} \cong \Com (4)$. We shall now describe an explicit
isomorphism, from which we will easily recover the above 
results concerning the spinor module $S$ of ${\rm Spin}(1,4)$. 
We consider the real vector space $\Hom^2$ and denote by 
$L_q : \Hom^2 \rightarrow \Hom^2$ 
the left-multiplication and by 
$R_q: \Hom^2 \rightarrow \Hom^2$ the right-multiplication by a 
quaternion $q\in \Hom$. We put $I := R_i$, $J := R_j$ and $K := IJ = -R_k$.
We define the following operators on $\Hom^2$:
\be \gamma^0 := ID\, ,\quad \gamma^1 := IEL_i\, ,\quad  \gamma^2 := 
IEL_j\, ,\quad  \gamma^3 := IEL_k\, ,\quad \gamma^4 := IED \, , 
\ee 
where 
\be E \left( \begin{array}{c}
q_1\\
q_2
\end{array}\right) := \left( \begin{array}{c}
q_1\\
-q_2
\end{array}\right)\, , \quad D \left( \begin{array}{c}
q_1\\
q_2
\end{array}\right) := \left( \begin{array}{c}
q_2\\
q_1
\end{array}\right)\, .\ee 
One can immediately check that the $\gamma^\mu$ satisfy the Clifford
relation for the 5-dimensional Minkowski metric $\eta$ and hence define on 
$\Hom^2$ the structure of a $Cl_{1,4}$-module. By dimensional reasons this
module is irreducible and provides a faithful representation of the
Clifford algebra. The $Cl_{1,4}$-invariant complex structure $I$ provides
the identifications $\Hom^2 = \Com^4$ and $Cl_{1,4} = \Com (4)$, where 
$\Kom (l)$ denotes the full matrix algebra of rank $l$ over $\Kom \in 
\{\R, \Com , \Hom\}$.  The even Clifford algebra  $Cl^0_{1,4}\subset 
Cl_{1,4}$ generated by the products $\gamma^\mu \gamma^\nu$ corresponds to 
$\Hom (2) \subset \Com (4)$, the centralizer of the quaternionic structure 
$J$ in $Cl_{1,4} = \Com (4)$. Now we can identify the 
spinor module $S = \Som$ with $\Hom^2 = \Com^4$ and the connected spinor group 
${\rm Spin}_0(1,4)$ with ${\rm USp}(2,2) = {\rm Sp}(4,\Com) \cap 
{\rm U}(2,2)$, where the complex symplectic structure corresponds to the
bilinear form $C$ introduced in \ref{Cliffordsubsection} and the 
indefinite Hermitian metric corresponds 
to the sesquilinear form $ih_\Som$ (note that $h_\Som$ is anti-Hermitian 
in the Minkowskian case $t = 1$). 
The latter is the complex part of the standard indefinite 
quaternionic-Hermitian
metric on $\Hom^2 = \Hom^{1,1} = \Com^{2,2} = \R^{4,4}$:
\be (q,q') \mapsto \overline{q}_1q_1' - \overline{q}_2q_2'\, ,\ee 
where now $q = (q_i)$, $q' = (q_i')$, $i =1,2$. The real part of that
form is the scalar product $g = \beta_0$ of signature $(4,4)$, 
the $i$-imaginary part is  
$\beta_1 = g(I\cdot , \cdot )$ the $j$-imaginary
part is $\beta_2 = g(J\cdot , \cdot )$ and the  $k$-imaginary
part is $\beta_3 = g(K\cdot , \cdot )$. The symmetry and 
type of these forms, indicated above, can be easily checked using this model. 

\pagebreak[3] 
\subsubsection*{The 5-dimensional supersymmetry algebra}

According to \cite{AC}, the minimal supersymmetry algebra in 
$(1,4)$-dimensions is  
given by $\Gamma_{\beta_1}$, where $\beta_1$ is the unique 
(up to scale) admissible real bilinear form on the spinor module $S$. 
In the model $S = \Hom^2$, described above, it is given by  
the symplectic form $\beta_1 = g(I\cdot , \cdot)$. Under the identifications  
$S = \Som$ and $S \cong \Som_{SM}$ it  corresponds to the forms 
${\rm Re}\, h_{\Som}$ and $b_{SM}(i_{SM}\cdot ,\cdot )$.  
In standard physics notation, the corresponding supersymmetry algebra  
with Dirac spinors as supercharges is given by: 
\be
\{ Q_{\alpha}, Q_{\beta } \} =    
{\rm Re}((\gamma^{\mu} A^{-1})_{\alpha {\beta}})P_{\mu} \;, 
\label{SusyAlgminimal}
\ee
where $A^{-1} = (A_{\alpha {\beta}})$ is the inverse of
$A = (A^{{\alpha}\beta})$.  
Thus we have four independent
complex (or eight independent real) supertransformations.
This is the smallest supersymmetry algebra in 5 dimensions.

We now move on to the standard form of the 5-dimensional supersymmetry algebra,
which is \cite{con2}:
\be
\{ Q_{i \alpha}, Q_{j \beta} \} = - \frac{1}{2} 
\epsilon_{ij} (\gamma^{\mu} \CC^{-1})_{\alpha \beta} P_{\mu} 
\label{SusyAlg1,4} \;.
\ee
Here $\CC^{-1} = (\CC_{\alpha \beta})$ is the inverse of the
charge conjugation matrix and the supercharges are subject to 
the symplectic Majorana condition (\ref{SymMajCon}).

Before imposing this constraint, the algebra (\ref{SusyAlg1,4})
corresponds, up to a factor, to the 
${\rm Spin}_0(1,4)$-invariant bracket
$\Gamma_b$ associated, by the universal formula (\ref{ACEqu}), to the 
admissible complex bilinear form $b = C \otimes \epsilon$ 
on the space of double spinors $\Som \otimes \Com^2$.     
The supersymmetry charges are, hence, pairs of Dirac spinors and 
we have eight independent complex supertransformations.
This is not the smallest supersymmetry algebra in 5 dimensions.

But when imposing the symplectic Majorana condition (\ref{SymMajCon})
on the supercharges $Q_{i \alpha}$, we are left with 
eight real or four complex supercharges and the algebra
(\ref{SusyAlg1,4}) becomes isomorphic to (\ref{SusyAlgminimal}).
We emphasize that the restriction of the bracket 
$\Gamma_b$ to the space 
$\Som_{SM} \subset \Som \otimes \Com^2$ is real-valued, although
the restriction of the bilinear form $b$ is purely imaginary. 
The reason is that the definition of the bracket $\Gamma_b$ involves 
the Clifford multiplication $\gamma^\mu$, which anticommutes with
the real structure $\rho$ due to (\ref{jgammaEqu}). More explicitly,
we can use the isomorphism 
$\Som_{SM} \ni  (\lambda^1, \lambda^2)^T = 
( \lambda^1 ,  j_{\Som}(\lambda^1))^T
\rightarrow \lambda^1 \in \Som$ to find 
$\left. \Gamma_b \right|_{SM} = - 2 \Gamma_{\beta_1}$, which
maps (\ref{SusyAlg1,4}) to (\ref{SusyAlgminimal}).

\subsubsection*{The R-symmetry group} 
Let $\gg = \gg_0 + \gg_1 = ({\rm so}(V) + V) + S$ 
be a super-Poincar\'e algebra, $V = \R^{t,s}$. 
According to \cite{AC} the Lie superbracket is a 
${\rm Spin}_0(t,s)$-equivariant symmetric bilinear map of the form 
$\Gamma_\beta : S\times S \rightarrow V$, see (\ref{ACEqu}),  
where $\beta$ is a linear combination of 
admissible bilinear forms. 
The {\it R-symmetry group} of $\gg$ is the group 
\be G_R := \{ \phi \in {\rm Aut}({\gg})|\;  \phi|_{{\gg}_0} = 
{\rm Id}\}
\ee  
of automorphisms of $\gg$ which act trivially on the even part. 
Notice that $G_R$ can be considered as a subgroup of ${\rm GL}(S)$, 
since the action on ${\gg}_0$ is trivial. More precisely,
\be G_R = G_R({\gg}) 
= \{ \phi \in {\rm GL}(S)|\; [\phi , {\rm so}(V)] = 0 \quad 
\mbox{and} \quad \Gamma_\beta (\phi \lambda , \phi \chi ) = 
\Gamma_\beta ( \lambda ,  \chi )\quad \mbox{for all}\quad \lambda ,  \chi\in 
S\}\, ,\ee 
which
is the centralizer of ${\rm Spin}_0(t,s)$ in the automorphism group of the
vector-valued bilinear form $\Gamma_\beta$. 

Next we describe the R-symmetry group in the case $V = \R^{t,s} = \R^{1,4}$ 
in terms of symplectic Majorana spinors $S \cong 
\Som_{SM} \subset \Som \otimes \Com^2$. 
This formalism has the advantage that
the R-symmetry group acts on the internal space $\Com^2$. 
Here $G_R$ is identified with a subgroup of the complex
Lie group ${\rm GL}(\Sigma)$, $\Sigma = \Som \otimes \Com^2$:   
\be 
G_R = \{ \phi \in {\rm GL}(\Sigma)|\;  
[\phi , {\rm so}(V)] = 0\, , \quad \phi\rho = \rho \phi \quad 
\mbox{and} \quad \Gamma_b (\phi \lambda , \phi \chi ) = 
\Gamma_b ( \lambda ,  \chi )\quad \mbox{for all}\quad \lambda ,  \chi\in 
\Sigma \}\,.
\ee 
Here $b = C \otimes \epsilon$ is the symmetric bilinear form on the
the space of double spinors $\Sigma$ 
and $\rho$ is the real structure which defines
the symplectic Majorana spinors. 
To determine this group, let us first remark that by  Schur's Lemma
the first condition $[\phi , {\rm so}(V)] = 0$ implies that $\phi = 
{\rm Id}\otimes \varphi$, $\varphi \in {\rm GL}(2, \Com )$. The 
reality condition $\phi\rho = \rho \phi$ is equivalent to
$\varphi j_{\Com^2} = j_{\Com^2}\varphi$, {\it i.e.}, to 
$\varphi \in {\rm GL}(1, \Hom ) \subset {\rm GL}(2,\Com )$. 
An element $\phi = {\rm Id} \otimes \varphi$ satisfies 
the third condition 
$\Gamma_b (\phi \cdot , \phi \cdot ) = \Gamma_b$ if and only if
$\varphi \in {\rm Sp}(2,\Com )$. Thus we have proven that the
R-symmetry group of (\ref{SusyAlg1,4}) subject to the reality constraint
(\ref{SymMajCon}) is given by 
\be   G_R = {\rm Id} \otimes ({\rm Sp}(2,\Com )\cap {\rm GL}(1, \Hom ))
= {\rm Id} \otimes {\rm SU}(2) \cong {\rm SU}(2) \;.\ee

\subsection{4-dimensional $\mathcal{N}=2$ 
supersymmetry from dimensional reduction \label{SectDRCliff}}

Next we discuss the relation of the $(1,4)$-dimensional 
supersymmetry algebra to the 
$\mathcal{N}=2$ supersymmetry algebras in
space-times of dimensions $(1,3)$ and $(0,4)$. Starting from
5 dimensions, these supersymmetry algebras are obtained
by dimensional reduction over a space-like or time-like
direction, respectively. We perform a standard Kaluza-Klein
reduction, {\it i.e.}, we take all dependencies on the reduced 
direction to be trivial. At the level
of the algebra this amounts to setting to zero one space-like
or time-like momentum operator ($=$ translation operator). 

It is convenient to take the 5-dimensional space-time indices
to be $\mu, \nu= 0,1,2,3,5$. The corresponding gamma-matrices
are 
\be 
\gamma^0, \, \gamma^1, \, \gamma^2, \, \gamma^3, \, \gamma^5, 
\ee
where, as already mentioned, $\gamma^0$ is anti-Hermitian 
whereas the other matrices are Hermitian. 
We use a representation where these matrices
are related by \cite{con1}:
\be
\gamma^5 = -i \gamma^0 \gamma^1 \gamma^2 \gamma^3\,,\quad \gamma^5\gamma^5 = \id  \;.
\ee
Mathematically, this corresponds to the fact that the 
Clifford algebra $\Com l_5 = \Com (4) \oplus \Com (4)$ has
two irreducible modules, which differ by the value of 
$i\gamma^0 \gamma^1 \gamma^2 \gamma^3 \gamma^5 = \pm \id$.
  
Upon dimensional reduction, one of the five Lorentz indices 
becomes an internal index, which we denote by $*$. 
When reducing over a space-like dimension we take $*=5$,
in reduction over time we have $* = 0$. The Clifford
algebras are generated by
\bea
\gamma^0, \, \gamma^1, \, \gamma^2, \, \gamma^3, &\qquad& 
\text{for $Cl_{1,3}$ and } \; \nonumber \\
\gamma^{1}, \, \gamma^{2}, \, \gamma^{3}, \, \gamma^{5}, &\qquad& 
\text{for $Cl_{0,4}$\;.}
\eea

The additional
generator $\gamma^{*}$ can be used to impose a chirality
constraint. If the number 
of space-time dimensions is even, the spinor module 
decomposes into two irreducible eigenspaces with respect to
the so-called volume element \cite{LawMic}, which is proportional
to the product of all Clifford generators. In 
physical  terminology this is the 
`generalized $\gamma^5$-matrix.'\footnote{
Taken literally, chirality refers to the distinction between
left-handed and right-handed frames in odd-dimensional space.
But in a slight abuse of language we will generally  
refer to the above decomposition as `chiral.'}
The minimal spinor representation of 5 dimensions 
becomes reducible in 4 dimensions, 
as Dirac spinors decompose into Weyl spinors.

In the $(1,3)$ theory $\gamma^{*} = \gamma^5$ is the
standard chirality matrix. For later notational convenience we set
$\Gamma_{*} := \gamma^5$. 
The corresponding projector can be used to decompose spinors:
\be
\lambda^i = \lambda^i_{+} + \lambda^i_{-} \;,\;\;\;
\mbox{where} \;\;\; \lambda_{\pm}^i  := \Gamma_{\pm} \lambda^i
= \ft12 ( \id \pm \Gamma_{*} ) \lambda^i \;.
\ee
The 5-dimensional symplectic Majorana constraint (\ref{SymMajCon})
can be written as
\be
(\lambda^i)^{*} = \CC \gamma_0 \epsilon_{ij} \lambda^j \;,
\ee
where we used $A=\gamma_0$. 
Chiral decomposition of the symplectic Majorana constraint gives
\be
(\lambda^i_{\pm})^* = \, \CC \, \gamma_0 \, \epsilon_{ij} \, 
\lambda_{\mp}^j \, .
\label{ChiralMink}
\ee
Note that the chiral projection is not compatible with the
symplectic Majorana condition, which is not surprising, since there are no 
symplectic Majorana-Weyl spinors for $(t,s)=(1,3)$. If one wants
to work with irreducible spinors, one can reformulate the theory either in 
terms of Weyl spinors, or, equivalently, in terms of 
(standard, not symplectic) Majorana spinors. 
For our purposes, however, it 
is more convenient to stick to symplectic Majorana spinors, 
because they 
exist in all three space-time signatures $(1,4)$, $(1,3)$ and $(0,4)$. 
Since a considerable part of the literature 
uses Majorana spinors, we briefly review the relation 
between standard and symplectic Majorana spinors in $(1,3)$ dimensions
in the appendix (see in particular (\ref{A.17})). 

Let us now 
consider the chiral decomposition 
for dimension $(0,4)$. Now
$\gamma^0$ can be used to impose a chirality constraint. 
Since $\gamma^0$ squares to $-\id$ rather than $\id$, we take
the chirality matrix to be\footnote{The choice of the overall sign
is suggested by the dimensional reduction of the 5-dimensional
supersymmetry transformations. See section \ref{SubSecSusyVar}.}
\be
\Gamma^\rmE_{*} := - i \gamma^0 \;,
\ee
with corresponding projector
\be
\Gamma^\rmE_{\pm} := \frac{1}{2} ( \id \pm \Gamma^\rmE_{*}) \;.
\ee
Then spinors decompose according to
\be
\lambda^i = \es^{i}_+ + \es^{i}_- \;,
\;\;\; \es^{i}_{\pm} := \Gamma^\rmE_{\pm} \lambda^i \;.
\ee
Using $B=C\gamma_0$, the symplectic Majorana constraint 
(\ref{SymMajCon}) becomes
\be
\left( \lambda^i \right)^* 
= - \, \ComplexI \, \CC \, \Gamma^{\rm E}_* \, \epsilon_{ij} \, \lambda^j \, .
\ee
Using the chiral projection we find
\be
(\es^{i}_{\pm})^* = - \, \ComplexI \, \CC \, \Gamma_*^{\rm E}  \,
\epsilon_{ij} \, \es_{\pm}^{j} \;,
\label{ChiralEuc}
\ee
which shows that in dimension $(0,4)$ the chiral and symplectic
Majorana constraints are compatible, {\it i.e.}, there are symplectic
Majorana-Weyl spinors. This is in accordance with the general analysis,
see \cite{con1,ACDV03}.

We now turn to the supersymmetry algebra. 
Since the supercharges in all three cases are symplectic Majorana spinors, 
the reduction is straightforward:
\be 
\left\{ Q_{i \alpha} , Q_{j \beta} \right\} = 
- \, \half \, \epsilon_{ij} \, 
\left( \gamma^\mu \CC^{-1}\right)_{\alpha \beta} \, P_\mu  = 
 - \, \half \, \epsilon_{ij} \, 
\left( \gamma^m \CC^{-1} \right)_{\alpha \beta}
\, P_m  - \, \half \, \epsilon_{ij} \, \left( 
\gamma^* \CC^{-1} \right)_{\alpha \beta} \, P_* \, .\ee 
Since $P_*f = i \ft{\der}{\der x^{*}}f = 0$ for fields 
$f$ which do not depend on the 
internal direction $x^{*}$, the resulting supersymmetry algebra 
in 4 dimension is    
\be \left\{ Q_{i \alpha} , Q_{j \beta} \right\} 
= - \, \half \, \epsilon_{ij} \, 
\left( \gamma^m \CC^{-1} \right)_{\alpha  \beta} \, P_m \, . 
\label{SusyAlgRed}
\ee
More generally, $P_{*}$ could act 
as a real central charge on the Fourier modes 
of the 5-dimensional fields.
But we discard all
non-constant Fourier modes, so that $P_{*}$ acts 
trivially.\footnote{We have also
omitted the real central charge of the $(1,4)$-dimensional 
supersymmetry algebra,
because we are only interested in massless states, on which it
acts trivially. If one includes the 5-dimensional central charge
in dimensional reduction and keeps $P_{*}$, they combine into the
complex central charge of the 4-dimensional $\mathcal{N}=2$ supersymmetry
algebra. Again, this central charge acts trivially on massless
states and is irrelevant for our purpose.} 

Notice that the super Lie algebra (\ref{SusyAlgRed}), 
subject to the symplectic Majorana constraint, is the 
standard ${\cal N}=2$ super-Poincar\'e algebra in dimension $(1,3)$. 
Using the formulae given in the appendix one can rewrite the 
supercharges in terms
of two Majorana spinors. This algebra is not the minimal one, which
is generated by just one Majorana spinor, or, equivalently,
one Weyl spinor of supercharges \cite{WesBag}.
 
In dimension $(0,4)$ we also get a real ${\cal N}=2$ super-Poincar\'e algebra
with odd part $\Som = \Som_+ + \Som_-$ and  
$\{ \Som_+, \Som_+\} = \{ \Som_-, \Som_-\} = 0$. 
In this case we do not need to distinguish between the 
real and complex spinorial modules: $S = \Som$ and $\Som_\pm = S_\pm$. 
In Euclidean 4-space the above super Lie algebra 
is a minimal super-Poincar\'e algebra, in contrast to the
Minkowski case. This is clear from the fact that only the
superbrackets between $\Som_+$ and $\Som_-$ are non-vanishing. 
Moreover, although the vector space 
$(\vee^2S^* \otimes V)^{{\rm Spin}(0,4)}$ of 
super-Poincar\'e algebra structures with odd part $S$ is 4-dimensional 
\cite{AC}, one can show that 
any other nontrivial ${\cal N}=2$ super-Poincar\'e algebra is isomorphic to 
(\ref{SusyAlgRed}) subject to the symplectic Majorana constraint. 
 
Obviously, the two 4-dimensional supersymmetry algebras inherit 
the R-symmetry group ${\rm SU}(2)$ of the 5-dimensional algebra.
But since $\gamma^{*}$ does not represent any element of the 
dimensionally reduced Clifford
algebra ($Cl_{1,3}$ or $Cl_{0,4}$), it now generates an internal
transformation.
This is a candidate for a new R-symmetry, because it commutes
with the even part 
$Cl^0_{t,s}$ of $Cl_{t,s}$,  for both $(t,s)=(1,3),(0,4)$.
Since the invariant bilinear form $C$ on the spinor  
module of ${\rm Spin}(5,\Com )$ has type $\tau (C) = +1$ and $\gamma^{*}$ 
anticommutes with all other gamma matrices, the transformation
\be
Q_i \rightarrow \exp ({ \gamma^{*} \Phi}) Q_i
\ee
leaves (\ref{SusyAlgRed}) invariant for arbitrary complex transformation
parameters $\Phi$.  We still have to check
the reality constraint
\be
(\exp( {\gamma^{*} \Phi} ) Q^i)^{*} = \CC \gamma_0 \epsilon_{ij}
\exp ({ \gamma^{*} \Phi}) Q^j \;.
\ee
In dimension $(1,3)$ we have $\gamma^{*} = \gamma^5$, which
is Hermitian and anticommutes with $\gamma^0$. Thus we find
\be
\exp ({(\gamma^5)^{*} \Phi^{*}} ) = 
\exp ({- (\gamma^5)^T \Phi}) \,.
\ee
Using again that $\gamma^5$ is Hermitian we learn that 
$\Phi$ must be imaginary. Thus the new R-symmetry is
\be
Q_i \rightarrow \exp ({i\gamma^5 \phi}) Q_i \;,\;\;\;
\phi \in {\R} \;.
\label{2.15a}
\ee
The group generated by $\gamma^5$ is isomorphic to ${\rm U}(1)$
and acts chirally, {\it i.e.}, the chiral components $Q_{i\,\pm}$ transform
with opposite phases $e^{\pm i \phi}$. Thus we find the
well-known R-symmetry group of $\mathcal{N}=2$ supersymmetry 
\cite{WesBag}: 
\be
G_R = {\rm U}(2) \simeq {\rm U}(1) \times {\rm SU}(2) \simeq {\rm U}(1) \times
{\rm USp}(2) \;.
\ee

In dimension $(0,4)$ we have $\gamma^{*} = \gamma^0$, which
is anti-Hermitian and commutes with $\gamma^0$.
We find
\be
\exp ({ (\gamma^0)^{*} \Phi^{*} }) = 
\exp ({(\gamma^0)^T \Phi}) \;,
\ee
and using that $\gamma^0$ is anti-Hermitian we see that
$\Phi$ must be imaginary, $\Phi = - i \phi$, with 
$\phi \in {\R}$. Since $(\gamma^0)^2 = - \id$ we get
a non-compact R-symmetry group. 
The R-symmetry transformation is 
\be
Q_i \rightarrow \exp ({-i \gamma^0 \phi}) Q_i =
\exp ({\Gamma^\rmE_{*} \phi}) Q_i \;,
\label{2.17a}
\ee
with real $\phi$. Since $(\Gamma^\rmE_{*})^2 = \id$, the generator
$\Gamma^\rmE_{*}$  has 
eigenvalues $\pm 1$ (rather than $\pm i$) and acts by chiral
scale transformations:
\be
Q_{i \, \pm} \rightarrow e^{\pm \phi} Q_{i \, \pm} \;.
\ee
Taking into account the obvious symmetry 
$Q_{i \, \pm} \rightarrow - Q_{i \, \pm}$ of (\ref{SusyAlgRed}), we find that 
the R-symmetry group contains an additional subgroup ${\rm SO}(1,1)$, which  
commutes with ${\rm SU}(2)$. 
Analyzing the spinor representation in dimension $(0,4)$, 
with the same methods that were used above for dimensions $(1,4)$, one 
can prove that the full R-symmetry group of 4-dimensional Euclidean 
$\mathcal{N}=2$ supersymmetry is
\be
G_R = {\rm SO}(1,1) \times {\rm SU}(2) \;.
\ee
The fact that the Abelian factor of the $\mathcal{N}=2$ R-symmetry
becomes non-compact for Euclidean signature has been observed
by various authors, starting from \cite{zumino}. In \cite{blau},
which studies the dimensional reduction from dimension $(1,5)$
to dimension $(0,4)$, the ${\rm SO}(1,1)$ factor was related to
the internal part of the\linebreak[3] 6-dimensional 
Lorentz group. Note that this
is different for our reduction, which starts in dimension $(1,4)$,
because there is no subgroup ${\rm SO}(1,1)$ of ${\rm SO}(1,4)$ which commutes
with ${\rm SO}(4)$.
Instead, we found that the new part of the R-symmetry group
arises from the Clifford algebra.

\subsection{Commuting versus anticommuting spinors \label{ExplGrass}}

In supersymmetric field theory one uses spinor fields with components
which are not real or complex numbers, but `Grassmann numbers', 
{\it i.e.}, elements 
of a Grassmann algebra defined by a system of anticommuting generators. 
In geometrical terms, this means to work with super vector 
spaces and super manifolds \cite{Freund,deWitt,Manin}.

Let us explain what this means for the cases we are interested 
in. If $S$ is the real spinor module underlying the supersymmetry
algebra of our field theory, then 
we replace it by $\Pi S$, which is the spinor module considered as 
a purely odd super vector space of dimension $(0\, |\,m)$, where
$m$ is the dimension of $S$. ($\Pi$ is called 
the parity change functor in \cite{Manin}.)  
The elements 
of $\Pi S$ are called anticommuting spinors. Since we want
to  consider spinor fields, we also need to identify the 
appropriate spinor bundle. For commuting spinors the spinor 
bundle is $S(\R^{t,s}) = \R^{t,s}  \times S \rightarrow \R^{t,s}$, 
the trivial bundle over
space-time $\R^{t,s}$ with fibre $S$. It is trivial since we only consider  
flat simply connected space-times. To define anticommuting 
spinor fields, it is not sufficient to replace the fibre 
$S$ by $\Pi S$, as the resulting super vector bundle 
$\R^{t,s} \times \Pi S \rightarrow 
\R^{t,s}$ of rank $(0|m)$ 
has no sections other than the zero section. The reason simply
is that the $m$ local components of a section must be odd superfunctions, 
which can be non-zero only if the base of the bundle is a 
supermanifold with a non-trivial odd part. 
Therefore 
one replaces the space-time $\R^{t,s}$ by 
the flat superspace 
$\R^{t,s| m} = \R^{t,s} \times N$, 
where $N$ is an internal, purely odd  
parameter space of dimension $m$.  
The super vector bundle $\Pi S( \R^{t,s|m} ) := \R^{t,s|m}\times 
\Pi S \rightarrow \R^{t,s|m}$ has non-trivial sections. 
An anticommuting spinor field 
is, by definition, a section of the bundle $\Pi S( \R^{t,s|m} )$.  
This is used in the field theoretic part 
of the paper. The above construction can be easily generalized to 
the case of space-times with non-trivial spinor bundle.

Note that going from 
commuting to anticommuting spinor components changes the symmetry 
properties of the bilinear forms in the obvious way. This should be kept in 
mind, because the actions and supersymmetry transformation rules 
appearing in the following sections
involve spinor bilinears, which are built out of anticommuting 
spinors. In contrast, we used commuting spinors in this section.

\end{section}

\begin{section}{Vector Multiplets in 5 Dimensions \label{Section4}}

The aim of this section is to construct the $\cN = 2$ rigidly
supersymmetric Lagrangian\footnote{As explained in the introduction 
we generally refer to theories with eight real supercharges as ${\cal N}=2$.
Other authors call this theory ${\cal N}=1$ supersymmetric, because
the underlying supersymmetry algebra is minimal.} 
for Abelian vector multiplets in dimension
(1,4). Our motivation is that we need
it as the starting point for dimensional reduction, in order
to explore the properties of the resulting Lagrangians in dimensions
$(1,3)$ and $(0,4)$.

5-dimensional supersymmetric field theories have been studied using string 
theory in \cite{5dgauge} and, subsequently, in other
papers including \cite{4d5drigid,Eguchi}. For Abelian vector multiplets
the Lagrangian is completely determined by a real cubic polynomial,
the prepotential \cite{5dgauge}. Since none of the above 
references specifies the explicit Lagrangian and supersymmetry
rules, we derive them in the following.
The construction makes essential use of the results of \cite{con2},
where the general Lagrangian of 5-dimensional 
rigid super{\em conformal} vector multiplets was constructed in the
framework of the superconformal calculus. We adapt this construction
to the case of {\em Poincar\'e} supersymmetry.  
In both cases one finds that all couplings are encoded in a real prepotential.
In the superconformal case
this function must be a {\em homogeneous polynomial of degree three}. 
Moreover, the superconformal invariance leads to additional terms in the
supersymmetry variations. This is due
to the fact that the superconformal algebra contains, besides the
standard supersymmetry transformation (Q-transformations), a second set
of so-called special supersymmetry transformations (S-transformations).
Since we are interested in a general super-Poincar\'e invariant Lagrangian, 
we do not require invariance under scale and special conformal
transformations, nor under S-transformations. We therefore have to  
reanalyze the constraints on the prepotential and we will find that it
is now allowed to be an arbitrary {\em cubic polynomial}. 
Higher order terms are ruled out by Abelian gauge invariance
\cite{5dgauge}.

The basic building block for our model is the $\cN = 2$ off-shell vector 
multiplet \cite{con2}, which has the following field content 
(our conventions are summarized in the appendix) :   
\be\label{1.1}
\left\{ A_\mu,\, \lambda^i, \, \sigma, \, Y^{ij} \right\}.
\ee
Here $A_\mu$ is a 5-dimensional one-form, which should be considered as the 
gauge potential of a connection in a line bundle. As is common in physics, 
we understand that one-forms and
vector fields have been identified using the space-time metric, 
which explains the name ``vector multiplet.'' The pair $\lambda^i$ is a 
symplectic Majorana spinor, and $\sigma$ is a real scalar field. In order to
have the correct number of off-shell degrees of freedom, the multiplet also
contains the auxiliary field $Y^{ij}$, which is a real, symmetric tensor
of the R-symmetry group ${\rm SU}(2)$: 
\be
Y^{ij} = Y^{ji} \;,\;\;\;
(Y^{ij})^* = Y_{ij} = \epsilon_{ik} \epsilon_{jl} Y^{kl} \;.
\ee
Note that the real structure on symmetric tensors over $\Com^2$ used here 
is simply the tensor square of the ${\rm SU}(2)$-invariant 
quaternionic structure on $\Com^2$. 

As usual in supersymmetric field theories, 
the spinors $\lambda$ are anticommuting 
(see section \ref{ExplGrass}).
Note that this changes the symmetry
properties of the bilinear forms discussed in the previous section,
where we used commuting spinors.
The formulae needed for handling expressions containing anticommuting
spinors are collected in the appendix.

The kinetic terms for this multiplet are given by
\be\label{1.2}
\Lag = -\forth\,F_{\mu\nu}F^{\mu\nu}
  -\half\,\lb \ds\, \lambda
  -\half\,\partial_\mu\sigma\,\partial^\mu\sigma
  +Y^{ij}Y_{ij} \, ,
\ee
where $\ds = \gamma^\mu\partial_\mu$ is the Dirac operator.   
The action corresponding to this Lagrangian is invariant under the following off-shell supersymmetry variations:
\bea\label{1.3}
\nonumber \delta A_\mu &=& \half \, \eb \, \gmu \, \lambda \,,\\
\nonumber \delta Y^{ij} &=& - \, \half \, \eb^{(i} \ds \lambda^{j)}\,, \\
\nonumber \delta \lambda^i &=& - \, \forth \, \gamma^{\mu\nu} \, F_{\mu\nu} \, 
\epsilon^{i} \, - \, \frac{\ComplexI}{2} \, \ds \sigma \, 
\epsilon^i \, - \, Y^{ij} \, \epsilon_j \,,\\
\delta \sigma &=& \, \frac{\ComplexI}{2} \, \eb \, \lambda \, .
\eea
Here $\epsilon = (\epsilon^i)$ is the parameter of the supersymmetry 
transformation, which is an anticommuting symplectic Majorana spinor, and 
$\eb$ is its Majorana conjugate, see (\ref{MajConjEqu}).
Working in an off-shell formulation has the advantage that the 
supersymmetry transformations do not depend on equations of motion 
derived from the Lagrangian. Hence they will retain their form when 
we add further terms to the Lagrangian.

We now take $N$ copies of (\ref{1.2}), and couple them by a 
symmetric matrix, $a_{IJ}(\sigma)$:\footnote{One might have 
expected that in the spinor term the partial derivative is
promoted to a covariant derivative with respect to the Levi-Civita
connection of $a_{IJ}$. However, the term containing the connection
is identically zero for anticommuting symplectic Majorana spinors.}
\be\label{1.4}
 \Lag_{\text{kin}} = \left(-\forth\,F^I_{\mu\nu}F^{J\,\mu\nu}
  -\half\,\lambdaB^I \ds\, \lambda^J
  -\half\,\partial_\mu\sigma^I\,\partial^\mu\sigma^J
  +Y^I_{ij}Y^{J\,ij}\right)a_{IJ}(\sigma) \,.
\ee
$N$ is the number of vector multiplets in our model, which are labeled by the
indices $I,J \in \{1, \ldots , N\}$. 
Note that the matrix $a_{IJ}(\sigma)$ is allowed to depend on the
scalar fields. In particular, the scalar kinetic term has now 
been promoted to a non-linear sigma model. The scalar fields
$\sigma^I$ can be interpreted as a map from space-time $\R^{1,4}$
into an $N$-dimensional Riemannian manifold ${\cal M}$ with
metric $a_{IJ}(\sigma)$. 

We require that the variations (\ref{1.3}) 
hold for every copy of the vector multiplet separately:
\bea\label{1.7}
\nonumber \delta \sigma^I &=& \, \frac{\ComplexI}{2} \, \eb \, 
\lambda^I \,,\\
\nonumber \delta A_\mu^I &=& \half \, \eb \, \gmu \, \lambda^I \,,\\
\nonumber \delta \lambda^{iI} &=& - \, \forth \gamma^{\mu\nu} \, 
F_{\mu\nu}^I \, 
\epsilon^{i} \, - \, \frac{\ComplexI}{2} \, \ds \sigma^I \, \epsilon^i \, - \, Y^{ij \, I} \, \epsilon_j \,,\\ 
\nonumber  \delta \lambdaB^{iI} &=& \forth\,\Bar{\epsilon}^i\,\gamma^{\mu\nu} F_{\mu\nu}^I
  -\frac{i}{2}\,\Bar{\epsilon}^i\ds\,\sigma^I - 
\Bar{\epsilon}_j\,Y^{ij\,I} \,,\\
\delta Y^{ij \, I} &=& - \, \half \, \eb^{(i} \ds \lambda^{j)I} \;  .
\eea

Calculating the supersymmetry variation of the action corresponding to the 
Lagrangian (\ref{1.4}), we find that the action is invariant up to terms which contain either a derivative $\partial_\mu \amet$ or a variation $\delta \amet$ of the metric. These can be combined by rewriting them as $(\frac{\partial}{\partial \sigma^K} \,  \amet) \, \partial_\mu \, \sigma^K$ and $(\frac{\partial}{\partial \sigma^K} \amet) \, \delta \sigma^K$, respectively. 
The non-vanishing terms of the variation are then of the form
\be\label{1.8}
\delta S_\text{kin} = \int \, d^5x \left( \ldots \right) \, \frac{\partial}{\partial \sigma^K} \, \amet \, .
\ee
Hence we find that the action arising from (\ref{1.4}) is 
supersymmetric if we impose the condition that $\amet$ is independent of 
$\sigma$. 

However, this is not the most general form of the 
Lagrangian if we can add further terms to the action, 
whose supersymmetry transformations cancel the terms left in (\ref{1.8}). 
This can indeed be accomplished 
by adding interactions of Chern-Simons type,  
if we require $\frac{\partial}{\partial \sigma^K} \,  \amet$ 
to be symmetric in all three indices. Then
$\amet$ can be expressed as the 
second derivative of a function $F(\sigma)$, the prepotential:
\be\label{1.6}
a_{IJ}(\sigma) = \frac{\partial}{\partial \sigma^I} \, \frac{\partial}{\partial \sigma^J} \, F(\sigma) \, .
\ee
Using the prepotential, 
we may then rewrite $\frac{\partial}{\partial \sigma^K} \,  \amet$ as 
\be\label{1.9}
F_{IJK}(\sigma) := \, \frac{\partial}{\partial \sigma^I} \, \frac{\partial}{\partial \sigma^J} \, \frac{\partial}{\partial \sigma^K}  \, F(\sigma)  \, . 
\ee
By construction, 
$F_{IJK}(\sigma)$ is totally symmetric in all indices.

Having imposed this condition, we now add the following Chern-Simons-like 
interactions to the Lagrangian: 

\be\label{1.10}
\Lag_{\text{CS}} = \left(-\frac{1}{24}\,\epsilon^{\mu\nu\lambda\rho\sigma}\,
  A^I_\mu F^J_{\nu\lambda} F^K_{\rho\sigma}
  -\frac{\ComplexI}{8}\,\lambdaB^I\gamma^{\mu\nu} F^J_{\mu\nu} \lambda^K
  -\frac{\ComplexI}{2} \bar{\lambda}^{iI}\lambda^{jJ}\,Y^K_{ij}\right) 
F_{IJK}(\sigma)\, .
\ee
Calculating the supersymmetry variation of the action corresponding to 
\be\label{1.11}
\Lag = \Lag_{\text{kin}} + \Lag_{\text{CS}},
\ee
we find that the action is invariant up to terms which are proportional to 
the fourth derivative of the prepotential:
\be\label{1.12}
\delta \left( S_\text{kin} + S_\text{CS} \right) = \int d^5x \, \left( \ldots \right) \, F_{IJKL}(\sigma) \; .
\ee
Here $F_{IJKL}(\sigma) := \frac{\partial}{\partial \sigma^I} \,  \frac{\partial}{\partial \sigma^J} \,  \frac{\partial}{\partial \sigma^K} \,  \frac{\partial}{\partial \sigma^L} \, F(\sigma)$. 
 
In order to ensure gauge invariance,   
the prepotential must be restricted to a 
{\em polynomial of degree at most 3} and this is sufficient 
to ensure that the action defined by the Lagrangian (\ref{1.11}) is 
supersymmetric.  
The crucial observation is that the Chern-Simons term 
$\epsilon^{\mu\nu\lambda\rho\sigma}\, A^I_\mu F^J_{\nu\lambda}
F^K_{\rho\sigma}$ is gauge invariant up to partial integration, only. If we 
allowed for $F_{IJK}$ to be a function of $\sigma$, 
this partial integration would generate a nontrivial term with the 
partial derivative acting on $F_{IJK}(\sigma)$. But such a term would break 
the gauge invariance of the action. This forces us to restrict the
prepotential $F$ to be a polynomial of  at most cubic degree. 
Since this restriction implies $F_{IJKL}(\sigma) = 0$, the remaining terms in 
the supersymmetry variation (\ref{1.12}) vanish identically.

As a result, we arrive at the following general 
$\cN = 2$ vector multiplet Lagrangian 
in dimension $(1,4)$:
\bea\label{1.13}
 \Lag &=& \left(-\forth\,F^I_{\mu\nu}F^{J\,\mu\nu}
  -\half\,\lambdaB^I \ds\, \lambda^J
  -\half\,\partial_\mu\sigma^I\,\partial^\mu\sigma^J
  +Y^I_{ij}Y^{J\,ij}\right)a_{IJ}(\sigma) \\
\nonumber
 && \, + \left(-\frac{1}{24}\,\epsilon^{\mu\nu\lambda\rho\sigma}\,
  A^I_\mu F^J_{\nu\lambda} F^K_{\rho\sigma}
  -\frac{\ComplexI}{8}\,\lambdaB^I\gamma^{\mu\nu} F^J_{\mu\nu} \lambda^K
  -\frac{\ComplexI}{2} \, \bar{\lambda}^{iI}\lambda^{jJ}\,Y^K_{ij}\right)
 F_{IJK} \;.
\eea
It is invariant under the supersymmetry transformations given in 
(\ref{1.7}) and 
the prepotential $F(\sigma)$ is restricted to be a polynomial of  
degree not greater than 3.

The analogous locally supersymmetric action, {\it i.e.}, the 
action of $N$ vector multiplets coupled to minimal 5-dimensional
supergravity was worked out long ago in \cite{GST}. In this case
the theory is fully determined by a {\em homogeneous} polynomial
${\cal V}(h^{\hat{I}})$
of degree 3 in $N+1$ variables, $\hat{I}=0,\ldots,N$.
The corresponding scalar manifold ${\cal M}_{\mscr{local}}$ is
still $N$-dimensional, because it is defined by the 
cubic hypersurface ${\cal V}(h^{\hat{I}}) = 1$. This is the
defining property of a 
{\em projective (or local) very special real manifold} \cite{GST,deWvP,ACDV}.
Above we found the scalar geometry of
the corresponding globally supersymmetric theories: the metric
$a_{IJ}(\sigma)$ must be the Hessian of a polynomial of degree at most three
\cite{5dgauge}. 
This can be taken as the
definition of an {\em affine very special real manifold}.
In the case of superconformal theories the cubic function must
be homogeneous \cite{VanPro,con2}. If one admits 5-dimensional
space-times with non-trivial topology or 
introduces charged fields, then
the coefficients of the cubic polynomial are subject to further
constraints \cite{ZagGun,5dgauge}.

An alternative way to derive (\ref{1.13}) would be to start with 
the locally supersymmetric action of \cite{GST} and to decouple
gravity by sending the Planck mass to infinity. Even without doing so
in detail, 
it is clear that this will give a Lagrangian of the form 
(\ref{1.13}).
In analogy to the rigid limit of 4-dimensional ${\cal N}=2$
vector multiplets (see for example \cite{it}), the rigid limit 
freezes one variable of the homogeneous cubic polynomial 
${\cal V}(h^{\hat{I}})$,
so that one is left with a general cubic polynomial 
${F}(\sigma^I)$ in 
the remaining variables. Roughly speaking, the frozen variable 
corresponds to graviphoton, which is the Abelian gauge field in the
supergravity multiplet. By inspection of \cite{GST}, one sees 
that the terms surviving the rigid limit have the form (\ref{1.13}).

Our formulation of the theory is not covariant with respect to
general coordinate transformations of the scalar manifold 
${\cal M}$, but only covariant with respect to affine transformations
$\sigma^I \rightarrow R^I_J \sigma^J + a^I$, with constant,
invertible $R^I_J$, and constant $a^I$. 
Thus the scalar fields $\sigma^I$ are {\em affine coordinates}.
In analogy to $(1,3)$-dimensional ${\cal N}=2$ vector multiplets
we will also  call them {\em special coordinates}.
There is no principal problem to reformulate the 
theory in terms of general coordinates, as has been done in 
(1,3)-dimensional case , see \cite{it,CRTV}. 
However, we prefer to work in special coordinates,
which are adapted to the symmetries of the theory,
since they are the lowest components of ${\cal N}=2$ 
vector supermultiplets. 

For completeness, we give a global (in the mathematical sense) 
characterization of
the scalar manifolds of 5-dimensional rigid 
vector multiplets: an {\em affine very special real
manifold} is a differentiable manifold equipped with (i)
a flat torsion-free connection $\nabla$, and with (ii) a
Riemannian (or, more generally, a pseudo-Riemannian) metric, which, 
when expressed in local affine
coordinates, is the Hessian of a polynomial of degree at most 3. 
The first condition ensures that the manifold is an affine
manifold, {\it i.e.}, it can be
covered with local affine coordinate systems
$\sigma^I$.  In each patch affine coordinates are characterized by
$\nabla d \sigma^I=0$, and coordinates in different patches
are related by affine transformations, as above.
Therefore the second condition makes sense.

\end{section}
\begin{section}{Dimensional reduction to 4 dimensions \label{SectDR}}
In this section we perform a standard Kaluza-Klein reduction of the 
Lagrangian \refeq{1.13} on a circle $S^1$, keeping only the massless 
modes. This corresponds to the limit where 
the $S^1$ is shrunk to zero radius, so that all excited 
Kaluza-Klein states (non-constant Fourier modes) become infinitely
heavy and decouple.
Taking the compact dimension to be either space- or time-like, we obtain 
$\cN = 2$ supersymmetric Lagrangians with Minkowskian and Euclidean signature,
respectively. We then identify the geometric structures 
underlying these theories and show that they  can be mapped to 
one another.

The section is organized as follows: We first dimensionally reduce
the bosonic terms of the Lagrangian (\ref{1.13}) to $(1,3)$ and $(0,4)$ 
dimensions and discuss the structures of the resulting scalar manifolds. We 
then determine the supersymmetry variations of the new Lagrangians, before we 
give the complete fermionic terms. Next we show how our results can be 
generalized to Euclidean theories not obtained by dimensional reduction
and display the general Lagrangian.
Finally we explain how the Abelian factor of the R-symmetry group
is related to the existence of a  complex or para-complex structure on 
the scalar manifold.

%
%
\subsection{The bosonic sector}
\begin{subsubsection}{Dimensional reduction}
We start with the dimensional reduction of  
the bosonic sector of our 5-dimensional Lagrangian (\ref{1.13}):
\begin{equation}\label{5.1}
  \begin{split}
    \mathcal{L}_{\text{bos}}^{(1,4)} =
    \left(-\forth\,F^I_{\mu\nu}F^{J\,\mu\nu}
      -\half\,\partial_\mu\sigma^I\,\partial^\mu\sigma^J
      +Y^I_{ij}Y^{J\,ij}\right)a_{IJ}(\sigma) 
    -\frac{1}{24}\,\epsilon^{\mu\nu\lambda\rho\sigma}\,
      A^I_\mu F^J_{\nu\lambda} F^K_{\rho\sigma}\;F_{IJK} \, .
  \end{split}
\end{equation}

We use the following conventions: $\mu,\nu = 0,1,2,3,5$ are 5-dimensional 
vector indices and $m,n,\dots$ are the corresponding quantities in four 
dimensions (see appendix A for a summary of conventions).  
The index $*$ takes the values $0$ or $5$ for compactifying the time-like and 
space-like coordinate, respectively.
The 5-dimensional gauge potential $A_\mu^I$ decomposes into its
4-dimensional counterpart $A_m^I$ and a new scalar field $b^I$:\footnote{
We are of course free to choose the sign and normalization of 
the scalar field $b^I$. The above choice will turn out to be convenient.}
\begin{equation}\label{5.2}
  A^I_\mu \quad\Longrightarrow\quad
  \Cases {
    \Big(A^I_m,\; b^I := A^{I5}=A^I_5\ \ \Big)\quad(1,3)\;,
  }{
    \Big(A^I_m,\; b^I := A^{I0}=-A^I_0\Big)\quad(0,4)\;.
  }
\end{equation}

\noindent 
Employing these conventions and \re{epsilonEqu}, 
the terms in \refeq{5.1} dimensionally reduce as follows:
\be\label{5.4}
\begin{array}{rclc}
  -\forth\,F^I_{\mu\nu}F^{J\,\mu\nu} & \Longrightarrow &
  \Cases{
    -\forth F^I_{mn}F^{J\,mn} - \half\, \partial_m  b^I\partial^m b^J 
    }{
    -\forth F^I_{mn}F^{J\,mn} + \half\, \partial_m  b^I\partial^m b^J 
    } & 
  \begin{array}{c} (1,3)\,, \\ (0,4)\,, \end{array} \\[1.5ex]
-\half\,\partial_\mu \sigma^I \, \partial^\mu \sigma^J  & \Longrightarrow & 
\Cases{
  -\half\,\partial_m \sigma^I \, \partial^m \sigma^J}{-\half\,\partial_m \sigma^I \, \partial^m \sigma^J} &
\begin{array}{c} (1,3)\,, \\ (0,4) \,,\end{array} \\[1.5ex]
  - \frac{1}{24} \, \epsilon^{\mu\nu\lambda\rho\sigma} 
  A^{(I}_\mu F^J_{\nu\lambda}F^{K)}_{\rho\sigma} 
  & \Longrightarrow &
  \Cases{
    + \, \frac{1}{8} \, \epsilon^{mnpq}b^{(I} F^J_{mn}F^{K)}_{pq}
    = + \, \frac{1}{4} \,b^{(I}\, \Tilde{F}^J_{mn}F^{K)\,mn}
    }{
    + \, \frac{1}{8}\, \epsilon^{mnpq}b^{(I} F^J_{mn}F^{K)}_{pq}
    = + \, \frac{1}{4} \,b^{(I}\, \Tilde{F}^J_{mn}F^{K)\,mn}} &
   \begin{array}{c} (1,3) \,, \\ (0,4) \,.\end{array}  \\ 
\end{array}
\ee
The term containing the auxiliary field $Y^{I\,ij}$ reduces trivially, since it does not contain any space-time derivatives. To obtain the result for the Chern-Simons term we integrated by parts and introduced the dual
4-dimensional field strength tensor:
\begin{equation}\label{5.6}
  \Tilde{F}_{mn} := \half\,\epsilon_{mnpq}\, F^{pq}\;.
\end{equation}

Hence the bosonic sector of the dimensionally reduced Lagrangian in 
4-dimensional Minkowski-space is then given by
\begin{equation}\label{L31R}
  \begin{split}
    \mathcal{L}_{\text{bos}}^{(1,3)} =&
    \left( -\forth F^I_{mn}F^{J\,mn} 
      -\half\,\partial_m\sigma^I \partial^m \sigma^J
      - \half\, \partial_m  b^I\partial^m b^J
    \right) a_{IJ}(\sigma)\\
    & +\forth\, b^I\, \Tilde{F}^J_{mn}F^{K\,mn} F_{IJK}
    +Y^I_{ij}Y^{J\,ij}a_{IJ}(\sigma) \; .
  \end{split}
\end{equation}
For Euclidean signature we obtain:
\begin{equation}\label{L40R}
  \begin{split}
    \mathcal{L}_{\text{bos}}^{(0,4)} =&
    \left( -\forth F^I_{mn}F^{J\,mn} 
      -\half\,\partial_m\sigma^I \partial^m \sigma^J
      + \half\, \partial_m  b^I\partial^m b^J
    \right)a_{IJ}(\sigma)\\
   & +\forth\, b^I\, \Tilde{F}^J_{mn}F^{K\,mn} F_{IJK}
   +Y^I_{ij}Y^{J\,ij}a_{IJ}(\sigma) \; .
  \end{split}
\end{equation}
As was already anticipated in the introduction, the new scalar kinetic term
obtained from dimensional reduction of a field strength has a minus sign
relative to the kinetic term of $\sigma^I$. Hence the metric of
the scalar manifold has split signature. 
\end{subsubsection}
\begin{subsubsection}{The scalar manifold in the Minkowskian case}
Let us now discuss the geometry underlying the Lagrangian \refeq{L31R}. It
is well known that the corresponding scalar manifold ${\cal M}_M$ must be an 
{\em affine
special K\"ahler manifold} \cite{Gates,ST,it,CRTV,skg,Freed,ACD}. 
We start to make this manifest by introducing complex fields
\begin{equation}\label{5.9}
  X^I := \sigma^I + ib^I\;,\quad \Xb^I := (X^I)^* = \sigma^I - ib^I \;,
\end{equation}
in terms of which the scalar kinetic term becomes
\begin{equation}\label{5.10}
    -\half\left(\partial_m\sigma^I\partial^m\sigma^J
      + \partial_m b^I\partial^m b^J
    \right)\,a_{IJ}(\sigma) =
    -\half\, \partial_m \Xb^I\partial^m X^J\,a_{IJ}(\sigma) \, .
\end{equation}
Since the matrix $(a_{IJ}(\sigma))$ is real, symmetric and positive 
definite, 
we find that the metric 
appearing in 
\refeq{5.10} is Hermitian with respect to the 
complex structure on the scalar manifold ${\cal M}_M$ specified by 
\re{5.9}. Next we construct a K\"ahler potential for the metric,
which shows that ${\cal M}_M$ is K\"ahler. 
In this course we first define a new holomorphic prepotential $F(X)$ by 
replacing the argument of the 5-dimensional real polynomial  
prepotential $F(\sigma)$ by the 
new holomorphic coordinate $X$:
\be\label{5.10a}
F(\sigma) \rightarrow F(X), \quad \mbox{by substituting 
$\sigma \rightarrow X$} \;.
\ee
Note that for this substitution to make sense, it is sufficient to assume that 
$F(\sigma)$ is real analytic, not necessarily polynomial. 
We further define
\be\label{5.17a}
  F_{IJ}(X) := \frac{\partial^2 F(X)}{\partial X^I\partial X^J} ,\quad
  \Fb_{IJ}(\Xb) :=
  \frac{\partial^2 \Fb(\Xb)}{\partial \Xb^I\partial \Xb^J}  \; ,
\ee
where $\Fb(\Xb) := \left( F(X) \right)^*$ is the complex conjugate
of $F(X)$. Since $F(\sigma)$ is of at most cubic degree, we can explicitly 
express $F_{IJ}(X)$ in terms of the real fields $\sigma^I, b^I$ by means of 
an exact first order Taylor expansion around $\sigma$:
\begin{equation}\label{5.18a}
    F_{IJ}(X) = a_{IJ}(\sigma) + i\,F_{IJK}\,b^K, \quad \Fb_{IJ}(\Xb) =
    a_{IJ}(\sigma) - i\,F_{IJK}\,b^K  \;.
\end{equation}
This relation can then be used to express $a_{IJ}(\sigma)$ in terms of $X^I$ 
and $\Xb^I$:
\be\label{5.16a}
 N_{IJ}(X,\Xb) := \half \left( F_{IJ}(X) + \Fb_{IJ}(\Xb)\right) = a_{IJ}(\sigma) \, .
\ee
We see that $N_{IJ}(X,\Xb)$ has the K\"ahler potential
\begin{equation}\label{5.21a}
  K(X,\Xb) =
    \half\,\big(F_I(X)\Xb^I + \Fb_I(\Xb)X^I\big) \, .
\end{equation}
In fact, the K\"ahler potential is not generic, because it
can be expressed in 
terms of the holomorphic prepotential $F(X)$. Therefore ${\cal M}_M$ is
not only a K\"ahler manifold, but an affine special K\"ahler manifold.  

In order to rewrite the remaining terms in \refeq{L31R} in terms of the 
complex fields \re{5.9}, 
we decompose the field strength into its selfdual and antiselfdual
parts:\footnote{In Minkowski signature the selfdual and antiselfdual
parts are complex, and they are related by complex conjugation. 
We choose $F_{+|mn}^I$ such that it is selfdual for Euclidean signature and require that  Minkowskian and Euclidean expressions take the same form. When comparing to \cite{bc,ClEtAl}
one needs to take into account that their $\epsilon$-tensor is 
defined by $\epsilon^{0123}=i$, whereas ours is defined by
$\epsilon_{0123} =1$, see also appendix A. \label{FN1}} 
\be\label{FEa}
F_{\pm|mn}^I := \frac{1}{2}\big(F_{mn}^I \pm \frac{1}{\ComplexI} \Tilde{F}_{mn}^I \big)\, .
\ee
Employing this decomposition, the gauge kinetic term and the Chern-Simons term combine to:
\be\label{5.14}
\begin{split}
&  -\forth F^I_{mn}F^{J\,mn}\,a_{IJ}(\sigma)
    + \forth\, b^I\, \Tilde{F}^J_{mn}F^{K\,mn}\,F_{IJK} \\   & \quad =  -\forth\,F^I_{-|mn}F_{-}^{J|mn}\,F_{IJ}(X)
    -\forth\,F^I_{+|mn}F_{+}^{J|mn}\,\Fb_{IJ}(\Xb)  \; .
\end{split}
\ee

This result then completes the rewriting of the Lagrangian \refeq{L31R} in term of the complex scalar fields \refeq{5.9}: 
\begin{equation}\label{5.19}
  \begin{split}
    \mathcal{L}^{(1,3)}_{\text{bos}} = &
    -\forth\,F^I_{-|mn}F_{-}^{J|mn}\,F_{IJ}(X)
    -\forth\,F^I_{+|mn}F_{+}^{J|mn}\,\Fb_{IJ}(\Xb)\\
    &-\half\, \partial_m \Xb^I\partial^m X^J\,N_{IJ}(X, \Xb)
    +Y^I_{ij}Y^{J\,ij}\,N_{IJ}(X, \Xb)
  \end{split}
\end{equation}

In order to make contact with the more recent literature, and also to the 
construction of affine special K\"ahler manifolds given in \cite{ACD}, 
we now change from the ``old conventions'' of \cite{deWvP2} to the 
``new conventions'' of \cite{DKLL} by rescaling the prepotential:
\begin{equation}\label{redM}
F^{\mscr{(new)}}(X) = \frac{1}{2 \ComplexI} F^{\mscr{(old)}}(X) \; .
\end{equation}
In terms of these conventions the Lagrangian \refeq{5.19} becomes
\begin{equation}\label{5.34}
  \begin{split}
    \mathcal{L}^{d=4}_{\text{bos}} = &
    \frac{\ComplexI}{2}\,F^I_{+|mn}F_{+}^{J|mn}\,\Fb_{IJ}(\Xb)
    -\frac{\ComplexI}{2}\,F^I_{-|mn}F_{+}^{-|mn}\,F_{IJ}(X)\\
    &-\half\, \partial_m X^I\partial^m \Xb^J\,N_{IJ}(X,\Xb)
    +Y^I_{ij} Y^{J\,ij}\,N_{IJ}(X,\Xb)\;,
  \end{split}
\end{equation}
where $N_{IJ}$ is the same as above, {\it i.e.}, 
\be N_{IJ} = i (F_{IJ} - \bar{F}_{IJ})\ee 
in terms of the new prepotential $F$, 
while the K\"ahler potential \refeq{5.21a} is replaced by:
\begin{equation}\label{5.32}
  K(X,\Xb) =
    i\big(F_I(X)\Xb^I - \Fb_I(\Xb)X^I\big)  
= \ComplexI \VecT{\Xb^I}{\Fb_I(\Xb)} \Matrix{0}{\umat}{-\umat}{0}
\Vec{X^I}{F_I(X)} \; .
\end{equation}

In order to further explain the geometrical structure of ${\cal N}=2$ vector
multiplets, let us recall that the scalar fields $X^I$ are the
components of a map $\varphi : M = \R^{1,3} \rightarrow {\cal M}_M$ 
from Minkowski space-time to an affine special K\"ahler manifold ${\cal M}_M$
with respect to a system of special local coordinates.
An intrinsic definition of affine 
special K\"ahler manifolds was given in \cite{Freed}:
An {\it affine special K\"ahler structure} on a K\"ahler manifold $(M,J,g)$ 
is a flat and  torsion-free connection $\nabla$, which satisfies 
(i) and (ii) of Definition \ref{Def9} in section \ref{SectPCGeom}. 
This definition is equivalent to the definition given in 
\cite{it,CRTV,skg}, as was shown in \cite{ACD}.  In fact, there
is a close analogy with the case of affine special para-K\"ahler manifolds, 
which was developed in section \ref{SectPCGeom}. 
Any simply connected affine special K\"ahler manifold ${\cal M}_M$ 
of complex dimension $N$ 
admits a (holomorphic) K\"ahlerian Lagrangian immersion 
\be \phi : {\cal M}_M \rightarrow
T^*\Com^N\, ,\quad p \mapsto \phi (p) = \Vec{z^I(\phi(p))}{w_I(\phi (p))} \;.
\ee 

This immersion is uniquely determined by the special K\"ahler
data $(J,g,\nabla )$ on ${\cal M}_M$ up to a complex affine transformation
of $T^*\Com^N$ with linear part in ${\rm Sp}(2N,\R)$. Here $(z^I,w_I)$ are
canonical coordinates on $T^*\Com^N = \Com^{2N}$. Up to an affine 
transformation as above, we can assume that the functions 
$\tilde{z}^I := z^I \circ \phi$ provide local holomorphic 
coordinates (called {\it special
coordinates}) 
in a neighborhood of a  
point in ${\cal M}_M$. The functions $\tilde{w}_I := w_I \circ \phi$
are then expressed in terms of 
\be \tilde{w}_I = F_I({\tilde{z}^1, \ldots , \tilde{z}^N}) \, ,\ee
where $F = F(z^1, \ldots , z^N)$ is the holomorphic prepotential, 
which locally generates the
holomorphic Lagrangian immersion.    
This shows that 
\be \Vec{X^I(x)}{F_I(X(x))} = \phi (\varphi (x)) = 
\Vec{\tilde{z}^I(\varphi (x))}{\tilde{w}_I(\varphi (x))}\, ,\ee
where $x \in M$ and $\varphi : M \ra {\cal M}_M$.

If the target manifold ${\cal M}_M$ 
is not simply connected, we can cover it by simply connected open sets
$U_\alpha$, such that we have K\"ahlerian Lagrangian immersions  
\be \phi_\alpha : U_\alpha \rightarrow
T^*\Com^N\, ,\quad p \mapsto \phi_\alpha (p) = 
\Vec{z^I(\phi_\alpha (p))}{w_I(\phi_\alpha (p))} =: 
\Vec{z_\alpha^I(p)}{w^\alpha_I(p)} \, .  
\ee
These are related by 
\be \Vec{z_\alpha^I}{w^\alpha_I} =  M_{\alpha \beta} 
\Vec{z_\beta^I}{w^\beta_I} +  v_{\alpha \beta}\, , \ee
where $M_{\alpha \beta} \in {\rm Sp}(2N,\R)$ and 
$v_{\alpha \beta} \in \Com^{2N}$. One can prove that it is possible to 
choose the $U_\alpha$ such that the
functions $x_\alpha^I := {\rm Re}\, z_\alpha^I$ and 
$y_I^\alpha := {\rm Re}\, w_I^\alpha$ define a (real) local 
affine coordinate system on $U_\alpha$ with respect to the flat
torsion-free special connection $\nabla$ \cite{ACD}. 
Then the data $(M_{\alpha \beta}, v_{\alpha \beta})$ automatically 
satisfy a cocycle condition.  
The real affine symplectic transformations 
$(M_{\alpha \beta}, {\rm Re}\, v_{\alpha \beta})$ are the transitions 
between the $\nabla$-affine coordinate systems on $U_\alpha$ and $U_\beta$.   
Note that the linear parts $(M_{\alpha \beta})$ are the   
constant transition functions of the tangent bundle endowed with the flat
connection $\nabla$.   

The flat torsion-free connection $\nabla$ is part of the intrinsic definition 
of an affine special K\"ahler manifold ${\cal M}_M$. 
Any such manifold admits an 
$S^1$-family of such special connections $\nabla^t := e^{tJ}\circ 
\nabla \circ e^{-tJ}$, 
where $J$ is the complex structure of ${\cal M}_M$ and $\circ$ denotes 
composition, {\it i.e.}, 
\be \nabla^t_XY := e^{tJ}\nabla_X(e^{-tJ}Y)\quad \mbox{for all
vector fields}\quad X,Y\, .\ee 
If we do not fix 
the connection $\nabla$ then the immersions $\phi_\alpha$ are only unique   
up to a complex affine transformation with linear part in 
${\rm U}(1)\cdot {\rm Sp}(2N,\R)$. This explains the additional phase factor
in the transition functions of \cite{it,CRTV,skg}.

If we choose $U_{\alpha}$ sufficiently small, such that the immersion 
$\phi_\a$ is an embedding, then we can identify $U_\a$ with
its image:  $U_{\alpha} \simeq \phi_{\alpha}(U_{\alpha})
\subset T^* \Com^N$. 
The embedding $\phi_{\alpha}$ provides us with $2N$ holomorphic functions 
$({z}^I_{\alpha}, {w}_J^{\alpha})$ 
on $U_{\alpha} \subset {\cal M}_M$, such that a point $p\in U_\a$ is 
completely determined by the values of these functions at $p$. Moreover, 
by further restricting $U_\a$ if necessary, 
we can choose $N$ of these functions 
to define a global holomorphic coordinate system on $U_\a$. 
If the submanifold $\phi_{\alpha}(U_{\alpha})
\subset T^* \Com^N$ is transverse to the fibers of the
bundle $\pi : T^* \Com^N\ra \Com^N$, then the  
${z}^I_{\alpha}$ can be taken as holomorphic coordinates 
of ${\cal M}_M$, as already mentioned. 
Geometrically, this corresponds to
projecting the submanifold onto the directions of the coordinates
$z^I$ of $T^* \Com^N$ via the projection $\pi$. 
{}From this picture it is immediately clear
that there are also non-generic
situations where the submanifold has a vertical tangent vector 
at some point $p$, so that the $z^I$ are not independent, 
and cannot be used as coordinates near $p$. In field theory this corresponds
to situations where the scalar fields $X^I$ are not independent, and where
the lower components $F_J$ of the vector $(X^I, F_J)^T $ are not 
the components of the gradient of a function $F$ \cite{ExPrep1}. This is  
then called a `symplectic basis without a prepotential.' 
Therefore it is often
advantageous to work in terms of the
embedding coordinates (also called symplectic vectors)
$(\tilde{z}^I, \tilde{w}_J)$ $(X^I, F_J)$, {\it etc.},  
which are provided naturally by the extrinsic
construction. Alternatively, one can always perform 
a linear symplectic transformation of $T^* \Com^N$ which 
makes the immersion transverse to the fibers of $\pi : T^* \Com^N\ra \Com^N$ 
and the upper components $\tilde{z}^I$ independent, in a neighborhood of 
a given point $p \in U_\a$. 
In other words, 
there is always a symplectic basis where a prepotential exists 
\cite{CRTV,ExPrep2}.

We now turn to the gauge fields. As is well known,
the ${\rm Sp}(2N,\R)$ transformations act on them as 
generalized electric-magnetic duality transformations \cite{dWvP}. 
In fact, these duality rotations are responsible for the
additional geometric structures on 
the scalar manifold ${\cal M}_M$, which make it special K\"ahler
rather than just K\"ahler.
To specify the action of ${\rm Sp}(2N,\R)$ on the gauge fields
one defines dual gauge fields by $G_{-I|mn} := F_{IJ}
F^J_{-|mn}$ and $G_{+I|mn} := G_{-I|mn}^*$. Then the combined
Bianchi identities and Euler-Lagrange equations take the form
\be
\der^m \left( \begin{array}{c}
F^I_{+|mn} - F^I_{-|mn} \\
G_{+J|mn} - G_{-J|mn} \\
\end{array} \right)= 0\, .
\ee
As such, the equations are invariant under ${\rm GL}(2N,\R)$. But
the fields $G_{\pm I|mn}$ are not independent of the
$F^I_{\pm | mn}$. In fact, they are completely determined by
the gauge fields and the scalars, and therefore transformations
of the $G_{\pm I|mn}$
must be induced by transformations of the independent fields.
Moreover, we are working within the Lagrangian formulation of
field theory. Hence the transformed
equations are required to be the Bianchi identities and 
Euler-Lagrange equations of a Lagrangian of the form 
(\ref{5.34}). For a generic prepotential $F$ this implies 
that one can only  
make ${\rm Sp}(2N,\R)$ transformations (or rescalings, which are 
not interesting because they do not mix different gauge fields). 
Moreover $F_{IJ}$ must also transform under ${\rm Sp}(2N,\R)$,
with a transformation law, which is precisely induced by 
a linear symplectic transformation
of the vector $(X^I, F_J(X))^T$. 
The action of ${\rm Sp}(2N,\R)$
extends to the full ${\cal N}=2$ supersymmetric equations of motion, 
including the fermions, if one defines the dual gauge fields
as $G_{-I|mn} = \delta S/\delta F^{-I|mn} = F_{IJ} 
F^{J}_{-|mn} + O_{-I|mn}$, where $O_{-I|mn}$ are those fermionic terms
in the action which couple to $F^I_{-|mn}$. 

Notice that 
$(F^I_{-|mn}, G_{- J|mn})^T$ is of the form 
$(V^I, F_{JK}V^K)^T$ and, hence, 
is tangent, along the map $\phi_\alpha \circ \varphi$, to the Lagrangian 
submanifold $\phi_{\alpha}(U_{\alpha})
\subset T^{*} \Com^N$,   
defined by the prepotential $F$.  
In other words, for fixed $x \in M$, 
the vector $(F^I_{-|mn}(x), G_{- J|mn}(x))^T$ 
is tangent to the submanifold 
$\phi_{\alpha}(U_{\alpha}) \subset T^{*} \Com^N$ at the point 
$\phi_{\alpha}(\varphi(x))$. It corresponds to a tangent vector
of ${\cal M}_M$ at $\varphi (x)\in U_{\alpha} \subset {\cal M}_M$. 
The corresponding  
vector field is a local section of the bundle $\varphi^*(T {\cal M}_M) \ra M$, 
the pullback by $\varphi : M \ra {\cal M}_M$ of the tangent bundle
$T {\cal M}_M \ra {\cal M}_M$. 

Recall that we are assuming that 
the holomorphic functions $z^I$, which correspond to the scalar fields $X^I$, 
are independent on the submanifold 
$\phi_{\alpha}(U_{\alpha}) \subset T^{*} \Com^N$ and therefore provide 
holomorphic coordinates $z_\a^I = z^I \circ \phi_\a$ on  $U_\a \subset 
{\cal M}_M$.  
 The components of the tangent 
vector $(F^I_{-|mn}(x), G_{- J|mn}(x))^T$ with respect to the coordinate 
vector fields $\partial/\partial z_\a^I$ are precisely the 
$F^I_{-|mn}(x)$, $I = 1,\ldots, N$.

The vector $(F^I_{+|mn}(x), G_{+ J|mn}(x))^T \in T^{*} \Com^N$ is 
perpendicular to the  submanifold 
$\phi_{\alpha}(U_{\alpha}) \subset T^{*} \Com^N$ at the point 
$\phi_{\alpha}(\varphi(x))$ with respect to the canonical pseudo-Hermitian
metric $\gamma := i\O (\cdot , \bar{\cdot})$, defined by the
complex symplectic form $\O$ and the complex conjugation on  $T^{*} \Com^N$. 
This follows from the fact that $(F^I_{+|mn}(x), G_{+ J|mn}(x))^T$ 
is the complex conjugate of the tangent vector 
$(F^I_{-|mn}(x), G_{- J|mn}(x))^T$. 

\end{subsubsection}
\begin{subsubsection}{The Euclidean scalar manifold in terms of adapted coordinates}

In the Euclidean signature case (\ref{L40R}) we can perform an analogous construction. Here we define real fields $X_+^I$ and $X_-^I$:
\begin{equation}\label{5.11}
  X_+^I := \sigma^I + b^I\;,\quad X_-^I := \sigma^I - b^I\;.
\end{equation}
As these correspond to the adapted coordinates $z_{\pm}^i$ introduced in
section \ref{SectPCGeom}, we will refer to them as ``adapted coordinates.''
The scalar kinetic term now takes the form
\begin{equation}\label{5.12}
  -\half\left(
    \partial_m\sigma^I\partial^m\sigma^J 
    - \partial_m b^I\partial^m b^J\right) \, a_{IJ}(\sigma) =
  -\half\, \partial_m X_-^I\partial^m X_+^J \, a_{IJ}(\sigma) \; .\\
\end{equation}

In order to rewrite all other quantities appearing in the Lagrangian \refeq{L40R} in terms of adapted coordinates, we now carry out a construction similar to the one used for Minkowskian signature. In this course we first introduce new prepotentials $F^+(X_+)$ and $F^-(X_-)$ by replacing the argument of the 
real-valued polynomial prepotential $F(\sigma)$ by $X_+$ and $X_-$, respectively:
\be\label{5.10b}
F(\sigma) \rightarrow F^+(X_+), \quad \mbox{substituting: $\sigma \rightarrow X_+$}, \qquad
F(\sigma) \rightarrow F^-(X_-), \quad \mbox{substituting: 
$\sigma \rightarrow X_-$}\, .
\ee
Note that the substitution makes sense for any real-valued 
function $F(\sigma)$ and that $F^+(X_+)$ and $F^-(X_-)$ are again 
{\em real-valued} functions.  
Analogous to eq. \refeq{5.17a}, we also 
define $F_{IJ}^+(X_+)$ and $F_{IJ}^-(X_-)$ as 
\be\label{5.17b}
   F_{IJ}^+(X_+) := \frac{\partial^2 F^+(X_+)}{\partial X_+^I\partial X_+^J} ,\quad 
  F_{IJ}^-(X_-) :=   
  \frac{\partial^2 F^-(X_-)}{\partial X_-^I\partial X_-^J} \; .
\ee
Since the prepotential is a polynomial 
of degree at most 3, we can relate $F^+_{IJ}(X_+)$ 
and $F^-_{IJ}(X_-)$ to the real scalar fields $\sigma^I$ and $b^I$ by:
\begin{equation}\label{5.18}
    F_{IJ}^+(X_+) = a_{IJ}(\sigma) +  \,F_{IJK}\,b^K, \quad F_{IJ}^-(X_-) 
= a_{IJ}(\sigma) -  \,F_{IJK}\,b^K  \; .
\end{equation}
The metric $a_{IJ}(\sigma)$ may then be rewritten as
\be\label{5.16b}
   N_{IJ}(X_+,X_-) :=   \half \left( F_{IJ}^+(X_+) + F_{IJ}^-(X_-)\right) =  
a_{IJ}(\sigma) \; .
\ee
Analogously to the Minkowskian case, we also obtain a potential
\begin{equation}\label{5.22}
  K(X_+,X_-) =  \half\,\big(F_I^+(X_+) X_-^I + F_I^-(X_-) X_+^I \big)\;,
\end{equation}
which satisfies:
\be\label{5.23}
 N_{IJ}(X_+, X_-) =
  \frac{\partial^2 K(X_+,X_-)}{\partial X_+^I \partial X_-^J} \;.
\ee
Thus $K(X_+,X_-)$ is a potential
for $N_{IJ}$, which can be obtained
from a prepotential $F^+(X_+)$, which only depends on $X_+$ but not
on $X_-$. In fact, the other function $F^-(X_-)$ which enters the 
definition of $K(X_+,X_-)$ is obtained by simply replacing the argument 
of $F^+(X_+)$ by $X_-$. As will become clear later, this reflects the 
fact that the underlying para-holomorphic prepotential is not the 
most general one, but is real-valued on real points, since it comes 
from the real-valued prepotential $F(\sigma)$. A general 
para-holomorphic prepotential can be described in terms of two
independent real-valued prepotentials $F^+(X_+)$ and $F^-(X_-)$, see 
\ref{SectPCGeom}.  
  
Comparing to section \ref{SectPCGeom} we see that
the scalar manifold ${\cal M}_E$ is an 
{\em (affine) special para-K\"ahler manifold}, parametrized in 
terms of adapted real coordinates. Hence we can also parameterize 
the manifold by para-holomorphic coordinates.
This will be done in the next section, where we will also discuss
the relation to section \ref{SectPCGeom} in more detail. 

We conclude this section by rewriting the terms containing the field strength 
in terms of the adapted coordinates $X_+^I, X_-^I$. To this end we introduce 
real selfdual and antiselfdual field strength tensors according to:
\begin{equation}\label{FE}
    F^{{\rm{E}}\,I}_{\pm|mn} := \frac{1}{2}\big(F^{I}_{mn}\pm 
\Tilde{F}^{I}_{mn}\big) \; .
\end{equation}
Substituting this decomposition into the gauge kinetic term and Chern-Simons term of eq. \refeq{L40R}, we obtain:
\begin{equation}\label{5.15}
  \begin{split}
    &-\forth F^I_{mn}F^{J\,mn}\,a_{IJ}(\sigma)
    + \forth\, b^I\, \Tilde{F}^J_{mn}F^{K\,mn}\,F_{IJK}
    \\
    &  \quad = -\forth\,F^{{\rm E} \,I}_{-|mn}F_{-}^{\rE\,J|mn}\, F_{IJ}^+(X_+)
    -\forth\,F^{\rE\,I}_{+|mn}F_{+}^{\rmE\,J|mn}\,F_{IJ}^-(X_-) \; .    \\
  \end{split}
\end{equation}

Combining the terms derived in this subsection we are then in a position to write down the bosonic Lagrangian \refeq{L40R} in terms of adapted coordinates:
\begin{equation}\label{5.20}
  \begin{split}
    \mathcal{L}^{(0,4)}_{\text{bos}} = &
    -\forth\,F^{\rmE\,I}_{-|mn}F_{-}^{\rmE\,J|mn}\,F_{IJ}^+(X_+)
    -\forth\,F^{\rmE\,I}_{+|mn}F_{+}^{\rmE\,J|mn}\, F_{IJ}^-(X_-)\\
    &-\half\, \partial_m X_-^I\partial^m X_+^J\,N_{IJ}(X_+, X_-)
    +Y^I_{ij}Y^{J\,ij}\,N_{IJ}(X_+, X_-) \, .
  \end{split}
\end{equation}

\end{subsubsection}
\begin{subsubsection}{The Euclidean scalar manifold in terms of 
para-complex coordinates}
In section 2 we found that para-complex manifolds can be parametrized 
by using either adapted coordinates $z^i_\pm$ (as we did in the last
subsection) or para-complex coordinates $z^i, \bar{z}^i$. 
Having studied the adapted coordinates in the last section, we now introduce
para-complex fields 
\begin{equation}\label{5.26}
  X^I := \sigma^I+eb^I\;,\quad \Xb^I := \sigma^I - e b^I\;.
\end{equation}
Here $e$ is the para-complex unit number introduced in \ref{SectPCGeom}, 
which satisfies $e^2 = 1$, $e \neq \id$ and  $\overline{e} = - e$. 

For these fields we now carry out a construction similar to the one in
eqs. \refeq{5.10a} to \refeq{5.21a}. We first define a para-holomorphic 
prepotential $F(X)$ by:
\be\label{5.10c}
F(\sigma) \rightarrow F(X), \quad \mbox{by substituting $\sigma 
\rightarrow X$}\, .
\ee
This substitution makes sense for any real-analytic function $F(\sigma)$. 
The first and second derivatives of $F(X)$ with respect to one of its 
arguments are again denoted by $F_I(X)$ and $F_{IJ}(X)$. In the case where 
$F(\sigma)$ is a polynomial of degree at most 3, we can again set:
\be\label{5.27a}
    F_{IJ}(X) = a_{IJ}(\sigma) + e\,F_{IJK}\,b^K, \quad \Fb_{IJ}(\Xb) = a_{IJ}(\sigma) - e\,F_{IJK}\,b^K \; .
\ee
The relation to the metric on the scalar manifold ${\cal M}_E$ is given by:
\be\label{5.27b}
   N_{IJ}(X,\Xb) :=  \half \left( F_{IJ}(X) + \Fb_{IJ}(\Xb)\right) =  a_{IJ}(\sigma)
\ee
Comparing eq. \refeq{5.27b} to \refeq{5.16a}, we can then immediately write
down a para-K\"ahler potential for the metric,
\be\label{5.27d}
 K(X,\Xb) =
    \half\,\big(F_I(X)\Xb^I + \Fb_I(\Xb)X^I\big) \;.
\ee
This shows again that ${\cal M}_E$ is a para-K\"ahler manifold
and since the para-K\"ahler potential comes from a para-holomorphic
prepotential $F(X)$, it is in fact an {\em (affine) special para-K\"ahler
manifold}. Note that the potential \re{5.22} is also a  
para-K\"ahler potential, since $\frac{\partial^2}{\partial X_+^I 
\partial X_-^J} = \frac{\partial^2}{\partial X^I 
\partial \Xb^J}$. In fact, using the decomposition 
\be F(X) = \frac{1}{2}(F^+(X_+) + F^-(X_-)) + e\frac{1}{2}(F^+(X_+) - 
F^-(X_-))  \,, \ee 
one can easily check that 
both potentials coincide: $K(X_+,X_-) = K(X,\Xb)$. 

In order to completely rewrite the Lagrangian \refeq{L40R} in terms of 
para-complex fields, we also introduce para-complex (anti-)selfdual field 
strength tensors
\begin{equation}\label{5.29}
  F_{\pm|mn}^I := \half\big(F_{mn}^I \pm \frac{1}{e} \, \Tilde{F}_{mn}^I \big) 
\end{equation}
and $\Tilde{F}^I_{\pm|mn}$ defined by the analogous formula.  
These satisfy \footnote{So far, the symbol $*$ denoted the
usual complex conjugation, while para-complex
conjugation was denoted by a $\overline{\cdot}$. Here and in the following we
use $*$ to denote the para-complex conjugation, in order to
emphasize the analogy of the two geometries.}:
\begin{equation}\label{5.30}
  \Tilde{F}^I_{\pm|mn} = 
\pm e \,F^I_{\pm|mn}\;,\quad (F^I_{\pm|mn})^* = F^I_{\mp|mn}\;. 
\end{equation}

The para-complex version of the Euclidean Lagrangian \refeq{L40R} 
is:
\begin{equation}\label{5.32b}
  \begin{split}
    \mathcal{L}^{(0,4)}_{\text{bos}} = &
    -\forth\,F^I_{-|mn}F_{-}^{J|mn}\,F_{IJ}(X)
    -\forth\,F^I_{+|mn}F_{+}^{J|mn}\,\Fb_{IJ}(\Xb)\\
    &-\half\, \partial_m \Xb^I\partial^m X^J\,N_{IJ}(X, \Xb)
    +Y^I_{ij}Y^{J\,ij}\,N_{IJ}(X, \Xb) \;.
  \end{split}
\end{equation}
Comparing this result to the Minkowskian Lagrangian given in eq. \refeq{5.19}
we see that both Lagrangians take the same form when written 
in (para-)holomorphic coordinates.

Again it is useful to redefine the prepotential:
\begin{equation}\label{redE}
  F^{\mscr{(new)}}(X)  = \frac{1}{2 \, e} F^{\mscr{(alt)}}(X) \, .
\end{equation}
With this rescaling our Lagrangian \refeq{5.32b} becomes
\begin{equation}\label{5.34b}
  \begin{split}
    \mathcal{L}^{(0,4)}_{\text{bos}} = &
    \frac{e}{2}\,F^I_{+|mn}F_{+}^{J|mn}\,\Fb_{IJ}(\Xb)
    -\frac{e}{2}\,F^I_{-|mn}F_{-}^{J|mn}\,F_{IJ}(X)\\
    &-\half\, \partial_m X^I\partial^m \Xb^J \, N_{IJ}(X,\Xb)
    +Y^I_{ij} Y^{J\,ij}\,N_{IJ}(X,\Xb)\;,
  \end{split}
\end{equation}
where 
\be
N_{IJ}(X,\overline{X}) = 
\frac{\der^2 K(X,\Xb)}{
\der X^I \der {\overline{X}}^{J}
} 
\label{PKmetric}
\ee
is a para-K\"ahler metric with para-K\"ahler potential
\begin{equation}\label{5.32c}
  K(X,\Xb) = e \VecT{\Xb^I}{\Fb_I(\Xb)} \Matrix{0}{\id}{-\id}{0}
\Vec{X^I}{F_I(X)} \; .
\end{equation}

We now connect this result to the mathematical description given in
section \ref{SectPCGeom}. In this course we proceed in the same
way as when relating the Minkowskian theory to 
\cite{ACD}: the para-holomorphic scalar fields $X^I$ are
the components of a map $\varphi: E=\R^{0,4} \rightarrow
{\cal M}_E$ from Euclidean 4-space into an (affine)
special para-K\"ahler manifold ${\cal M}_M = \bigcup_{\alpha} U_{\alpha}$. This manifold can be locally immersed into the cotangent bundle
$T^{*} C^N$ of the para-complex vector space $C^N$ by
$\phi_{\alpha}: U_{\alpha} \rightarrow T^* C^N$. If the immersion
is generic, it 
induces local para-holomorphic coordinates $z^I_{\alpha} = z^I 
\circ \phi_{\alpha}$ (and also local adapted coordinates
$z^I_{\pm | \alpha}$) and we have $X^I(x) = z^I_{\alpha} \circ \varphi(x)$
(and $X^I_{\pm}(x) = z^I_{\pm | \alpha} \circ \varphi(x)$).

Through (\ref{PKmetric}) and (\ref{5.32c}) 
the prepotential $F(X)$ determines the para-Hermitian form
\be
N_{IJ}(z_\alpha,\zb_\alpha) dz^I_\alpha \otimes d \zb^J_\alpha 
= 
\left( 
2 \;{\rm Im} \ft{ \der^2 F}{\der z^I_\alpha \der z^J_\alpha} \right)
dz^I_\alpha \otimes d \zb^J_\alpha \, ,
\ee
which equals (up to an overall sign)\footnote{In order to obtain 
the same sign, it 
suffices to define $\g_V := -e\O (\cdot , \tau \cdot )$ instead of
$\g_V := e\O (\cdot , \tau \cdot )$, cf.\ section 2.}  the 
para-Hermitian form $\gamma = \phi_\a^*\gamma_V$ 
induced by the immersion $\phi_{\alpha}=\phi_F : U_\a \ra V = T^*C^N$,
see (\ref{Prep2Metric}).

Thus we see that all the structures of the bosonic part
of the Minkowskian theory carry over to the Euclidean theory,
by just replacing holomorphic by para-holomorphic quantities. 
In particular symplectic transformations play the same role in
both theories. A difference occurs when
one does not fix the special connection $\nabla$, but considers
families of such connections, which are generated by conjugation
with $e^{tJ}$ and $e^{tI}$. Here $J$ and $I$ are the complex and 
para-complex structure
of ${\cal M}_M$ and ${\cal M}_E$, respectively. In both cases the structure generates an Abelian 
group, which is compact for $J$, but non-compact for $I$.
This reflects itself in the symmetries of the (para-)K\"ahler
potential: while (\ref{5.32}) is invariant under 
phase transformations $(X^I, F_I)^T \rightarrow \exp({i \alpha})
(X^I, F_I)^T$, eq. (\ref{5.32c}) is invariant under
para-complex phase transformations  
$(X^I, F_I)^T \rightarrow \pm  \exp(e \alpha) (X^I,F_I)^T$. 
The corresponding
groups are ${\rm U}(1)$ and ${\rm SO}(1,1)$. In the language of
\cite{skg} this implies a change in the Abelian factor of the structure group of the affine 
bundle characterizing the special geometry. We have 
already seen that a similar replacement happens for the R-symmetry
groups of the underlying supersymmetry algebras. As we will see
in more detail in section \ref{R-Sym}, this is intimately related to the 
fact that the Euclidean scalar geometry has to be para-complex
rather than complex.
\end{subsubsection}
%
\subsection{The supersymmetry variations \label{SubSecSusyVar}}

After reducing the bosonic sector of (\ref{1.13}), let us proceed to the 
5-dimensional supersymmetry variations (\ref{1.7}). In their reduction we utilize the 
dimensional reduction of the $(1,4)$ dimensional Clifford algebra,
which was presented in section \ref{SectDRCliff}.


{}From experience with $\cN = 2$ supersymmetry, we expect that the
supersymmetry variations decompose into a ``holomorphic'' and an 
``antiholomorphic'' piece. Like the gauge fields,
the spinors $\lambda^I$ carry an index of the scalar target manifold and thus
have a geometrical interpretation as tangent vectors. One therefore 
expects that the introduction of (para-)holomorphic or adapted coordinates 
will induce a ``chiral decomposition'' of the fermions, as was already 
anticipated in section \ref{SectDRCliff}.

In order to identify the projectors of this decomposition, we consider the
bosonic part of the supersymmetry variations \refeq{1.7}. Employing
eq. \refeq{5.2}, the dimensional reduction of the scalar and vector fields
yields:
\begin{equation}\label{5.35}
 \delta A^I_m =  \half\,\Bar{\epsilon}\gamma_m\lambda^I, 
   \qquad 
 \delta \sigma^I = \frac{i}{2}\,\Bar{\epsilon}\lambda^I, 
   \qquad 
 \delta b^I = 
   \Cases{
     \half\,\epsilonB\gamma^5\lambda^I \quad(1,3) \,,
  }{
    \half\,\epsilonB\gamma^0\lambda^I \quad(0,4) \; .
  } 
\end{equation} 
Rewriting these variations in terms of the (para-)holomorphic and adapted coordinates, \refeq{5.9}, \refeq{5.26} and \refeq{5.11}, we obtain:
\bea\label{5.37}
\nonumber 
\delta X^I = \frac{i}{2} \, \epsilonB \big( \id + \gamma^5\big) \lambda^I 
\; , \qquad & 
\delta \Xb^I 
=  \frac{i}{2} \, \epsilonB \left( \id - \gamma^5 \right) 
\lambda^I \,, & (1,3), 
\\ 
\delta X^I  = \frac{i}{2} \, \epsilonB  \, \left( \umat - \ComplexI \,  e \gamma^0 \right) \lambda^{I} \; , 
\qquad &
\delta \Xb^I = \frac{i}{2} \, \epsilonB  \, \left( \umat + \ComplexI \, e \gamma^0  \right) \lambda^{I} \;, & (0,4) , 
\\ \nonumber
    \delta X_+^I =
     \frac{i}{2} \,\epsilonB\big(\id -\ComplexI \gamma^0 \big)\lambda^I, \;
\qquad &
\delta X_-^I = 
     \frac{i}{2} \,\epsilonB\big(\id + \ComplexI \gamma^0 \big)\lambda^I \,, 
& (0,4). 
\eea
This motivates introducing the following matrices:
\begin{equation}\label{5.38}
  \Gamma^{\rm M}_* := \gamma^5, \qquad
   \Gamma_* := -i \, e \, \gamma^0, \qquad 
   \Gamma^\rmE_*  := -i\gamma^0 \;.
\end{equation}
The relation between these $\gamma$-matrices and the Clifford algebra in 
the 5-dimensional theory was worked out at the beginning of 
section \ref{SectDRCliff}. 
Note that all three, $\Gamma_*^{\rm M}$, $\Gamma_*$ and $\Gamma_*^\rmE$,
square to $+\id$. Additionally,  $\Gamma_*^{\rm M}$ and $\Gamma_*^\rmE$ are
Hermitian, while $\Gamma_*$ is Hermitian with respect to the complex
structure and anti-Hermitian with respect to the para-complex structure. 
Since they anticommute with
the $\gamma$-matrices forming the respective 4-dimensional 
Clifford algebra, we can use them to
define chiral projectors:
\bea\label{5.39}
\nonumber
    \Gamma_\pm & := \half\,\big(\id\pm\Gamma_*^{\rm M} \big)\,& 
\quad(1,3)\,, \\
    \Gamma_\pm & := \half \, \big( \id \pm \Gamma_* \big) \,  & 
\quad(0,4) \,, \mbox{w.r.t. para-complex coordinates} \,,\\ 
\nonumber
    \Gamma^\rmE_\pm & := \half\,\big(\id\pm\Gamma^\rmE_*\big) 
&\quad(0,4)\,, \; \mbox{w.r.t. adapted coordinates} \: .
\eea
The relation to para-complex and adapted coordinates will become explicit
in equation \re{5.40} below. 
Using these projectors, we decompose our spinors according to
\bea\label{5.41}
\nonumber   \lambda = \lambda_+ + \lambda_-, &\qquad& \lambda_\pm := \Gamma_\pm \lambda \, \quad (1,3),(0,4) \; \mbox{w.r.t. (para-)holomorphic coordinates} \\
 \lambda  = \es_{+} + \es_{-}, &\qquad& \es_\pm := \Gamma^{\rm{E}}_\pm \lambda \,     \quad (0,4) \; \mbox{w.r.t. adapted coordinates} .
\eea
Here and henceforth we will use $\lambda_\pm$, $\epsilon_{\pm}$ to denote 
the chiral components of $\lambda$ and $\epsilon$ 
with respect to (para-)complex coordinates, while we use $\es_{\pm}$ and
$\esv_{\pm}$ for the chiral projections of $\lambda$ and $\epsilon$ 
with respect to adapted coordinates. 

Let us briefly comment on this decomposition. First we observe that in the 
Euclidean case $\Gamma_\pm$ becomes  $\Gamma_{\mp}$ under the 
combined Hermitian conjugation with respect to both the complex and
the para-complex structure, while in the Minkowskian case the
projectors are invariant: 
\begin{equation}\label{5.53}
  \big(\Gamma_\pm \big)^\dagger = \Cases {
    \Gamma_\pm\quad(1,3) \,,
    }{
      \Gamma_\mp\quad(0,4) \; .
    }
\end{equation}
Looking at the projected symplectic Majorana conditions \refeq{ChiralMink},
\refeq{ChiralEuc}, 
we see that by introducing an explicit factor $e$ into the projector
we have managed to write the reality constraint for spinors in 
a uniform way:
\be
( \lambda^i_{\pm} )^* = B \epsilon_{ij} \lambda^j_{\mp} \,, \;\;\;
B = C \gamma_0 = -i C \Gamma_*^E \;,
\ee
where in the Euclidean case `$*$' denotes simultaneous complex and
para-complex conjugation. Thus by introducing para-complex valued
chiral projections we can compensate for the fact that standard 
chiral projections in Euclidean signatures are real, in the sense that
$(\xi^i_{\pm})^* = B \epsilon_{ij} \xi^j_{\pm}$.

Of cause we also have to check that $\Gamma_\pm$ project onto complementary
subspaces for
Euclidean signature. In terms of the spinors \refeq{5.41} the Euclidean
projector $\Gamma_\pm$ has eigenvalues $(1+e)$ and $(1-e)$. This can also be
deduced
from the identification $\Gamma_* \equiv e\,\Gamma^\rmE_*$. This looks 
peculiar, but, due to $(1+e)(1-e) = 0$, the projector identity $\Gamma_\pm \,
\Gamma_\mp = 0$ still holds. At this point it is crucial that the ring of 
para-complex numbers has zero divisors, in order for $\Gamma_\pm$ being a well-defined projector.

Having established the decomposition \refeq{5.41}, we now rewrite the supersymmetry variations \refeq{5.37} in terms of the chirally projected spinors:
\begin{equation}\label{5.40}
  \begin{split}
    \delta X^I   &= i\,\epsilonB\,\Gamma_+\lambda^I = 
    i\,\epsilonB_+\,\lambda_+^I, \qquad 
    \delta \Xb^I = i\,\epsilonB\,\Gamma_-\lambda^I =
    i\,\epsilonB_-\,\lambda_-^I  \;,  \\
    \delta X_+^I   &= i\,\epsilonB\,\Gamma^\rmE_+\lambda^I =
    i\,\esvB_+\,\es_+^{I}, \qquad
    \delta X_-^I = i\,\epsilonB\,\Gamma^\rmE_-\lambda^I =
    i\,\esvB_-\,\es_-^{I} \;.
  \end{split}
\end{equation}
Here the first line holds for both signatures in terms of
complex and para-complex quantities, respectively. We further observe 
that the supersymmetry variations indeed split into a
holomorphic and antiholomorphic sector. 

It is now straightforward to reduce the remaining supersymmetry variations 
and to rewrite them in terms of (para-)holomorphic and adapted coordinates, 
respectively. We start with the 5-dimensional supersymmetry variations of the spinor fields $ \delta \lambda^{iI} = -\forth\,\gamma^{\mu\nu} F_{\mu\nu}^I\epsilon^i - \frac{i}{2}\,\ds\,\sigma^I\epsilon^i - Y^{ij\,I}\epsilon_j$. Using the identity
\begin{equation}\label{5.42}
  \gamma^{\mu\nu}F^I_{\mu\nu} = 
\left\{
  \begin{array}{ll}
    \!\gamma^{mn}F^I_{mn} + 2\gamma^m\gamma^5 \partial_m b^I
    = \gamma^{mn}F^I_{mn} + 2\gamma^m\Gamma_*^{\rm M} \, \partial_m b^I & 
(1,3) \,,\\
    \gamma^{mn}F^I_{mn} - 2\gamma^m\gamma^0 \partial_m b^I
    = \gamma^{mn}F^I_{mn} - 2 i e \gamma^m \Gamma_* \, \partial_m b^I & 
(0,4) \,,\\
    \gamma^{mn}F^I_{mn} - 2\gamma^m\gamma^0 \partial_m b^I
    = \gamma^{mn}F^I_{mn} - 2i\gamma^m\Gamma^\rmE_* \, \partial_m b^I & 
(0,4) \,, \\
  \end{array}
\right.  
\end{equation}
one observes that the reduced terms containing the scalars $\sigma^I$ and $b^I$ combine into the fields $X^I$, $\Xb^I$ and $X_+^I$, $X_-^I$ as follows:
\begin{equation}\label{5.43}
  \begin{split}
    -\half\,\gamma^m\partial_m b^I\,\Gamma_*^{\rm M} 
    -\frac{i}{2}\,\gamma^m\partial_m\sigma^I
    &= -\frac{i}{2}\,\big(\ds \Xb^I \Gamma_+ + \ds X^I\Gamma_-\big)\quad(1,3) \; ,\\
 +\frac{i e }{2}\,\gamma^m\partial_m b^I\,\Gamma_*
    -\frac{i}{2}\,\gamma^m\partial_m\sigma^I
    &= -\frac{i}{2}\,\big(\ds \Xb^I \Gamma_+ + \ds X^I\Gamma_-\big)       \quad(0,4) \; ,\\
    +\frac{i}{2}\,\gamma^m\partial_m b^I\,\Gamma^\rmE_*
    -\frac{i}{2}\,\gamma^m\partial_m\sigma^I
    &= -\frac{i}{2}\,\big(\ds X_-^I \Gamma^\rmE_+ + \ds X_+^I\Gamma^\rmE_-\big)\quad(0,4) \; .\\
  \end{split}
\end{equation}

\noindent 
The dimensional reduction of the auxiliar field $Y^{I\,ij}$ is trivial. The only change that occurs in its supersymmetry variations is $\ds = \gamma^\mu \, \partial_\mu \longrightarrow \ds = \gamma^m \, \partial_m$, since the spinors $\lambda^i$ no longer depend on the coordinate $x^*$. Using the identity \refeq{A.15} one can further check that  $\gamma^{mn} F^I_{mn}$ can be decomposed into (anti-)selfdual terms according to
\begin{equation}\label{5.49}
  \gamma^{mn}F^I_{mn} = \gamma^{mn} F^{I}_{-|mn} \, \Gamma_+ 
  +\gamma^{mn} F^{I}_{+|mn} \, \Gamma_- \; ,
\end{equation}
which holds for all three (anti-)selfdual field strength tensors \refeq{FEa},
\refeq{5.29} and \refeq{FE} and the corresponding projectors \refeq{5.39},
respectively.
 
We can now write down the complete dimensionally reduced supersymmetry
variations. For (para-) holomorphic fields they can be uniformly written as:
\begin{equation}\label{5.44}
  \begin{split}
    \delta X^I   &= i\,\Bar{\epsilon}_+\,\lambda^I_+ \,,\\
    \delta \Xb^I &= i\,\Bar{\epsilon}_-\,\lambda^I_- \,,\\
    \delta \lambda^{iI}_+ &= -\forth\,\gamma^{mn} F_{-\,mn}^I\epsilon^i_+
    -\frac{i}{2}\,\ds X^I\epsilon^i_-
    -Y^{ij\,I}\epsilon_{+\,j} \,,\\
    \delta \lambda^{iI}_- &= -\forth\,\gamma^{mn} F_{+\,mn}^I\epsilon^i_-
    -\frac{i}{2}\,\ds \Xb^I\epsilon^i_+
    -Y^{ij\,I}\epsilon_{-\,j} \,,\\
    \delta A^I_m &=  \half\,\Big(
     \Bar{\epsilon}_+\gamma_m\lambda_-^I
    +\Bar{\epsilon}_-\gamma_m\lambda_+^I
    \Big) \,,\\
    \delta Y^{ij\,I} &=
    -\half\,\Big(
     \Bar{\epsilon}^{(i}_+ \ds\lambda^{j)\,I}_-
    +\Bar{\epsilon}^{(i}_- \ds\lambda^{j)\,I}_+
    \Big)\;.
  \end{split} 
\end{equation} 

\noindent 
For completeness, we also give the supersymmetry variations in terms of the adapted coordinates $X_+^I,X_-^I$ and the corresponding chirally projected spinors $\es_\pm$. These read:
\begin{equation}\label{MS}
  \begin{split}
    \delta X_+^I   &= i\,\esvB_+\,\es^{I}_+ \,,\\
    \delta X_-^I &= i\,\esvB_-\,\es^{I}_- \,,\\
    \delta \es^{iI}_+ &=
    -\forth\,\gamma^{\rmE\,mn} F_{-|mn}^{\rmE\,I}\esv^{i}_+
    -\frac{i}{2}\,\ds X_+^I \esv^{i}_-
    -Y^{ij\,I}\esv_{+\,j} \,,\\
    \delta \es^{iI}_- &=
    -\forth\,\gamma^{\rmE\,mn} F_{+|mn}^{\rmE\,I}\esv^{i}_-
    -\frac{i}{2}\,\ds X_-^I\esv^{i}_+
    -Y^{ij\,I} \esv_{-\,j} \,,\\
    \delta A^I_m &=  \half\,\Big(
     \esvB_+ \gamma_m \es_-^{I}
    +\esvB_- \gamma_m \es_+^{I}
    \Big) \,,\\
    \delta Y^{ij\,I} &=
    -\half\,\Big(
     \esvB^{(i}_+ \ds\es^{j)\,I}_-
    +\esvB^{(i}_- \ds\es^{j)\,I}_+
    \Big) \; .
  \end{split}  
\end{equation}
\subsection{The fermionic sector}
We now turn to the fermionic part of the 
5-dimensional Lagrangian \refeq{1.13} 
\begin{equation}\label{5.57}
  \mathcal{L}^{(1,4)}_{\text{ferm}} = 
   -\half\,\lambdaB^I \ds\, \lambda^J\,a_{IJ}(\sigma)
     -\frac{i}{8}\,\lambdaB^I\gamma^{\mu\nu} F_{\mu\nu}^J\,\lambda^K F_{IJK}
  -\frac{i}{2}\,\bar{\lambda}^{iI}\lambda^{jJ}\,Y^K_{ij}\,F_{IJK} \, .
\end{equation} 
With the results of the previous section
is now straightforward to reduce \refeq{5.57}. The corresponding terms become:
\bea\label{5.57z}
 - \half \lb^I \ds \lambda^J \, a_{IJ}(\sigma)
&\longrightarrow& - \half \lb^I \ds \lambda^J \, a_{IJ}(\sigma) = - \half \lb^I \ds \lambda^J \, a_{IJ}(\sigma) - \frac{1}{4} \lb^I (\ds \, \sigma^K) \lambda^J \, F_{IJK} \\
\label{5.57a}  -\frac{i}{8}\,\lambdaB^I\gamma^{\mu\nu} F_{\mu\nu}^J\,\lambda^K\,F_{IJK} 
  &\longrightarrow&
\left\{ 
\begin{array}{cc}
    -\frac{i}{8}\,\lambdaB^I\gamma^{mn} F_{mn}^J\,\lambda^K\,F_{IJK} -
    \frac{i}{4}\,\lambdaB^I\ds b^J \Gamma_*^{\rm M} \lambda^K\,F_{IJK} & (1,3) 
\\
     -\frac{i}{8}\,\lambdaB^I\gamma^{mn} F_{mn}^J\,\lambda^K\,F_{IJK} -
    \frac{e}{4}\,\lambdaB^I\ds b^J \Gamma_* \lambda^K\,F_{IJK} & (0,4)
\\
    -\frac{i}{8}\,\lambdaB^I\gamma^{mn} F_{mn}^J\,\lambda^K\,F_{IJK} -
    \frac{1}{4}\,\lambdaB^I\ds b^J \Gamma^{\rm E}_* \lambda^K\,F_{IJK} & (0,4) 
\end{array}
\right.
\eea 
The last term in \refeq{5.57} does not change its form. To obtain the
reduction of the first term, observe that the piece proportional to $F_{IJK}$
vanishes identically by virtue of eq. (\ref{A.2}). It turns out, however, that
this contribution is needed in order to recast the action in a completely
(anti-)holomorphic form. The sign of this term has no invariant meaning,
as changing it can always be compensated by a spinor rearrangement. 
We choose `$-$', since in 
this case the resulting expression on the r.h.s. of \refeq{5.57z}
involves the covariant Dirac operator with respect to the
Levi-Civita connection.

We now rewrite the dimensionally reduced terms using 
the (para-)complex and adapted coordinates. Here we find that the terms containing $\ds \sigma^I$ and $\ds b^I$ can be combined as follows: 
\bea\label{5.59}
  -\forth\,\lambdaB^I \ds\sigma^J \lambda^K\,F_{IJK}
  -\frac{i}{4}\,\lambdaB^I\ds b^J\Gamma_*^{\rm M} \lambda^K\,F_{IJK} 
 & = 
  -\forth\,\Bar\lambda^I\big(\ds F_{IJ}\big)\Gamma_+\lambda^J
  -\forth\,\Bar\lambda^I\big(\ds \Bar{F}_{IJ}\big)\Gamma_-\lambda^J & (1,3) \; , \\ \nonumber
 -\forth\,\lambdaB^I \ds\sigma^J \lambda^K\,F_{IJK}
  -\frac{e}{4}\,\lambdaB^I\ds b^J\Gamma_* \lambda^K\,F_{IJK} 
  &  = -\forth\,\Bar\lambda^I\big(\ds F_{IJ}\big)\Gamma_+\lambda^J
  -\forth\,\Bar\lambda^I\big(\ds \Bar{F}_{IJ}\big)\Gamma_-\lambda^J & (0,4)_{PC} \; ,
\\
 \nonumber   -\forth\,\lambdaB^I \ds\sigma^J \lambda^K\,F_{IJK}
  -\frac{1}{4}\,\lambdaB^I\ds b^J\Gamma_*^\rmE \lambda^K\,F_{IJK} 
  &  = 
  -\forth\,\Bar\lambda^I\big(\ds F_{IJ}\big)\Gamma_+^\rmE\lambda^J
  -\forth\,\Bar\lambda^I\big(\ds \Bar{F}_{IJ}\big)\Gamma_-^\rmE\lambda^J & (0,4)_{AC} \;.
\eea
The final form of the fermionic sectors of the dimensional reduced Lagrangians
it then obtained by decomposing the field strength according to \refeq{5.49}
and introducing the chiral spinors \refeq{5.41}. Again the expressions
are independent of space-time signature when using (para-)holomorphic
coordinates. Combining them with
the bosonic terms (\ref{5.19}) in the Minkowskian or 
in the Euclidean case we find the following Lagrangian, which 
applies to both signatures: 
\begin{equation}\label{Lag:comp}
  \begin{split}
  \mathcal{L}^{d=4} =&
    -\frac{1}{4}\,\Big(
      F^I_{-|mn}F_{-}^{J|mn}\,F_{IJ}(X)
      + F^I_{+|mn}F_{+}^{J|mn}\,\Fb_{IJ}(\Xb)
    \Big)\\
    &- \, \half\, \partial_m X^I\partial^m \Xb^J\,N_{IJ}(X,\Xb)
    +Y^I_{ij} Y^{J\,ij}\,N_{IJ}(X,\Xb)\\
    & -\half\,\Big(\lambdaB_-^I \ds\, \lambda_+^J
                 +\lambdaB_+^I \ds\, \lambda_-^J\Big)\,N_{IJ}(X,\Xb)\\
    &-\frac{1}{4}\,\Big(\Bar\lambda_-^I\big(\ds F_{IJ}(X)\big)\lambda_+^J +
     \Bar\lambda_+^I\big(\ds \Bar{F}_{IJ}(\Xb)\big)\lambda_-^J\Big)\\
    &- \frac{\ComplexI}{8}\,\Big(
     \lambdaB_+^I\gamma^{mn} F_{-|mn}^J\,\lambda_+^K\,F_{IJK}
     + \lambdaB_-^I\gamma^{mn} F_{+|mn}^J\,\lambda_-^K\,\Fb_{IJK}
    \Big)\\
    & - \frac{\ComplexI}{2} \,\Big(
    \bar{\lambda}_+^{iI}\lambda_+^{jJ}\,Y^K_{ij}\,F_{IJK}
    + \bar{\lambda}_-^{iI}\lambda_-^{jJ}\,\Yb^K_{ij}\,\Fb_{IJK}
    \Big) \;.
  \end{split}
\end{equation}
The supersymmetry variations of this expression are given by \refeq{5.44}, 
again for both signatures. We summarize the relations between 
the Minkowskian and Euclidean fields in table
\ref{2.one}, which together with \refeq{Lag:comp} and \refeq{5.44} 
is one of the main results of this paper. 
\begin{table}[t]
\begin{tabular*}{\textwidth}{@{\extracolsep{\fill}} ccc} \hline \hline
complex coord. &  para-complex coord.  & adapted coord. \\ \hline
$X^I = \sigma^I + \ComplexI b^I$ & $X^I = \sigma^I + e b^I$ & $X_+^I = \sigma^I +  b^I$ \\
$\Xb^I = \sigma^I - \ComplexI b^I$ & $\Xb^I = \sigma^I - e b^I$ & $X_-^I = \sigma^I -  b^I$ \\
K\"ahler potential \refeq{5.21a} & para-K\"ahler potential \refeq{5.27d} &  
para-K\"ahler potential \refeq{5.22} \\
$F^I_{\pm} = \half \left( F^I \pm \frac{1}{\ComplexI} \tilde{F}^I \right) $ & 
$F^I_{\pm} = \half \left( F^I \pm \frac{1}{e} \tilde{F}^I \right)   $ &
$F^{{\rm E}\,I}_{\pm} = \half \left( F^I \pm  \tilde{F}^I \right) $ \\
$ \Gamma_{\pm} = \half \left( \umat \pm \Gamma_*^{\rm M} \right) $ &
$ \Gamma_{\pm} = \half \left( \umat \pm  \Gamma_* \right) $ &
$ \Gamma_{\pm}^{\rm E} = \half \left( \umat \pm \Gamma_*^{\rm E} \right) $ \\
$ \lambda_{\pm} = \Gamma_{\pm} \lambda $ &
$ \lambda_{\pm} = \Gamma_{\pm} \lambda $ &
$ \es_{\pm} = \Gamma_{\pm}^{\rm E} \lambda $ \\
 \hline \hline
\end{tabular*}
\caption{\label{2.one} Summary of the field definitions occurring in the Minkowskian and Euclidean version of the Lagrangian \refeq{Lag:comp}. For completeness we also summarize the corresponding field definitions in adapted coordinates.}
\end{table}
For future reference we also give the complete Lagrangian of the $(0,4)$
theory in terms of adapted real fields:
\begin{equation}\label{Lag:lcc}
\begin{split}
    \mathcal{L}^{(0,4)}_{\text{adapted}} =&
     -\frac{1}{4}\,\Big(
      F^{{\rm E}\, I}_{-|mn}F_{-}^{{\rm E}\,J|mn}\, F_{IJ}^+( X_{+} )
      + F^{{\rm E}\, I}_{+|mn}F_{+}^{{\rm E}\,J|mn}\, F_{IJ}^-( X_{-} )
    \Big)\\
    &- \, \half\, \partial_m X_+^I \partial^m X_-^J \, N_{IJ}( X_+ , X_- )
    +Y^I_{ij} Y^{J\,ij}\, N_{IJ}(X_+,X_-)\\
    & -\half\,\Big(
     \esB_-^{I} \ds\, \es_+^{J}
    +\esB_+^{I} \ds\, \es_-^{J} \Big) \,N_{IJ}(X_+,X_-)\\
    &-\frac{1}{4} \, \Big(
     \esB^{I}_- \big(\ds F_{IJ}^+\big) \es_+^{J}
    +\esB^{I}_+ \big(\ds F_{IJ}^-\big) \es_-^{J}
    \Big)\\
    &-\frac{i}{8}\,\Big(
      \esB_+^{I}\gamma^{mn} F_{-|mn}^{{\rm E}\, J} \es_+^{K} \, F_{IJK}^+
    + \esB_-^{I}\gamma^{mn} F_{+|mn}^{{\rm E}\, J} \es_-^{K} \, F_{IJK}^-
    \Big)\\
    &-\frac{i}{2}\,\Big( 
      \esB_+^{iI} \es_+^{jJ} \, Y^K_{ij} \,F_{IJK}^+
    + \esB_-^{iI} \es_-^{jJ} \, Y^K_{ij} \,F_{IJK}^-
    \Big)\;.
\end{split}
\end{equation} 
The supersymmetry variations of this Lagrangian are given in \refeq{MS}.

Finally we can express the (para-)holomorphic Lagrangian 
using the ``new conventions" by substituting $F^{\mscr{(new)}}(X)
= \frac{1}{2 \Imag} \, F^{\mscr{(old)}}(X)$ in \refeq{Lag:comp},
where $\hat{\rm{i}}=i$ for Minkowskian and $\hat{\rm{i}}=e$
for Euclidean signature:
\begin{equation}\label{Lag:new}
\begin{split}
    \mathcal{L}^{d=4}_{\text{new}} =&
     -\frac{\Imag}{2}\,\Big(
      F^I_{-|mn}F_{-}^{J|mn}\, F_{IJ}(X)
      - F^I_{+|mn}F_{+}^{J|mn}\,\Fb_{IJ}(\Xb)
    \Big)\\
    &- \, \half\, \partial_m X^I\partial^m \Xb^J \,N_{IJ}(X,\Xb)
    +Y^I_{ij} Y^{J\,ij}\,N_{IJ}(X,\Xb)\\
    & -\half\,\Big(\lambdaB_-^I \ds\, \lambda_+^J
                 +\lambdaB_+^I \ds\, \lambda_-^J\Big)\,N_{IJ}(X,\Xb)\\
    &-\frac{\Imag}{2} \, \Big(\Bar\lambda_-^I\big(\ds F_{IJ}(X)\big)\lambda_+^J
     - \Bar\lambda_+^I\big(\ds \Bar{F}_{IJ}(\Xb)\big)\lambda_-^J\Big)\\
    &- \frac{\ComplexI\, \Imag}{4}\,\Big(
     \lambdaB_+^I\gamma^{mn} F_{-|mn}^J\,\lambda_+^K\,F_{IJK}
     - \lambdaB_-^I\gamma^{mn} F_{+|mn}^J\,\lambda_-^K\,\Fb _{IJK}
    \Big)\\
    & - \ComplexI \, \Imag \,\Big(
    \bar{\lambda}_+^{iI}\lambda_+^{jJ}\,Y^K_{ij}\,F_{IJK}
    - \bar{\lambda}_-^{iI}\lambda_-^{jJ}\, \Yb^K_{ij}\,\Fb_{IJK}
    \Big) \; .
\end{split}
\end{equation} 
Its supersymmetry
variations are given by \refeq{5.44}.

\subsection{Extension to non-cubic prepotentials \label{SectNonCubic}}

We started our construction with a 5-dimensional vector multiplet
Lagrangian and therefore we obtained 4-dimensional
Lagrangians with a cubic prepotential with purely real 
(or, in new conventions, purely imaginary) coefficients.
Since every (para-)holomorphic prepotential defines a
(para-)holomorphic Lagrangian K\"ahler immersion, all terms
in the Lagrangian (\ref{Lag:new}) maintain their geometric
meaning when we allow non-cubic (para-)holomorphic prepotentials.
For the Euclidean theory this is a consequence 
of the results of section 2 on special para-K\"ahler manifolds.

Moreover, the Lagrangian is still real. When writing down
(\ref{Lag:new}), we already anticipated that we wanted 
to replace the cubic prepotential by a general (para-)holomorphic
one. Therefore we systematically used $F_{IJK}$ and 
$\Fb_{IJK}$, despite that in dimensional reduction
$F_{IJK}$ is purely real (or, in new conventions, purely imaginary).
Note that in our formalism, which uses chiral projections of
symplectic Majorana spinors, some of the relative signs between terms 
are different
from those one would get when using 
chiral projections of Majorana spinors instead, as in \cite{DLP,bc,it}.
Using the symplectic Majorana condition it is easy to check
that the fermionic terms
in (\ref{Lag:lcc}) and (\ref{Lag:new}) are real.
For example, the fifth line of (\ref{Lag:new}) is real,
for the case of Minkowski signature, because
$(\lambdaB_+^I\gamma^{mn} F_{-|mn}^J\,\lambda_+^K\,F_{IJK})^* =
- \lambdaB_-^I\gamma^{mn} F_{+|mn}^J\,\lambda_-^K\,\Fb _{IJK}$.
Note that in the Euclidean case one has to check reality
with respect to complex and para-complex conjugation 
separately, because `reality' with respect to the combined operation 
would admit expressions proportional to $ie$, which are not
real numbers. But since (\ref{Lag:new}) and (\ref{Lag:lcc})
are equal, invariance under para-complex conjugation is guaranteed,
so that (\ref{Lag:new}) is real if (\ref{Lag:lcc}) is invariant
under complex conjugation. In adapted coordinates
$F^+(X_+)$, $F^-(X_-)$, $F^{E\;I}_{+|mn}$ and
$F^{E\;I}_{-|mn}$ are real. Therefore verifying that the fifth
line of (\ref{Lag:lcc}) is real boils down to checking the spinor 
identity $( \overline{\xi}^i_+ \gamma^{mn} \xi^k_+ \epsilon_{ki})^*
= - \overline{\xi}^i_+ \gamma^{mn} \xi^k_+ \epsilon_{ki}$, which
easily follows from the symplectic Majorana condition and the 
properties of $\gamma$-matrices under complex conjugation.

If the prepotential is generalized to 
a {\em non-cubic} (para-)holomorphic function,
then the Lagrangian (\ref{Lag:new}) is not 
invariant under the supersymmetry transformations (\ref{5.44})
any more, because 
the supersymmetry variations of terms containing the third
derivative of the prepotential yield additional terms,
which  are proportional to the fourth derivative:
$\delta F_{IJK}(X) = F_{IJKL} \delta X^L$. 
Since we want to 
allow that $F_{IJKL} \not=0$, we need to add further terms
to (\ref{Lag:new}). 
Inspection of the supersymmetry rules
(\ref{5.44}) suggests to add a four-fermion term of the form 
\be
\Delta {\cal L}_{\mscr{new}}^{d=4} = 
- \, \frac{\Imag}{6} \left( F_{IJKL} \lambdaB^{iI}_+ \lambda^{jJ}_+ 
\lambdaB^{K}_{+i} \lambda^{L}_{+j} -
\Fb_{IJKL} \lambdaB^{iI}_- \lambda^{jJ}_- 
\lambdaB^{K}_{-i} \lambda^{L}_{-j} \right) \;.
\label{extra}
\ee 
In order to verify supersymmetry of the combined 
Lagrangian 
${\cal L}_{\mscr{new}}^{d=4} + \Delta {\cal L}_{\mscr{new}}^{d=4}$ 
we use the 
4-dimensional Fierz identity (\ref{4dFierz}) and the 
symmetry properties of  spinor bilinears. Note that these
relations take the same form for both signatures,
like the Lagrangian and the supersymmetry transformation. 
It is then straightforward to verify that 
the variation of the fermions $\lambda^{iI}_{\pm}$
in $\Delta {\cal L}_{\mscr{new}}^{d=4}$ precisely cancels the
terms containing  $F_{IJKL}$ and $\Fb_{IJKL}$, which appear in the 
supersymmetry variation of (\ref{Lag:new}). Since the five-fermion
term, which is generated by the 
variation of the scalars in $\Delta {\cal L}_{\mscr{new}}^{d=4}$, 
vanishes identically, the combined Lagrangian
${\cal L}_{\mscr{new}}^{d=4} + \Delta {\cal L}_{\mscr{new}}^{d=4}$ 
is supersymmetric for every (para-)holomorphic prepotential.
This proof of supersymmetry is independent of
the space-time signature. The only property of the prepotential 
which enters the calculation is that it is a (para-)holomorphic
function, $\der F / \der \overline{X}^I =0$.

For Minkowski signature the general Lagrangian for ${\cal N}=2$
vector multiplets is of course well known. Its locally
supersymmetric version was constructed in 
\cite{DLP} 
using the superconformal tensor calculus.\footnote{
See also \cite{DVDV,DVV} for the construction of the relevant
supermultiplets and their transformation rules.} 
The rigid version 
of this Lagrangian is given explicitly in \cite{DDKV} and \cite{bc}.
An alternative derivation, based on the rheonomic method
was given in \cite{it2}, see \cite{it} for a summary.
We have compared our results, specialized to Minkowski signature,
to \cite{DLP,DDKV,bc}, who also 
work in special coordinates. Since we use symplectic
Majorana spinors, we need to rewrite $\lambda^{iI}_{\pm}$ and
$\epsilon^i_{\pm}$ in terms of 
Majorana spinors. This is briefly explained in the appendix,
see in particular (\ref{A.17}). Moreover the auxiliary field
$Y^{ij}$ is an ${\rm SU}(2)$ tensor, and therefore it is also  subject to
a non-trivial field redefinition. Apart from this one has to take
into account different (conventional) normalizations of the fields,
which means that our fields differ from the fields use in 
\cite{DDKV,bc} by constant real factors. When comparing to 
\cite{DLP}, we also need to rescale the prepotential in order 
to convert from old to new conventions and we must 
take the rigid limit of the supergravity Lagrangian given in \cite{DLP}.
Taking all these details into account, 
we find that our supersymmetry transformation rules
(\ref{5.44}) and Lagrangian (\ref{Lag:new}), (\ref{extra})
completely agree with those of \cite{DLP,DDKV,bc}.

\begin{subsection}{R-symmetry \label{R-Sym}}

In this section we will study the R-symmetry properties of our
theories. 
We focus on the Abelian factors 
${\rm U}(1)_R$ and ${\rm SO}(1,1)_R$, which are the additional
structure occurring when going from 5 to 4 dimensions. They were 
already introduced in section 3.3.
The main result of this subsection
is that the operation of these R-symmetries on 
the fermions already exhibits the 
(para-)complex structure on 
${\cal M}_M$ and ${\cal M}_E$, respectively. This implies that
the indefiniteness of the scalar kinetic terms is a consequence of
implementing the Euclidean supersymmetry algebra.\footnote{Since 
one obtains definite scalar kinetic terms in the Osterwalder-Schrader
framework \cite{Nicolai}, it is clear that supersymmetry  is implemented
in a different way in this approach. 
We intend to address this in  a future publication.} 
We will also show
that the general Lagrangian is only invariant under the
subgroup $\Zom_2 \times {\rm SU}(2)$ of the R-symmetry group.

So far, we discussed R-symmetry as a property of the
supersymmetry algebra. Since the generators $Q_{i\alpha}$ transform 
the components of a supermultiplet into one another, the fields
inherit their behaviour from them. 
The supersymmetry transformations \refeq{5.44} imply that
the members of a vector multiplet differ by one unit of 
R-charge. However, the
absolute value of the R-charge is undetermined \cite{weinbergIII}.
Following the literature
we take the natural assignment that the fermions $\lambda$ carry
the same R-charge as the supercharges, {\it i.e.}, charge $\pm 1$.
Then the gauge fields carry charge 0, while the scalars carry
charge $\pm 2$, see below.

Let us start with the $(0,4)$ theory. The transformation of the
supersymmetry generators under ${\rm SO}(1,1)_R$ is given by
\refeq{2.17a}. Since we take the fermions to carry the same
R-charge as the supercharges, this implies that their
chiral components transform as
\be\label{2.24}
\delta \es_{\pm} = \pm \phi \, \es_{\pm} \;,\;\;\;
\delta \lambda_{\pm} = \pm e \, \phi \, \lambda_{\pm} 
\ee 
under infinitesimal and 
\be\label{2.25}
\es_{\pm} \rightarrow  
\exp({\pm \phi}) \,  \es_{\pm}  \;,\;\;\;
\lambda_{\pm} \rightarrow  
\exp({\pm e \, \phi}) \, \lambda_{\pm}  
\ee
under finite R-transformations in the connected component
$SO(1,1)_0$ of the R-symmetry group $SO(1,1)_R$.\footnote{
By exponentiation of infinitesimal transformations we only
generate the connected component of $\mathbbm{1} \in SO(1,1)_R$.
The transformation in the other connected component take
the form $\es_{\pm} \rightarrow - \exp({\pm \phi}) \,  \es_{\pm}$ 
and $\lambda_{\pm} \rightarrow  +
\exp({\pm e \, \phi}) \, \lambda_{\pm}$.} 
This chiral transformation is consistent with the reality
condition \refeq{ChiralEuc}.

For the $(1,3)$ theory \refeq{2.15a} implies 
\be
\delta \lambda_{\pm} = \pm i \, \phi \lambda_{\pm} \; \Rightarrow \;
\lambda_{\pm} \rightarrow 
\exp({\pm i \, \phi}) \lambda_{\pm} \,,
\ee
so that in terms of (para-)holomorphic fields we can write
\be\label{2.25b}
\lambda_+ \rightarrow 
\exp({\Imag \phi}) \, \lambda_+ , \quad \lambda_- \rightarrow 
\exp( - \Imag \phi) \, \lambda_- \; .
\ee
Thus the compact R-symmetry group ${\rm U}(1)_R$ of the Minkowski theory is 
mapped to a non-compact group ${\rm SO}(1,1)_R$ by replacing 
$ \ComplexI \rightarrow e$ in the transformation.
Since all the fields are related to the spinors by the supersymmetry 
variations\refeq{5.44}, the R-transformations on spinors induce 
an action of the R-symmetry group on variations of the fields. This
action can be integrated to a linear action on the fields, once 
the scalar fields $X^I$ are specified, which corresponds 
to a choice of special coordinates on the target manifold. 
First we observe by looking at the supersymmetry variations of the scalars 
\refeq{5.44}, that these also transform under chiral 
rotations:
\be\label{2.26}
X^I \rightarrow \exp({ 2 \, \Imag \, \phi}) \, 
X^I, \quad \Xb^I \rightarrow \exp({- 2 \, \Imag \,  \phi}) \, \Xb^I \,.
\ee
Again we find that (para)-holomorphic fields carry a definite
R-charge. 
The supersymmetry transformations further tell us that the fields $A_m^I,
F_{\pm |mn}^{I}$, and $Y^{ij\,I}$ do not transform. 

We can now read off the behaviour 
of the Lagrangian \refeq{Lag:new} under R-symmetry. Since all terms are 
${\rm SU}(2)$-scalars, invariance with respect to this factor of the 
R-symmetry group is manifest. 
With the transformation properties under the Abelian factor at hand
we see that for a generic  
choice of the prepotential the
${\rm U}(1)_R$ (${\rm SO}(1,1)_R$) is broken to the discrete subgroup ${\Zom}_2$
acting by $\lambda_{\pm} \rightarrow - \lambda_{\pm}$.

Let us recall that we have fibrewise (para-)complex structures on 
$T{\cal M}$, on spinors ($\Gamma_*^{M}$ and $\Gamma_*$, respectively)
and on two-forms (minus the Hodge star operator). It is 
remarkable that supersymmetry 
acts chirally, in the sense that it is consistent
with the type decomposition defined by these three (para-)complex structures,
{\it i.e.}, with the decomposition into sections of type $(1,0)$ and $(0,1)$. 
In fact, supersymmetry relates 
$X^I\rightarrow \lambda_+^I 
\rightarrow F^I_{-|mn}$ and 
$\Xb^I \rightarrow \lambda_-^I \rightarrow F^I_{+|mn}$. 
This ties R-symmetry to the 
\mbox{(para-)}complex structure of ${\cal M}$. We will now show
that R-symmetry acts on the fermions by multiplication with
the (para-)complex structure of the target manifold. 

For definiteness, we consider the Euclidean theory. We assume
that the local immersion $\phi_{\alpha}$ of 
${\cal M}$ into $T^* C^N$ is generic,
so that it induces local para-holomorphic coordinates\footnote{The 
assumption that $z^I, \bar{z}^I$ form a local system
of coordinates can be avoided when working with the
symplectic vector $(\lambda^I_+, F_{IJ} \lambda^J_+)^T$, 
which is a tangent vector of the immersed manifold.} 
${z}^I_{\alpha} = z^I \circ \phi_{\alpha}$ and local
adapted coordinates $z^I_{+|\alpha} = z^I_+ \circ \phi_{\alpha}, 
z^I_{-|\alpha} = z^I_- \circ \phi_{\alpha}$. 
For notational convenience we suppress the index $\alpha$ in the
following.
The spinor fields $\lambda(x)$ are sections of 
the spinor bundle $\Pi(\Som_{SM}) \rightarrow \R^{0,4|8}$
over superspace $\R^{0,4|8}$ with
even part $E = \R^{0,4}$, which was 
introduced in subsection \ref{ExplGrass}. Using adapted coordinates
$z^I_{\pm}$ on ${\cal M}_E$, the anticommuting spinor field
$\lambda$ evaluated at the point $x\in E$ takes the form
\be
\lambda(x) = \left( 
\xi_+^I(x) \frac{\der}{\der z^I_+} +
\xi_-^I(x) \frac{\der}{\der z^I_-}
\right)_{|\varphi(x)} \in \Pi \Som_{SM} \otimes T_{\varphi(x)} {\cal M}_E \;.
\ee
Here $\varphi : E \rightarrow {\cal M}_E$ is the map
which has the scalar fields $X^I$ as its components.
In terms of adapted coordinates $z^I_{\pm}$ we have 
the decomposition $T_{\varphi(x)} {\cal M}_E
= T^+_{\varphi(x)} {\cal M}_E \oplus 
T^-_{\varphi(x)} {\cal M}_E $ of the tangent space, which can now
be identified with $\R^N \oplus \R^N$, equipped with its
standard basis, as in example 6 of section \ref{SectPCGeom}.
Therefore the para-complex structure $I$ acts on $\lambda(x)$ by
\be
I \lambda(x) = \left( \xi^I_+(x) \frac{\der}{\der z^I_+} -
\xi^I_-(x) \frac{\der}{\der z^I_-} \right)_{|\varphi(x)} \;,
\ee
which, according to (\ref{2.24}), is an infinitesimal
R-symmetry transformation. 

Alternatively we can use para-holomorphic coordinates.
This requires to work with the para-complexified tangent space,
which has the decomposition
$(T_{\varphi(x)} {\cal M}_E)^C = T_{\varphi(x)}^{(1,0)} {\cal M}_E
\oplus T_{\varphi(x)}^{(0,1)} {\cal M}_E$. We can identify
$(T_{\varphi(x)} {\cal M}_E)^C$ with $C^N \oplus C^N$, where
both summands are related by para-complex conjugation. Now 
$\lambda(x)$ takes the form 
\be
\lambda(x) = \left( \lambda^I_+(x) \frac{\der}{\der z^I}
+ \lambda^I_-(x) \frac{\der}{\der \overline{z}^I} \right)_{|\varphi(x)}\,,
\ee
and the para-complex structure acts by
\be
I \lambda(x)  = \left(e \lambda^I_+(x) \frac{\der}{\der z^I}
-e \lambda^I_-(x) \frac{\der}{\der \overline{z}^I} 
\right)_{|\varphi(x)} \;,
\ee
which, according to  (\ref{2.24}), 
again is an infinitesimal R-symmetry transformation.

As discussed in section \ref{SectPCGeom}, the para-complex
structure generates the group $G = \{ \exp(\alpha I) | \alpha
\in \R \} \simeq {\rm SO}(1,1)_0$. This group acts on the spinors
by finite R-symmetry transformations (\ref{2.25}). We also
know from section \ref{SectPCGeom} that, for all points $p\in {\cal M}_E$, 
$\{ \exp(\alpha I_p) | \alpha
\in \R \}$ is a closed subgroup of the pseudo-orthogonal group 
${\rm O}(T_p{\cal M}_E, g_p)$  
defined by the para-K\"ahler metric $g_p$ at $p$.\footnote{
This is true although 
$I_p$ itself is an anti-isometry, and therefore is not an element
of the above group.}  Since $G$ is not compact,
it follows that the metric of ${\cal M}_E$ cannot be definite. 
Moreover, the eigenspaces of $I_p$ are isotropic and of the same
dimension. This shows that the metric is of split type, {\it i.e.}, 
${\rm O}(T_p{\cal M}_E, g_p) = {\rm O}(N,N)$. The integrability
of the para-complex structure follows from the description in terms
of para-holomorphic coordinates, the para-K\"ahler condition from
the existence of the para-K\"ahler potential.

In the Minkowskian theory the discussion is analogous, but this
time both the Abelian factor of the R-symmetry group and the 
group generated by the complex structure are compact and isomorphic
to ${\rm U}(1)$. Thus we see that the differences between the scalar
geometries of the Minkowskian and Euclidean theories are 
rooted in their different R-symmetries. In particular, the fact that
the scalar metric of the Euclidean theory is indefinite  
is a consequence of the non-compactness of
its R-symmetry group and the consistency of the
R-symmetry with the scalar metric.

We also note that the linear action\refeq{2.26}  of the
R-symmetry group on the special coordinates of the scalar manifold 
coincides with the square of the scalar multiplication (with
$\exp(\Imag \phi)$)
in $\Com^N$ and $C^N$, 
respectively. Recall that the linear R-symmetry action
on special coordinates is non-canonical from a geometric point of view, 
because it is coordinate dependent. It is remarkable that it
coincides (up to the square) 
with the (para-)complex scalar multiplication, the latter 
also being defined by the choice of (para-)holomorphic coordinates.
The link between the R-symmetry operation on the coordinates $X^I$ 
and the (para-)complex structure, 
as an endomorphism of the tangent bundle of the target, 
is again provided by supersymmetry, through the variation
\refeq{5.40}, which ties $\delta X^I$ to $\lambda_+^I$. 
It may be surprising   
that although $X^I$ is a coordinate and not a tangent vector, 
it transforms linearly under R-symmetry. The reason is 
that $X^I$ is a special coordinate, which sits in the
same supermultiplet as the tangent vector $\lambda^I_+$.

Finally we remark that our results fix the ambiguities which
occur when one tries to apply the $i \rightarrow e$
substitution rule of \cite{GibGrePer} naively. As we have seen
this rule means that one has to replace the complex structure
of the scalar manifold by a para-complex structure. However,
the Lagrangian also contains factors of $i$ which have a different
origin, namely the complex structure of the spinor module.
Such factors of $i$ remain unchanged. 
Also note that the fields $\lambda(x)_{\pm}$ of the
Euclidean theory are complex as spinors, but para-complex as
tangent vectors. This explains the geometric meaning of
expressions like $e \gamma^m$, where the para-complex
unit is multiplied with a complex matrix: $e \gamma^m$ acts as 
multiplication by $e$ on 
$(T_{\varphi(x)}{\cal M}_E)^C$
and as Clifford multiplication on $\Som_{SM}$.

\end{subsection}
\end{section}
\subsection*{Acknowlegements}
V.C.\ and T.M.\ thank C.\ Herrmann for his collaboration in an
early stage of this project. We also thank
B.\ de Wit,
J.\ L\"ange,
T.\ Strobl,
U.\ Theis,
S.\ Vandoren,
P.\ van Nieuwenhuizen,
A.\ Van Proeyen,
and A.\ Wipf 
for useful discussions. This work is supported by
the DFG within the `Schwerpunktprogramm Stringtheorie.' 
F.S.\ acknowledges financial support by the 
`Studienstiftung des deutschen Volkes.'

\begin{appendix}
\section{Notations and Conventions}
In this paper we follow the conventions of Refs.~\cite{con1,con2}. 
Most of the relevant details are given in section \ref{Section3},
where we also explain the relation to \cite{AC}, which treats 
supersymmetry from the mathematicians point of view. Here we collect
several further formulae, which are needed for the calculations 
in sections \ref{Section4} and \ref{SectDR}. 

\subsubsection*{Spinor bilinears and $\gamma$ matrix identities}

We summarize our conventions for indices in table \ref{A.one}.
\begin{table}[t]
\begin{tabular*}{\textwidth}{l@{\hspace*{.5in}}l@{\hspace*{.5in}}l}\hline
  Indices & Description & Range\\\hline
  $\mu,\nu,\dots$ & space-time indices in dimension (1,4) & 0,1,2,3,5\\
  $m,n,\dots$     &  $\!\!\!\!$\Big\{\begin{tabular}{@{}ll}
      space indices in  dimension (0,4)  \\
      space-time indices in  dimension (1,3) 
    \end{tabular} & \begin{tabular}{@{}ll}
    1,2,3,5\\
    0,1,2,3
  \end{tabular}
  \\
  $i,j,\dots$     & ${\rm USp}(2)$ indices               & 1,2\\
  $I,J,\dots$     & vector multiplet labels        & 1,\dots,$N$\\
  $\alpha,\beta,\dots$ & spinor indices            & 1,2,3,4\\\hline
\end{tabular*}
\caption{\label{A.one} Summary of our index conventions.\hspace*{\textwidth}}
\end{table}
In the following $\lambda^i, \chi^i$ are anticommuting 
(Grassmann-valued) symplectic Majorana spinors in dimension (1,4),
see section \ref{ExplGrass}.
They satisfy (\ref{SymMajCon})
\be
(\lambda^i)^* = - B\lambda^j \epsilon_{ji} \;.
\ee
The matrix $B$ is defined in section 3 and satisfies
$B B^* = - \id$. 
Here, $\epsilon_{ij}$ is an antisymmetric two-by-two matrix, which we
take to be (\ref{EpsMat})
\be
(\epsilon_{ij}) =
 \left( \begin{array}{cc} 0 & 1 \\ -1 & 0 \end{array} \right)  \;.
\ee
The indices $i,j, \ldots=1,2$ are raised and lowered according 
to the so-called NW-SE convention:
$\lambda_i := \lambda^j\epsilon_{ji}$ and 
$\lambda^i = \epsilon^{ij} \lambda_j$, where $\epsilon^{ij} =
\epsilon_{ij}$ and therefore 
$\epsilon^{ik} \epsilon_{kj} = - \delta^i_j$.

We now list useful spinor identities valid in dimension
(1,4). Note that these can be re-interpreted in dimensions
(1,3) and (0,4), as discussed in section \ref{Section3}.

We define
\begin{equation}
  \gamma^{(p)} =\gamma^{\mu_1\cdots\mu_p}= 
  \gamma^{[\mu_1}\gamma^{\mu_2}\cdots\gamma^{\mu_p]} = \frac{1}{p!} 
\left( \gamma^{\mu_1}\ldots \gamma^{\mu_p} \pm \text{cyclic}\right) \;.
\end{equation}
Changing the order of spinors in a bilinear leads to the following
signs:
\begin{equation}
  \lambdaB^i\gamma_{(p)}\chi^j = t_p\,\chiB^j\gamma_{(p)}\lambda^i\;,\quad
  \left\{\begin{matrix}
      t_p = -1\quad\text{for}\quad p=2,3\\
      t_p = +1\quad\text{for}\quad p=0,1
    \end{matrix}\right.
\end{equation}
In particular this implies the useful identities
\begin{equation}\label{A.2}
  \lambdaB^{i(I}\lambda_i^{\ J)} = 0\;,\quad
  \lambdaB^{i(I}\gamma^\mu\lambda_i^{\ J)} = 0\; . 
\end{equation}
In order to check that the $(1,4)$ dimensional Lagrangian is supersymmetric, we made use of
the following Fierz rearrangements:
\begin{equation}
  \begin{split}
    \lambda_j\etaB^i &= -\frac{1}{4}\,\etaB^i\lambda_j
    -\frac{1}{4}\,\gamma_\mu\big(\etaB^i\gamma^\mu\lambda_j\big)
    +\frac{1}{8}\,\gamma_{\mu\nu}\big(\etaB^i\gamma^{\mu\nu}
        \lambda_j\big)\;,\\
    \lambdaB^{[i}\eta^{j]} &= -\half\,\lambdaB\eta\,\epsilon^{ij} \;.
  \end{split}
\label{A.6}
\end{equation}
Using the Clifford algebra, the following identities can be obtained:
\begin{equation}
  \gamma^{(q)}\gamma_{(p)}\gamma_{(q)} = c_{p,q}\gamma_{(p)}\;,\qquad\quad
   \begin{tabular}{c}
     \begin{tabular}{|c|r|r|}\hline
       $c_{p,q}$ & $q=1$ & $q=2$\\\hline
       $p=0$ &  5& -20 \\\hline
       $p=1$ & -3&  -4 \\\hline
       $p=2$ &  1&   4 \\\hline
       $p=3$ &  1&   4 \\\hline
     \end{tabular}\\
   \end{tabular}
 \end{equation}
The totally antisymmetric $\epsilon$-symbols are defined as
\begin{equation}\label{epsilonEqu}
  \begin{split}
    \epsilon_{01235} &= -1 = -\epsilon^{01235}\;,\quad (1,4)\\
    \epsilon_{0123} &= 1 = -\epsilon^{0123}\;,\quad (1,3)\\
    \epsilon_{1235} &= 1 = +\epsilon^{1235}\;,\quad (0,4)
  \end{split}
\end{equation}
These satisfy the following contraction identity ($t+s=n$):
\begin{equation}
  \epsilon_{\rho_1\dots\rho_p\mu_1\dots\mu_q}
  \epsilon^{\rho_1\dots\rho_p\nu_1\dots\nu_q}
  = (-1)^t\,p!q!\,
  \delta_{[\mu_1}^{[\nu_1}\dots\delta_{\mu_q]}^{\nu_q]} \;.
\end{equation}
In even dimensions, we can relate $\gamma_{(p)}$ to $\gamma^{(n-p)}$ 
by the following identity 
\begin{equation}\label{A.15}
  \gamma_{\mu_1\dots\mu_p} = \frac{i^{n/2+t}}{(n-p)!}\,
  \epsilon_{\mu_1\dots\mu_n}\,\Gamma_*\,\gamma^{\mu_n\dots\mu_{p+1}} \;.
\end{equation} 
Here, $\Gamma_*$ is proportional to the product of all gamma matrices
\begin{equation}\label{A.16}
  \Gamma_* = (-i)^{n/2+t}\,\gamma_0\dots\gamma_{n-1}\;,\quad
  \Gamma_*\Gamma_* = \id\;.
\end{equation}
Note that $\Gamma_*$ defined here is $-\Gamma^\rmE_*$ as used in the 
main part of the paper.

\subsubsection*{Verification of supersymmetry in 4 dimensions}

To verify supersymmetry for a general (para-)holomorphic
prepotential  in 4 dimensions, we need to use the symmetry
properties of spinor bilinears and a 4-dimensional Fierz
identity. Both follow from the corresponding 5-dimensional
expressions by dimensional reduction (using the formulae
derived in sections 3.3 and 5.2).
{}From the 5-dimensional Fierz identity we obtain
4-dimensional Fierz identity
\bea
\label{4dFierz}
\lambda^j_{\pm} \overline{\eta}^i_{\pm} &=&
- \ft12 ( \overline{\eta}^i_{\pm} \lambda^j_{\pm}) \Gamma_{\pm}
+ \ft18 (\overline{\eta}^i_{\pm} \gamma^{mn} \lambda^j_{\pm} )
\gamma_{mn} \;, \nonumber \\
\lambda^j_{\mp} \overline{\eta}^i_{\pm} &=&
- \ft12 ( \overline{\eta}^i_{\pm} \gamma^m \lambda^j_{\mp}) 
\gamma_m \Gamma_{\mp} \;,
\eea
which takes the same form for both signatures ($\Gamma_{\pm}$
are the projectors adapted to the (para-)holomorphic
parametrization, see section 5.2). Moreover, the symmetry
properties of spinor bilinears are the same as in 5 dimensions.
This shows that all formulae needed to verify supersymmetry
in 4 dimensions can be written in a form which is 
independent of the signature.

\subsubsection*{Majorana and symplectic Majorana spinors in dimension
(1,3)}

While we use symplectic Majorana spinors in this paper, most of the literature
on ${\cal N}=2$ supersymmetry in dimension (1,3) uses Majorana spinors.
Therefore we review how both formulations are related.
In 5 dimensions $\sigma = -1$ and 
$\tau = 1$ is the only consistent choice in (\ref{T1}). But in 
$n=(1,3)$ one can find matrices $\CC_+$ and $\CC_-$ which 
satisfy (\ref{T1}) with 
$\sigma (\CC_+) = -1$, $\tau (\CC_+) = -1$ and 
$\sigma (\CC_-) = -1$, $\tau (\CC_-) = +1$. They are related by 
$\CC_+ = \CC_- \gamma_5$. Only the representation $\CC_-$ can consistently 
be ``lifted'' to the 5-dimensional Clifford algebra. Using 
$B^{\rm T}_\pm \equiv \CC_\pm A^{-1}$ one finds $B^*_+ B_+ =  \umat$, 
while $B^*_- B_- =  - \umat$. Hence the ``+''-representation admits Majorana 
spinors, $\Omega^* = B_+ \Omega $, while the ``$-$''-representation allows 
symplectic Majorana spinors, $(\lambda^i)^* = \epsilon_{ij} B_- \lambda^j$, 
only. A symplectic Majorana spinor $(\lambda^1, \lambda^2)^T$ can be written in
terms of Majorana spinors $\Omega^{(1)}, \Omega^{(2)}$ as
\bea
\label{A.17}
\lambda^1 &=& \Omega^{(1)} - i \Omega^{(2)} \nonumber \\
\lambda^2 &=& - B^*_- ( \Omega^{(1)} + i \Omega^{(2)} ) 
\eea
This formula has been obtained
by using the relation (\ref{SympMaj2Dirac}) between symplectic Majorana and Dirac spinors and then decomposing
the Dirac spinor into two Majorana spinors. 
The transformation (\ref{A.17}) is not the most general one:
it is possible to rescale the $\Omega^{(i)}$ 
by an overall real factor and to apply a real 
rotation to  the vector $(\Omega^{(1)}, \Omega^{(2)})^T$. 
In fact, in order to relate our results to those of
\cite{DLP,DDKV,bc}, one needs to use both a rotation and a 
rescaling.

{}From (\ref{A.17}) one easily derives the relation between
the chirally  projected spinors $\lambda^i_{\pm}$ and
$\Omega^{(i)}_{\pm}$. Note that in the conventions of
\cite{DLP,DDKV,bc} the chirality is encoded in the position
of the ${\rm SU}(2)$ index. They denote the right- and left-handed
projections of $\Omega^{(i)}$ by
$\Omega_i$ and $\Omega^i$, respectively.

%
%
\end{appendix}

\end{document}